\documentclass[12pt]{article}
\usepackage{epsfig}
\newcommand{\pp}{\phantom{-}}
\newcommand{\IP}{\mbox{I}\!\mbox{P}}
\newcommand{\Li}{\mbox{Li}_2}
\newcommand{\Ds}{\displaystyle}
\newcommand{\Ss}{\scriptstyle}
\newcommand{\slashp}{p \hspace{-2mm} /}
\newcommand{\slashs}{s \hspace{-2mm} /} 
\newcommand{\slashq}{q \hspace{-2mm} /}

\newcommand{\Strutb}{\rule{0in}{4ex}}

\newenvironment{Eqnarray}{\arraycolsep 0.14em \begin{eqnarray}}{\end{eqnarray}}
\newenvironment{Eqnarray*}{\arraycolsep 0.14em \begin{eqnarray*}}
{\end{eqnarray*}}
\renewcommand{\theequation}{\mbox{\arabic{equation}}}
\newcounter{saveeqn}
\newcommand{\alpheqn}{\setcounter{saveeqn}{\value{equation}}
\stepcounter{saveeqn}\setcounter{equation}{0}%
\renewcommand{\theequation}{\mbox{\arabic{saveeqn}.\alph{equation}}}}
\newcommand{\reseteqn}{\setcounter{equation}{\value{saveeqn}}%
\renewcommand{\theequation}{\mbox{\arabic{equation}}}}

\begin{document}
\thispagestyle{empty}
\begin{flushright}
        MZ-TH/99-10\\
        hep-ph/0101322\\
        August 2001\\
\end{flushright}
\vspace{0.5cm}
\begin{center}
 {\Large\bf Complete angular analysis of polarized top}\\[.3cm]
 {\Large\bf decay at \protect \boldmath $ O(\alpha_s) $}\\[0.7cm]
 {\large M.~Fischer, S.~Groote, J.G.~K\"orner and M.C.~Mauser}\\[0.7cm]
         Institut f\"ur Physik, Johannes-Gutenberg-Universit\"at\\[.2cm]
         Staudinger Weg 7, D-55099 Mainz, Germany
\end{center}


\begin{abstract}
\noindent We calculate the full $ O(\alpha_s) $ radiative corrections to the
 three spin independent and five spin dependent structure functions that
 describe the angular decay distribution in the decay of a polarized top quark
 into a $ W $-boson (followed by the decay $ W^{+} \rightarrow l^{+} + \nu_l $
 or by $ W^+ \rightarrow \bar q + q $) and a bottom quark. The angular decay
 distribution is described in cascade fashion, i.e. the decay $ t(\uparrow)
 \rightarrow W^+ +X_b $ is analyzed in the top rest system while the subsequent
 decay $ W^+ \rightarrow l^+ + \nu_l $ (or $ W^+ \rightarrow \bar{q} + q $) is
 analyzed in the $ W $ rest frame. Since the structure function ratios depend
 on the ratio $ m_W/m_t $ we advocate the use of such angular decay measure%
 ments for the determination of the top quark's mass.  Our results for the
 eight $ O(\alpha_s) $ integrated structure functions are presented in analy%
 tical form keeping the mass of the bottom quark finite. In the limit
 $ m_b \rightarrow 0 $ the structure function expressions reduce to rather
 compact forms. We also present results on the $ m_b = 0 $ unpolarized and
 polarized $ O(\alpha_s) $ scalar structure functions relevant to the semi%
 inclusive decay of a polarized top quark into a charged Higgs boson
 $ t(\uparrow) \rightarrow H^{+} + X_{b} $ in the Two Higgs Doublet Model when
 $ m_b = 0 $ .
\end{abstract}

\newpage


\section{\bf Introduction} 

 In the decay of an unpolarized or polarized top quark to the $ W $-gauge boson
 and a bottom quark the $ W^{+} $ is strongly polarized, or, phrased in a 
 different language, the $ W^{+} $ has a nontrivial spin density matrix.
 Furthermore, the spin density matrix of the $ W $ can be tuned by changing the
 polarization of the top quark. The polarization of the $ W^{+} $ will reveal
 itself in the angular decay distribution of its subsequent decays $ W^{+}
 \rightarrow l^{+} + \nu_{l} $ (or $ W^{+} \rightarrow \bar{q} + q $)
 \footnote{
 From this point on we shall drop explicit reference to the $ W^{+} \rightarrow
 \bar{q} + q $ decay channel since it has the same angular decay distribution as
 $ W^{+} \rightarrow l^{+} + \nu_{l} $. In fact the branching fraction into the
 two hadronic channels $ (\bar{d}+u) $ and $ (\bar{s}+c) $ exceeds that of the
 sum of the three leptonic channels by a factor of approximately two because of
 the colour enhancement factor. Although not explicitly mentioned further on, the
 existence of the hadronic decay mode of the $ W^{+} $ is always implicitly 
 assumed in the following.}.

 In the first stage one will aim to analyze the decay of unpolarized top quarks
 (or average over its polarization). The decay distribution of unpolarized top
 quark decay is governed by three structure functions which we shall refer to as
 $ H_U $ (``unpolarized-transverse''), $ H_L $ (``longitudinal'') and $ H_F $
 (``forward-backward-asymmetric''). In fact, the CDF collaboration has already
 presented some results on the measurement of the longitudinal component of the
 $ W $ based on the limited RUN I data~\cite{cdf}. The measurement has confirmed
 the expected dominance of the longitudinal mode.
 The error on this measurement is quite large ($ \approx 45 \, \% $) but is
 expected to be reduced significantly during RUN II at the TEVATRON to start in
 the spring of 2001. In RUN II one will produce $ (5 - 6) \times 10^3 $ top quark
 pairs per year and detector.
 This number will be boosted to $ 10^7 - 10^8 $ top quark pairs per year and
 detector at the LHC starting in 2006/2007. It is conceivable that the errors on
 the structure function measurements can be reduced to the $ 1-2 \, \% $ level
 in the next few years~\cite{willenbrock}. If such an accuracy can, in fact, be
 achieved and, having in mind that the $ O(\alpha_s) $ corrections to the top
 decay rate amount to $ 8.5 \, \% $~\cite{c14,c15,c16,c17,c18,ghinculov}, it is
 quite evident that one needs to improve on the known theoretical Born level
 predictions for the above three structure functions by calculating their
 next-to-leading order radiative corrections. 

 At a later stage, when the data sample of polarized top quarks has become
 sufficiently large, one will be able to also analyze the decays of polarized
 top quarks. The top quark is very short-lived and therefore retains its full
 polarization content when it decays. Polarized top decay brings in five
 additional polarized structure functions which can be measured through an
 analysis of spin-momentum correlations between the polarization vector of the
 top quark and the momenta of its decay products. 

 Polarized top quarks will become available at hadron colliders through single
 top production which occurs at the $ 33 \, \% $ level of the top quark pair
 production rate~\cite{c7}. Future $ e^{+} e^{-} $ colliders will also be
 copious sources of polarized top quark pairs~\cite{c1,c2,c3,c4,c5,c6}. For
 example, at the proposed TESLA collider one expects rates of $ (1 - 4) \times
 10^5 $ top quark pairs per year. The polarization of these can be easily tuned
 through the availability of polarized beams (see e.g. \cite{fischer99}).
 Further, there is a high degree of correlation between the polarization of top
 and anti-top quarks produced in pairs either at $ e^{+} e^{-} $
 colliders~\cite{c8,c9,c10,c11} or at hadron colliders~\cite{c12} which can be
 probed through the joint decay distributions of the top and the anti-top quark.

 In this paper we study momentum-momentum and spin-momentum correlations in
 the cascade decay process $ t \!\rightarrow\! W^+ \!+\! b $ followed by $ W^{+} 
 \!\rightarrow\! l^{+} \!+\! \nu_{l} $. The step-one decay $ t \!\rightarrow\!
 W^{+} \!+\! b $ is analyzed in the $ t $-rest frame where we study the
 spin-momentum correlation between the spin of the top and the momentum of the
 $ W $. In step two we go to the rest frame of the $ W $ and analyze the
 correlation between the momentum of the lepton (or antiquark) and the initial
 momentum direction of the $ W $. In technical terms this means we analyze the
 double density matrix of the decaying top quark and the produced W-gauge boson.
 This must be contrasted with the {\it center of mass} analysis of polarized top
 decay where the spin-momentum correlations are all analyzed in the rest system
 of the top quark (for an $O(\alpha_s)$ analysis of this kind see \cite{cjkk}).
 Experimentally such a correlation measurement is easier, but from a
 theoretical point of view the cascade-type of analysis is advantageous because
 one can then better isolate the contribution of the longitudinal mode of the 
 $ W $-gauge boson  which is of relevance for the understanding of the
 electroweak symmetry breaking sector in the Standard Model. The results of the
 two analysis' are of course related through a Lorentz boost along the $ W $
 direction. However, the azimuthal correlations to be discussed later are not
 affected by such a Lorentz boost and are thus identical in both
 types of analysis.

 The complete angular decay distribution is governed by altogether eight
 structure functions which we calculate analytically including their full
 $ O(\alpha_s) $ radiative corrections. One of the motivations for calculating
 the $ O(\alpha_s) $ radiative corrections is the fact that the radiative QCD
 corrections populate helicity configurations that are not accessible at the
 Born level. Take for example unpolarized top decay where, at the Born level,
 the $ W^{+} $ cannot be right handed, i.e.~cannot have positive helicity, due
 to angular momentum conservation when $ m_{b} = 0 $. This implies that strictly
 forward $ l^{+} $ production does not occur at the Born level. However, when
 radiative corrections are taken into account, right-handed $W$'s do occur and
 strictly forward $ l^{+} $ production is allowed. As we shall see in Sec.~4
 technically this means that the structure function combination $ (H_U + H_F)/2 $
 vanishes at the Born term level but becomes nonzero at
 $ O(\alpha_s) $~\cite{FGKM}. We shall, however, see that the $ O(\alpha_s) $
 population of the right-handed $ W $ is rather small \cite{FGKM}. The same
 statement holds true for the other structure function combinations that vanish
 at the Born term level.

 In order to retain full control over the $ b $ mass dependence, and having also
 other applications in mind, we have kept a finite mass value for the $ b $
 quark in our calculation. This improves on our earlier calculation of polarized
 top decay where the $ b $ quark mass was neglected and where we limited our
 attention to the six (diagonal) structure functions that govern the polar
 angle distribution in the cascade decay~\cite{fischer99}. The additional two
 (non-diagonal) structure functions calculated in this paper describe the
 azimuthal correlation of the plane of the top quark's polarization and the
 plane defined by the final leptons.
 In addition we determine the unpolarized and polarized scalar structure
 functions which are of relevance in the analysis of top decay into a
 bottom quark and a charged Higgs boson~\cite{czarnecki}. We mention that our
 calculations have been done in the zero width approximation of the $ W $-boson.
 Finite width effects will be addressed in a forthcoming paper \cite{dgkm}
 (see also \cite{jk93}).
 
 Most of the results in this paper are new. They have been checked against 
 limiting cases and partial results obtained in other papers. We have checked
 our analytical $ O(\alpha_s) $ result for the total rate against the
 corresponding analytical rate result of Denner and Sack who also kept the $ b $
 quark mass finite~\cite{c14}. We find agreement. We took the zero $ b $-quark
 mass limit of the six diagonal structure functions and obtained agreement with
 our previous results in~\cite{fischer99}.
 These had already been checked against the analytical results on the total rate
 obtained in ~\cite{c15,c16,c17,c18} and on the longitudinal/transverse
 composition obtained in~\cite{c19}. All six (mass zero) diagonal structure
 functions had also been checked against the corresponding numerical results
 given in~\cite{c19,c20,c21}. The unpolarized scalar structure function has
 been checked against the results of \cite{czarnecki}.

 The central topic of this paper is the analysis of polarized top decay. We
 therefore  mostly limit our attention to results valid in the limit $ m_{b}
 \rightarrow 0 $ in the main part of our paper. This leads to enormous
 simplifications in the analytical rate formulas. The quality of the $ m_{b} =
 0 $ approximation may be judged from the Born term rate which increases by
 $ 0.27 \, \% $ going from $ m_{b} = 4.8 \mbox{ GeV} $ to $ m_{b} = 0 $. The
 full $ m_{b} \neq 0 $ structure is given in Sec.~8 and the Appendices.
 Apart from retaining full control over $ m_b \ne 0 $ effects the finite mass
 results are needed e.g. in the theoretical analysis of semileptonic
 $ b \rightarrow c $ decays where the $ c $-quark mass can certainly not be
 neglected.     

 Our paper is structured as follows. In Sec.~2 we define a set of three spin 
 independent and five spin dependent structure functions through the covariant
 expansion of the decay tensor resulting from the product of the two relevant
 current matrix elements. The eight invariant structure functions are related
 to eight helicity structure functions which form the angular coefficients of the
 angular decay distribution. In order to facilitate the calculation of the tree
 graph contributions we define a set of five covariant projection operators and a
 covariant representation of the spin vector of the top. These projectors can be
 used to covariantly project the requisite helicity structure functions from the
 hadron tensor. The advantage is that one thereby obtains the appropriate
 helicity structure functions and scalarizes the tensor integrands needed for
 the tree graph integration in one go. In Sec.~3 we derive the explicit form
 of the angular decay distribution in terms of the eight helicity structure
 functions for top decay. We also specify the changes in the angular decay
 distribution needed for antitop decay.
 Sec.~4 contains our Born term results. In Sec.~5 we list our results for the 
 $ m_b = 0 $ one-loop contributions. In Sec.~6 we provide expressions for the
 $ O(\alpha_s) $ tree graph contributions and discuss technical details of how
 we have handled the necessary tree graph integrations. We mention that the
 infrared divergencies are regularized by a finite small gluon mass. In Sec.~7 we
 take the $ m_b \rightarrow 0 $ limit of the $ m_b \ne 0 $ results in Sec.~8 and
 present rather compact analytical $ O(\alpha_s) $ formulas for the various
 structure functions. Sec.~7 also contains our numerical results in the
 $ m_{b} = 0 $ approximation. Sec.~8 gives our analytical results on the tree
 graph integrations plus the one-loop contributions for $ m_b \ne 0 $. Sec.~9
 provides a summary and our conclusions. In particular, we emphasize
 that angular measurements as advocated in this paper can be utilized
 to measure the mass of the top quark.  In Appendix A we provide a complete
 list of $ m_b \ne 0 $ basis integrals that appear in the calculation of the
 tree graph contributions. This set of basis integrals should also be useful
 for other $ O(\alpha_s) $ or $ O(\alpha) $ radiative correction calculations.
 The requisite coefficient functions that multiply the basic integrals in the
 structure function expressions are listed in Appendix B. Appendix C, finally,
 contains the one-loop contribution in the $ m_b \ne 0 $ case.


\section{\bf Invariant and helicity structure functions}

 The dynamics of the current-induced $ t \rightarrow b $ transition is embodied
 in the hadron tensor $ H^{\mu \nu} $ which is defined by

 \begin{Eqnarray} 
 \label{hadron-tensor1}
   H^{\mu \nu}(q_0, q^2=m_W^2, s_t) & = & (2 \pi)^3 
   {\textstyle{\sum}} \hspace{-4mm} \int\limits_{X_b}
   d\Pi_f \, \delta^4(p_t - q - p_{X_b}) \times \frac{1}{2m_t}
   \nonumber \\ & \times & 
   \langle t(p_t,s_t) |J^{\nu +}| X_b \rangle
   \langle X_b |J^{\mu}| t(p_t,s_t)   \rangle \, ,
 \end{Eqnarray}

 \noindent where $ d\Pi_f $ stands for the Lorentz-invariant phase space factor.
 In the Standard Model the weak current is given by
 $ J^\mu = \bar{q}_b \gamma^\mu P_L \bar{q}_t $ with 
 $ P_L = \frac{1}{2} ( 1 - \gamma_5 ) $.   

 We are working in the narrow resonance approximation of the $ W $-boson and set
 $ q^2 = m_W^2 $ as indicated in the argument of the hadron tensor. Thus the
 hadron tensor is a function of the energy $ q_0 $ of the $ W $ alone. Since we
 are not summing over the top quark spin the hadron tensor also depends on the
 top spin $ s_t $ as indicated in Eq.(\ref{hadron-tensor1}). The structure of
 the hadron tensor can be represented by a standard set of invariant structure
 functions defined by the expansion

 \begin{Eqnarray} 
 \label{tensor-expansion}
   H^{\mu \nu} & = & \Big( - g^{\mu \nu} \, H_1 + p_t^{\mu} p_t^{\nu} \, H_2 -
   i \epsilon^{\mu \nu \rho \sigma} p_{t,\rho} q_{\sigma} \, H_3 \Big) + 
   \nonumber \\ & - &
   (q \!\cdot\! s_t) \Big( - g^{\mu \nu} \, G_1 + p_t^{\mu} p_t^{\nu} \, G_2 -
   i \epsilon^{\mu \nu \rho \sigma} p_{t,\rho} q_{\sigma} \, G_3 \Big) +
   \\ & + &
   \Big(s_t^{\mu} p_t^{\nu} + s_t^{\nu} p_t^{\mu} \Big) \, G_6 +
   i \epsilon^{\mu \nu \rho \sigma} p_{t \rho} s_{t \sigma} \, G_8 +
   i \epsilon^{\mu \nu \rho \sigma} q_{\rho} s_{t \sigma} \, G_9 \, , \nonumber
 \end{Eqnarray}  

 \noindent where the $ H_i $ (i=1,2,3) and $ G_i $ (i=1,2,3,6,8,9) denote
 unpolarized and polarized structure functions, respectively.

 In the expansion (\ref{tensor-expansion}) we have kept only those structure
 functions that contribute in the zero lepton mass case. We have thus omitted
 covariants built from $ q^{\mu} $ and/or $ q^{\nu} $. We have also dropped
 contributions from invariants that are fed by {\it T-odd} or imaginary
 contributions which are both absent in the present case.

 In the expansion (\ref{tensor-expansion}) one has still overcounted by one
 term since there is a relationship between the three parity conserving (p.c.)
 spin dependent covariants appearing in (\ref{tensor-expansion}) due to the
 identity of Schouten. The identity between the three covariants reads

 \begin{equation} 
 \label{schouten-identity}
   q \!\cdot\! s_t \, \epsilon^{\mu \nu \rho \sigma} p_{t,\rho} q_{\sigma} -
   q^2 \epsilon^{\mu \nu \rho \sigma} p_{t,\rho} s_{t \sigma} +
   q \!\cdot\! p_t \, \epsilon^{\mu \nu \rho \sigma} q_{\rho} s_{t,\sigma} = 0
 \end{equation}

 \noindent We shall, however, keep the overcounted set of nine invariant
 structure  functions in (\ref{tensor-expansion}) for reasons of computational
 convenience.

 In this paper we shall only be concerned with two types of intermediate states
 in (\ref{hadron-tensor1}), namely $ | X_{b} \rangle = | b \rangle $ (Born term
 and $ O(\alpha_s) $ one-loop contributions) and $ | X_{b} \rangle = | b + g 
 \rangle $ ($ O(\alpha_s) $ tree graph contribution). The Feynman diagrams
 contributing to the respective processes are drawn in Fig.~1.

 The angular decay distribution that we are aiming for is given in terms of a
 set of angular decay coefficients which are linearly related to the set of
 unpolarized structure functions $ H_i $ and polarized structure functions
 $ G_i $ defined in Eq.~(\ref{tensor-expansion}).
 The relevant linear combinations are given by 

 \alpheqn 
 \begin{eqnarray} 
 \label{lincombeg}
   H_U   & = & H_{++} + H_{--} = H_{11} + H_{22}, \\
   H_L   & = & H_{\mbox{oo}} = H_{33}, \\
   H_F   & = & H_{++} - H_{--} = i (H_{12} - H_{21}), \\
   H_{U^P} & = & H_{++}(s_t^l) + H_{--}(s_t^l) =
   H_{11}(s_t^l) + H_{22}(s_t^l), \\
   H_{L^P} & = & H_{\mbox{oo}}(s_t^l) = H_{33}(s_t^l), \\
   H_{F^P} & = & H_{++}(s_t^l) - H_{--}(s_t^l) =
   i (H_{12}(s_t^l) - H_{21}(s_t^l)) , \\
   H_{I^P} & = & \frac{1}{4} \Big( H_{+\mbox{o}}(s_t^{tr}) +
   H_{\mbox{o}+}(s_t^{tr}) - H_{-\mbox{o}}(s_t^{tr}) -
   H_{\mbox{o}-}(s_t^{tr}) \Big) \\ & = &
   - \frac{1}{2 \sqrt{2}} (H_{13}(s_t^{tr}) + H_{31}(s_t^{tr})), \\
   H_{A^P} & = & \frac{1}{4} \Big( H_{+\mbox{o}}(s_t^{tr}) +
   H_{\mbox{o}+}(s_t^{tr}) + H_{-\mbox{o}}(s_t^{tr}) +
   H_{\mbox{o}-}(s_t^{tr}) \Big) \\ & = &
   \frac{i}{2\sqrt{2}} (H_{23}(s_t^{tr}) - H_{32}(s_t^{tr})),
 \label{lincomend}   
 \end{eqnarray}
 \reseteqn

 \noindent where $ H_{\lambda_W;\lambda_W'} = H_{\mu \nu} \epsilon^{\ast \mu}
 (\lambda_W) \epsilon^{\nu}(\lambda_W') $ are the helicity projections of the
 polarized and unpolarized pieces of the structure functions $ H^{\mu \nu} $.
 The $ \epsilon^{\ast \mu}(\lambda_W) $ and $ \epsilon^{\nu}(\lambda_W) $ are
 the usual spherical components of the polarization vector of the $ W $ gauge
 boson. In the top quark rest system with $ q^{\mu} = (q_0;0,0,|\vec{q} \,|) $
 and $| \vec{q} \,| = (q_0^2 - m_W^2)^{1/2} $ they read

 \begin{Eqnarray} 
 \label{helicity-projections}
   \epsilon^{\mu}(0) & = & \frac{1}{m_W} (|\vec{q} \,|;0,0,q_0), \\
   \epsilon^{\mu}(\pm) & = & \frac{1}{\sqrt{2}} (0;\mp 1,-i,0).
 \end{Eqnarray}

 \noindent In Eqs.(\ref{lincombeg}--\ref{lincomend}) we have also included the
 Cartesian components of the helicity structure functions in the $ W $-boson rest
 frame which are useful to have for some applications. For notational
 convenience we shall often refer to the set of helicity structure functions
 by their generic names. Thus we shall frequently use $ U $ for $ H_U $ and
 $ U^P $ for $ H_{U^P} $, etc..

 The rest frame components of the longitudinal (``l'') and transverse (``tr'')
 polarization vector of the top are simply given by $ s_t^l = (0;0,0,1) $
 and $ s_t^{tr} = (0;1,0,0) $. For the unpolarized helicity structure functions
 one sums over the the two diagonal spin configurations of the top while one
 takes the differences of these for the polarized  helicity structure functions
 (in the $ z $-basis for $ s_t^l $ and in the $ x $-basis for $ s_t^{tr} $).
 When computing the polarized structure functions from the relevant Dirac trace
 expressions one thus has to replace $ (\slashp_t + m_t) $ in the unpolarized
 Dirac string by $ (\slashp_t + m_t)(1 + \gamma_5 \slashs_t) $. Note that
 the longitudinal component contributes only to the diagonal helicity structure
 functions $ U,L $ and $ F $ while the transverse component contributes only to
 the non-diagonal structure functions $ I $ and $ A $. The physics behind this
 will become clear when we write down the angular decay distribution in Sec.~3.

 It turns out that it is rather convenient from the computational point of view
 to represent the helicity projections in (\ref{lincombeg}--\ref{lincomend})
 (defined by the gauge boson polarization vectors and the top polarization
 vector) in covariant form. One has

 \alpheqn
 \begin{Eqnarray}
   H_i^{\phantom{P}} & = & H_{\mu \nu} \mbox{I} \! \mbox{P}^{\mu \nu}_i
   \hspace{2.7cm} i = U,L,F, \\ \Strutb
   H_{i^P} & = & H_{\mu \nu}(s_t^l) \mbox{I} \! \mbox{P}^{\mu \nu}_i
   \hspace{2.0cm} i = U,L,F, \\ \Strutb
   H_{i^P} & = & H_{\mu \nu}(s_t^{tr}) \mbox{I} \! \mbox{P}^{\mu \nu}_i
   \hspace{2.0cm} i = I,A. \Strutb
 \end{Eqnarray}
 \reseteqn

 \noindent The covariant projectors onto the diagonal density matrix elements
 are given by

 \alpheqn
 \begin{Eqnarray}
 \label{kovprojanfang}
   \IP^{\mu \nu}_{L \phantom{+L}} & = &
   \frac{m_W^2}{m_t^2} \frac{1}{| \vec{q} \, |^2} 
   \Big(p_t^{\mu} - \frac{p_t \cdot q}{m_W^2} q^{\mu} \Big)
   \Big(p_t^{\nu} - \frac{p_t \cdot q}{m_W^2} q^{\nu} \Big), \\
   \IP^{\mu \nu}_{U+L} & = & - g^{\mu \nu} + \frac{q^{\mu} q^{\nu}}{m_W^2}, \\
   \IP^{\mu \nu}_{F \phantom{+L}} & = & \frac{1}{m_t} \frac{1}{| \vec{q} \,|}
   i \epsilon^{\mu \nu \alpha \beta} p_{t,\alpha} q_{\beta},
 \end{Eqnarray}

 \noindent where $ \epsilon^{0123} = -1 $.
 We do not write out the projector for the unpolarized-transverse
 component $ U $ but note that it can be obtained from the combination
 $ \IP^{\mu \nu}_{U+L} - \IP^{\mu \nu}_{L} $.

 The projectors onto the
 transverse-longitudinal non-diagonal density matrix elements are given by

 \begin{Eqnarray}
   \IP^{\mu \nu}_I & = & + \frac{1}{2 \sqrt{2}} \frac{m_W}{m_t}
   \frac{1}{| \vec{q} \, |} \Big\{ \epsilon^{\mu}(x)
   \Big(p_t^{\nu} - \frac{p_t \cdot q}{m_W^2} q^{\nu} \Big) +
   \mu \leftrightarrow \nu \Big\}, \\
   \IP^{\mu \nu}_A & = & - \frac{1}{2 \sqrt{2}} \frac{m_W}{m_t^2}
   \frac{1}{| \vec{q} \,|^2} \Big\{ i \epsilon^{\mu \alpha \beta \gamma} 
   \epsilon_{\alpha}(x) p_{t,\beta} q_{\gamma} 
   \Big(p_t^{\nu} - \frac{p_t \cdot q}{m_W^2} q^{\nu} \Big) -
   \mu \leftrightarrow \nu \Big\}.
 \label{kovprojende}
 \end{Eqnarray}
 \reseteqn

 \noindent They involve the the transverse polarization vector of the $ W $-gauge
 boson $ \epsilon_{\alpha}\!(x) \!=\! (0;1,0,0) $ pointing in the $x$-direction.

 The covariant representation of the longitudinal component of the polarization
 vector of the top spin vector $ s_t^l $ is given by

 \begin{equation} 
 \label{polvektor1} 
   s_t^{l, \mu}  =  \frac{1}{| \vec{q} \,|} 
   \Big( q^{\mu} - \frac{p_t \!\cdot\! q}{m_t^2} p_t^{\mu} \Big),
 \end{equation}

 \noindent whereas its transverse component $ s_t^{tr}  $ reads

 \begin{equation} 
 \label{polvektor2}
   s_t^{tr, \mu} = (0;1,0,0).
 \end{equation}

 Note the inverse powers of $ | \vec{q} \,| = \sqrt{q_0^2 - m_W^2} $ that enter
 the $ L,T,F,I $ and $ A $ projectors and the longitudinal polarization vector.
 They come in for normalization reasons.
 These inverse powers of $ | \vec{q} \,| $ will make the necessary tree graph
 integrations to be dealt with in Sec.~6 and in the Appendices A and B somewhat
 more complicated than the total $ (U+L) $ rate integration which has a rather
 simple projector as Eq.~(8.2) shows.

 As mentioned in the Introduction, the covariant forms of the projection
 operators (\ref{kovprojanfang}--\ref{kovprojende}) and the polarization vectors
 (\ref{polvektor1}) and (\ref{polvektor2}) are quite convenient for the
 calculation of the $ O(\alpha_s) $ tree graph contributions to be dealt with in
 Sec.6 . The covariant projectors allow one to scalarize the tree graph tensor
 integrands and to project onto the requisite helicity structure functions in
 one go.

 Although we shall mostly work in the helicity representation of the structure
 functions, it is sometimes convenient to have available the set of linear
 relations between the helicity and invariant structure functions. These can
 easily be worked out from the expansion (\ref{tensor-expansion}), the
 projectors (\ref{kovprojanfang}--\ref{kovprojende}) and the polarization
 vectors  (\ref{polvektor2}). One has

 \alpheqn
 \begin{Eqnarray}
 \label{system2anfang}
   \phantom{m_W^2} H_{U} & = & 2 H_1, \\ \Strutb 
   m_W^2 H_L & = & m_W^2 H_1 + |\vec{q} \,|^2 m_t^2 \, H_2, \\ \Strutb
   \phantom{m_W^2} H_{F} & = &  2 |\vec{q} \,| m_t \, H_3, \\ \Strutb
   \phantom{m_W^2} H_{U^P} & = &  2 |\vec{q} \,| \, G_1, \\ \Strutb
   m_W^2 H_{L^P} & = & |\vec{q} \,| ( m_W^2 \, G_1 + |\vec{q} \,|^2 m_t^2 \, 
   G_2 - 2 q_0 m_t G_6), \\ \Strutb
   \phantom{m_W^2} H_{F^P} & = & 2 \, |\vec{q}\,|^2 m_t \, G_3 - 2 \, m_t \,
   G_8 - 2 \, q_0 \, G_9, \\ \Strutb
   \phantom{m_W^2} H_{I^P} & = & \frac{1}{\sqrt{2}} \frac{m_t}{m_W} |\vec{q}\,| 
   \, G_6, \\
   \phantom{m_W^2} H_{A^P} & = & - \frac{1}{\sqrt{2}} \frac{m_t q_0}{m_W} \,
   G_8 - \frac{1}{\sqrt{2}} m_W G_9.
 \label{system2ende}
 \end{Eqnarray}
 \reseteqn

 Note that the three structure functions $ G_3 $, $ G_8 $ and $ G_9 $ always 
 contribute in the two combinations $ (m_W^2 G_3 + G_8) $ and $ (q_0 m_1 G_3 -
 G_9) $ proving again that there are only eight independent combinations of
 structure functions. If desired, Eqs.(\ref{system2anfang}--\ref{system2ende})
 can be inverted such that the invariant structure functions can be expressed
 in terms of the helicity structure functions. The inversion has to be done in
 terms of the two above linear combinations of $ G_3 $, $ G_8 $ and $ G_9 $.
 Since our later results will always be presented in terms of the helicity
 structure functions, we shall not write down the inverse relations here.


\section{\bf Angular decay distribution}

 We are now in the position to write down the full angular decay distribution 
 of polarized top decay into $ W^+ $ and $ b $ followed by the decay of the the
 $ W^{+} $ into $ (l^{+} + \nu_{l}) $. As noted before, the full angular decay
 distribution of the decay $ t(\uparrow) \rightarrow W^{+} (\rightarrow l^{+}
 + \nu_{l}) + X_{b} $, including polarization effects of the top quark, is
 completely determined by the three unpolarized and the five polarized helicity
 structure functions. Although the necessary manipulations to obtain the angular
 decay distribution involving Wigner's $ D_{m m^{\prime}}^J(\theta,\phi) $-%
 functions are standard (see e.g. \cite{c22}), it is quite instructive to
 reproduce the results here. To this end, it is useful to define helicity
 structure functions $ H_{\lambda_W \lambda^{\prime}_W}^{\lambda_t \,\,
 \lambda^{\prime}_t} $ where the helicity label of the top quark is made
 explicit. Put in a different language the four-index object
 $ H_{\lambda_W \lambda^{\prime}_W}^{\lambda_t \,\, \lambda^{\prime}_t} $
 is the unnormalized double density matrix of the top and the $ W $.
 The double density matrix is Hermitian, i.e. it satisfies

 \begin{equation}
  \Big( H_{\lambda_W \lambda^{\prime}_W}^
  {\lambda_t \,\,\lambda^{\prime}_t} \Big)^{\ast} =
  \Big( H_{\lambda^{\prime}_W \lambda_W}^
  {\lambda^{\prime}_t \,\,\lambda_t} \Big). 
 \end{equation}

 As has been remarked on before the elements of the double density matrix are
 real in the present application. The double density matrix is therefore
 symmetric. The relation of the components of the double density matrix to the
 previously defined unpolarized and polarized helicity structure functions is
 given by

 \alpheqn
 \begin{Eqnarray}
 \label{system3anfang}
   H_U & = & \phantom{\frac{1}{4} (}
   H_{++}^{++} + H_{++}^{--} + H_{--}^{++} + H_{--}^{--}, \\
   H_L & = & \phantom{\frac{1}{4} (}
   H_{\mbox{oo}}^{++} + H_{\mbox{oo}}^{--}, \\
   H_F & = & \phantom{\frac{1}{4} (}
   H_{++}^{++} + H_{++}^{--} - H_{--}^{++} - H_{--}^{--}, \\
   H_{U^P} & = & \phantom{\frac{1}{4} (}
   H_{++}^{++} - H_{++}^{--} + H_{--}^{++} - H_{--}^{--}, \\
   H_{L^P} & = & \phantom{\frac{1}{4} (}
   H_{\mbox{oo}}^{++} - H_{\mbox{oo}}^{--}, \\
   H_{F^P} & = & \phantom{\frac{1}{4} (}
   H_{++}^{++} - H_{++}^{--} - H_{--}^{++} + H_{--}^{--}, \\
   H_{I^P} & = & \frac{1}{4} (H_{+\mbox{o}}^{+-} + H_{\mbox{o}+}^{-+} -
   H_{-\mbox{o}}^{-+} - H_{\mbox{o}-}^{+-}) =
   \frac{1}{2} (H_{+\mbox{o}}^{+-} - H_{-\mbox{o}}^{-+}), \label{HIP} \\
   H_{A^P} & = & \frac{1}{4} (H_{+\mbox{o}}^{+-} + H_{\mbox{o}+}^{-+} +
   H_{-\mbox{o}}^{-+} + H_{\mbox{o}-}^{+-}) =
   \frac{1}{2} (H_{+\mbox{o}}^{+-} + H_{-\mbox{o}}^{-+}). \label{HAP}
 \end{Eqnarray}
 \reseteqn

 For ease of notation we have used $ (\pm) $-labels for both the helicities
 of the top $ (\lambda_t = \pm 1/2) $ and the transverse helicities of the
 $ W $ gauge boson $ (\lambda_W = \pm 1) $. In the case of the non-diagonal
 structure functions $ H_{I^P} $ and $ H_{A^P} $ one can make use of the fact
 that the double density matrix is symmetric (for real coefficients !) to
 simplify the structure functions as indicated in the last two lines of 
 Eqs.~(\ref{HIP}--\ref{HAP}). From the fact that we are not observing the spin of
 the $ X_b $ system in our semi-inclusive measurement one has $ \lambda_{X_b} =
 \lambda'_{X_b} $ leading to the constraint $ \lambda_W - \lambda^{\prime}_W =
 \lambda_t - \lambda^{\prime}_t $. From this constraint it is immediately clear
 that the polarized structure functions $ U,L $ and $ F $ are associated with the
 longitudinal spin of the top and the structure functions $ I $ and $ A $ are
 associated with the transverse spin of the top.

 The angular decay distribution can be obtained from the master formula

 \begin{equation}
   W(\theta_P,\theta,\phi) \propto
   \sum\limits_{\lambda_W - \lambda^{\prime}_W = \lambda_t - \lambda^{\prime}_t}
   e^{i (\lambda_W - \lambda^{\prime}_W) \phi} \,
   d^1_{\lambda_W 1}(\theta) \, d^1_{\lambda^{\prime}_W 1}(\theta) \,
   H_{\lambda_W \lambda^{\prime}_W}^{\lambda_t \: \lambda^{\prime}_t} \, 
   \rho_{\lambda_t \: \lambda^{\prime}_t} (\theta_P),
 \label{masterformula}
 \end{equation}

 \noindent where $ \rho_{\lambda_t \: \lambda^{\prime}_t} (\theta_P) $ is the
 density matrix of the top quark which reads

 \begin{equation}
   \rho_{\lambda_t \: \lambda^{\prime}_t} (\theta_P) =
   \frac{1}{2} \pmatrix{
   1 + P \cos \theta_P & P \sin \theta_P \cr
   P \sin \theta_P & 1 - P \cos \theta_P }.
 \end{equation}

 \noindent $ P $ is the magnitude of the polarization of the top quark. The sum
 in Eq.~(\ref{masterformula}) extends over all values of $ \lambda_W,
 \lambda^{\prime}_W, \lambda_t $ and $ \lambda^{\prime}_t $ compatible with
 the constraint $ \lambda_W - \lambda^{\prime}_W = \lambda_t - \lambda^{\prime}_t
 $. The second lower index in the small Wigner $ d(\theta) $-function
 $ d^1_{\lambda_W 1} $ is fixed at $ m = 1 $ for zero mass leptons because the
 total $ m $-quantum number of the lepton pair along the $ l^{+} $ direction is
 $ m = 1 $. Because there exist different conventions for Wigner's
 $ d $-functions we explicate the requisite components that enter
 Eq.~(\ref{masterformula}): $ d^1_{11} = (1 + \cos\theta) / 2 $,
 $ d^1_{01} = \sin\theta / \sqrt{2} $ and $ d^1_{-11} = (1 - \cos \theta) / 2 $.
 
 Including the appropriate normalization factor the four-fold decay distribution
 is given by
 
 \begin{Eqnarray}
 \label{DiffRate}
  \frac{d \Gamma}{dq_0 d \cos \theta_P d \cos \theta d \phi} & = &
  \frac{1}{4 \pi} \frac{G_F |V_{tb}|^2 m_W^2}{\sqrt{2} \, \pi}
  \, |\vec{q}\,| \Big\{
  \frac{3}{8} (H_U + P \cos \theta_p H_{U^P}) (1 + \cos^2 \theta) +
  \hspace{5mm} \nonumber
  \\ & & \hspace{-1.5cm} +
  \frac{3}{4} (H_L + P \cos \theta_p H_{L^P}) \sin^2 \theta +
  \frac{3}{4} (H_F + P \cos \theta_p H_{F^P}) \cos \theta
  \\ & & \hspace{-1.5cm} +
  \frac{3}{2 \sqrt{2}} P \sin \theta_p H_{I^P} \sin 2 \theta \cos \phi +
  \frac{3}{\sqrt{2}} P \sin \theta_p H_{A^P} \sin \theta \cos \phi \Big\}
  \nonumber
 \end{Eqnarray}

 \noindent We take the freedom to normalize the differential rate
 such that one obtains the total $ t \rightarrow b + W^{+} $ rate upon
 integration {\it and not} the total rate multiplied by the branching ratio of
 the respective $ W^{+} $ decay channel.

 The polar angles $ \theta_P $ and $ \theta $, and the azimuthal angle $ \phi $
 that arise in the full cascade-type description of the two-stage decay process
 $ t(\uparrow) \rightarrow W^{+} (\rightarrow l^{+} + \nu_{l}) + X_{b} $ are
 defined in Fig.~2. For better visibility we have oriented the lepton plane with
 a negative azimuthal angle relative to the hadron plane. For the hadronic
 decays of the $ W $ into a pair of light quarks one has to replace $ (l^{+},
 \nu_{l}) $ by $ ( \bar{q}, q) $ in Fig.~2. We mention that we have checked the
 signs of the angular decay distribution Eq.~(\ref{DiffRate}) using covariant
 techniques.

 As Eq.~(\ref{DiffRate}) shows the non-diagonal structure functions $ H_{I^P} $
 and $ H_{A^P} $ are associated with azimuthal measurements. This necessitates
 the definition of a hadron plane which is only possible through the availability
 of the $ x $-component of the polarization vector of the top (see Fig.~2). This
 is the physical explanation of why the two structure functions $ H_{I^P} $ and
 $ H_{A^P} $ are functions only of the transverse component of the polarization
 vector of the top quark. For similar reasons the polarization dependent
 structure functions $ H_{U^P}, H_{L^P} $ and $ H_{F^P} $ depend only on the
 longitudinal component of the polarization vector.

 Setting $P=0$ in Eq.~(\ref{DiffRate}) one obtains the decay distribution
 for unpolarized top decay. If desired, the transverse part of the unpolarized
 angular decay distribution can also be sorted in terms of decays into
 transverse-plus and transverse-minus $ W $-bosons given by the structure
 function combinations $ (U + F)/2 $ and $ (U - F)/2 $ which multiply the
 angular factors $(1 + \cos\theta)^2$ and $(1 - \cos\theta)^2$, resp., as
 done e.g. in \cite{FGKM}. 

 If there were an imaginary part in the one-loop contribution one would have
 two additional contributions to the angular decay distribution proportional
 to $ \sin \phi $. This can be easily seen with the help of
 Eq.~(\ref{masterformula}).
 We concentrate on those terms in the angular decay distribution that are
 proportional to the off-diagonal terms $ \rho_{+-} $ in the density matrix
 of the top. The relevant terms read

 \alpheqn
 \begin{Eqnarray}
  H_{+ \mbox{o}}^{+-} \, e^{+i \phi} +
  H_{\mbox{o} +}^{-+} \, e^{-i \phi} =
  2 \, \Big( \mbox{Re}(H_{+ \mbox{o}}^{+-}) \cos \phi -
  \mbox{Im}(H_{+ \mbox{o}}^{+-}) \sin \phi \Big), \\
  H_{\mbox{o} -}^{+-} \, e^{+i \phi} +
  H_{- \mbox{o}}^{-+} \, e^{-i \phi} =
  2 \, \Big( \mbox{Re}(H_{\mbox{o} -}^{+-}) \cos \phi -
  \mbox{Im}(H_{\mbox{o} -}^{+-}) \sin \phi \Big).
 \end{Eqnarray}
 \reseteqn 

 The real contributions multiplying the angular factor $ \cos \phi $ have been
 included in the angular decay distribution (\ref{DiffRate}) while the imaginary
 part contributions $ \mbox{Im}(H_{+ \mbox{o}}^{+-}) $ and $ \mbox{Im}
 (H_{\mbox{o} -}^{+-}) $ multiplying $ \sin \phi $ do not appear in
 Eq.~(\ref{DiffRate}) since the $ O(\alpha_s) $ contributions calculated in this
 paper are purely real. The helicity structure functions $ \mbox{Im}
 (H_{+ \mbox{o}}^{+-}) $ and $ \mbox{Im}(H_{\mbox{o} -}^{+-}) $ are
 conventionally called {\it T-odd} structure functions and are contributed to by
 the imaginary parts of loop contributions and/or by CP-violating contributions
 which, as has been emphasized before, are not present in this calculation.

 Of interest is also the corresponding angular decay distribution for polarized
 anti-top decay $ \bar{t}(\uparrow) \rightarrow W^{-} (\rightarrow l^{-} +
 \bar{\nu}_l) + X_{\bar{b}} $.
 The angular decay distribution is changed due to the fact that the total
 $ m $-quantum number of the lepton pair in the $ l^- $ direction is now
 $ m = - 1 $. The relevant components of the small Wigner $ d $-function are now
 $ d^1_{1-1} = (1 - \cos \theta) / 2 $, $ d^1_{0-1} = -\sin \theta / \sqrt{2} $
 and $ d^1_{-1-1} = (1 + \cos \theta) / 2 $. This can be seen to result in a
 sign change for the angular factors multiplying the $ F $, $ F^P $ and $ A^P $
 terms (and no sign change for the other terms). The structure functions of
 anti-top decay are related to those of top decay by $ CP $-invariance. 
 The p.v.\ structure functions $ F, U^P, L^P $ and $ I^P $ will undergo a sign
 change whereas the p.c.\ structure functions $ U,L,F^P $ and $ A^P $ keep their
 signs. Overall this means that the signs of the unpolarized terms in
 Eq.~(\ref{DiffRate}) will not change their signs while the polarized terms will
 change signs when going from top decay to anti-top decay. To be quite explicit,
 if one wants to use the results of this paper to describe anti-top decay, the
 only required effective change is to change the signs of the terms multiplying
 the $ U^P,L^P,F^P,I^P $ and $ A^P $ structure functions in the angular decay
 distribution Eq.~(\ref{DiffRate}), using, however, the same structure functions
 as written down in this paper.


\section{\bf Born term results}

 The Born term tensor is calculated from the square of the Born term amplitude
 (see Fig.~2(a)) given by

 \begin{equation} 
  M^{\mu} = V_{tb}\, \frac{g}{ \sqrt{2}} 
  \bar{u}_{b} \gamma^{\mu} \frac{1}{2} (1 - \gamma_5) u_t.
 \end{equation}

 \noindent We omit the coupling factor
 $ V_{tb}\, g /\sqrt{2} = 2 m_W V_{tb} (G_F/\sqrt{2})^{1/2} $
 and write for the Born term tensor (the spin of the $b$ quark is summed) 

 \begin{equation} 
 \label{born}
  B^{\mu \nu} = \frac{1}{4} \mbox{Tr} (\slashp_b + m_b) \gamma^{\mu}
  (1 - \gamma_5) (\slashp_t + m_t) (1 + \gamma_5 \slash{s}_t)
  \gamma^{\nu} (1 - \gamma_5).
 \end{equation}

 Since only even-numbered $ \gamma $-matrix strings survive between the two 
 $ (1 - \gamma_5) $-factors in (\ref{born}) one can compactly write

 \begin{equation} 
 \label{bornspur}
  B^{\mu \nu} = 2 ( \bar{p}_t^{\nu} p_b^{\mu} + \bar{p}_t^{\mu} p_b^{\nu} -
  g^{\mu \nu} \, \bar{p}_t \!\cdot\! p_b + i \epsilon^{\mu \nu \alpha \beta}
  p_{b,\alpha} \bar{p}_{t,\beta}),
 \end{equation}

 \noindent where

 \begin{equation}
  \bar{p}_t^{\mu} = p_t^{\mu} - m_t s_t^{\mu}.
 \end{equation}

 It is not difficult to obtain the Born term helicity structure functions from
 (\ref{bornspur}). This can be done in two ways. One can either read off the
 invariant structure functions according to the covariant expansion 
 Eq.~(\ref{tensor-expansion}).
 The nonvanishing elements are given by $ B_{H_1} = m_t^2 (1-x^2+y^2) $,
 $ B_{H_2} = - 2 B_{H_3} = 4 $ for the unpolarized invariants and by
 $ B_{G_1} = B_{G_6} = B_{G_8} = - B_{G_9} = - 2 m_t $ for the
 polarized invariants (the notation is self-explanatory).
 These can then be converted to the helicity structure functions using the
 linear relations (\ref{system2anfang}--\ref{system2ende}). Or, one can directly
 compute the helicity structure functions from (\ref{bornspur}) by using the
 covariant projectors defined in Sec.2  ({\it cf.} Eq.(\ref{kovprojanfang}--
 \ref{kovprojende})).
 
 In order to find the relation of the Born term tensor $ B^{\mu \nu} $ to the
 hadron tensor $ H^{\mu \nu} $ defined in Sec.~2 one has to insert the
 appropriate one-particle $ b $-quark state into Eq.~(\ref{hadron-tensor1})
 and then one has to do the requisite one-particle phase space integration.
 Technically this is done by rewriting the one-particle phase space as

 \begin{equation} 
  \label{1pps}
  \int \!\! d\Pi_{b} =
  \int \!\! \frac{d^3 \vec{p}_{b}}{2 \, E_b} =
  \int \!\! d^4 p_b \, \delta(p_b^2 - m_b^2).
 \end{equation}

 \noindent One can easily do the four-dimensional $ d^4 p_b $ integration in
 Eq.(\ref{hadron-tensor1}) with the help of the four-dimensional
 $ \delta $-function $ \delta^4(p_t - q - p_b) $. This converts $ p_b^2 $ in
 the argument of the $\delta$-function in Eq.(\ref{1pps}) into
 $ (p_t - q)^2 $. Rewriting the argument of the $ \delta $-function in terms
 of $ q_0 $ one finally arrives at

 \begin{equation} 
 \label{born+hadron-tensor}
  H^{\mu \nu}(\mbox{Born}) = \frac{1}{4 m_t^2}
  \delta(q_0 - \frac{m_t^2 + m_W^2 - m_b^2}{2 \, m_t}) B^{\mu \nu}.
 \end{equation}

 We will present our results in table form where we use the scaled variables
 $ x = m_W/m_t $ and $ y = m_b/m_t $ as well as the abbreviation $ | \vec{q} \,|
 = (m_t/2) \sqrt{ \lambda } $ with $ \lambda = \lambda (1,x^2,y^2) = 1 + x^4 +
 y^4 - 2x^2 y^2 - 2x^2 - 2y^2 $. The first column in Table~1 contains the $ m_b
 \neq 0 $, or equivalently, $ y \neq 0 $ results. In the second column we have
 set $ m_b = 0 $ $ (y = 0) $. In order to assess the quality of the $ m_b = 0 $
 approximation for the various rate functions we have listed the percentage
 increments when going from the $ m_b \ne 0 $ case to the $ m_b = 0 $ case
 {\it including} the phase space factor $ | \vec{q} \,| $ that multiplies the
 helicity structure functions in the rate formula Eq.~(\ref{DiffRate}). In this
 comparison we have used $ m_b = 4.8 \mbox{ GeV} $~\cite{pivovarov} together with
 $m_t = 175$ GeV and $m_W=80.419$ GeV. The
 increment due to the phase space factor $ | \vec{q} \,| $ alone amounts to
 $ 0.15 \, \% $. Note that one may have overestimated the mass effect since
 a fixed pole mass, rather than a running mass which is smaller at the high scale
 of the top mass, is used. For example, taking one-loop running and the same
 bottom pole mass as above one has $ \bar{m}_b(m_t) = 1.79 $ GeV. The increment
 in the total rate going from $ \bar{m}_b(m_t) = 1.79 $ GeV to $ m_b = 0 $
 would then only be $ 0.04 \, \% $ as compared to the $ 0.26 \, \% $ given in
 Table 1.

 \begin{table}[htbp]
 \begin{center}
 \begin{tabular}{|l|c|c|l|}\hline
  Born  & $ m_b \neq 0 $ & $ m_b = 0 $ & increment \\
  term  &                &             &           \\ \hline \hline
  $ \Ds{B_{U+L}} $ & $ \Ds{ m_t^2 \frac{1}{x^2} ((1 \!-\! y^2)^2 \!+\! x^2
  (1 \!-\! 2 x^2 \!+\! y^2))} $ &
  $ \Ds{ m_t^2 \frac{1}{x^2} (1 \!-\! x^2)(1 \!+\! 2 x^2)
  \phantom{\Bigg(}} $ & +0.27 \% \\ \hline
  $ \Ds{B_{U^P+L^P}} $ & $ \Ds{ m_t^2 \sqrt{\lambda} \frac{1}{x^2} (1 \!-\!
  2 x^2 \!-\! y^2)} $ & $ \Ds{ m_t^2 \frac{1}{x^2} (1 \!-\! x^2)(1 \!-\! 2 x^2)
  \phantom{\Bigg(}} $ & +0.42 \% \\ \hline
  $ \Ds{B_U} $ & $ \Ds{2 m_t^2 (1 \!-\! x^2 \!+\! y^2)} $ & $ \Ds{2 m_t^2
  (1 \!-\! x^2) \phantom{\Bigg(}} $ & +0.05 \% \\ \hline
  $ \Ds{B_{U^P}} $ & $ \Ds{-2 m_t^2 \sqrt{\lambda}} $ & $ \Ds{- 2 m_t^2
  (1 \!-\! x^2) \phantom{\Bigg(}} $ & +0.29 \% \\ \hline
  $ \Ds{B_L} $ & $ \Ds{ m_t^2 \frac{1}{x^2} ((1 \!-\! y^2)^2 \!-\! x^2 
  (1 \!+\! y^2))} $ & $ \Ds{ m_t^2 \frac{1}{x^2} (1 \!-\! x^2)
  \phantom{\Bigg(}} $ & +0.36 \% \\ \hline
  $ \Ds{B_{L^P}} $ & $ \Ds{ m_t^2 \sqrt{\lambda} \frac{1}{x^2} (1 \!-\! y^2)} $ &
  $ \Ds{ m_t^2 \frac{1}{x^2} (1 \!-\! x^2) \phantom{\Bigg(}} $ &
  +0.37 \%  \\ \hline
  $ \Ds{B_F} $ & $ \Ds{- 2 m_t^2 \sqrt{\lambda}} $ & $ \Ds{- 2 m_t^2 (1 \!-\!
  x^2) \phantom{\Bigg(}} $ & +0.29 \% \\ \hline
  $ \Ds{B_{F^P}} $ & $ \Ds{2 m_t^2 (1 \!-\! x^2 \!+\! y^2)} $ &
  $ \Ds{2 m_t^2 (1 \!-\! x^2) \phantom{\Bigg(}} $ & +0.05 \% \\ \hline
  $ \Ds{B_S} $ & $ \Ds{ m_t^2 \frac{1}{x^2} ((1 \!-\! y^2)^2 \!-\! x^2 
  (1 \!+\! y^2))} $ & $ \Ds{ m_t^2 \frac{1}{x^2} (1 \!-\! x^2)
  \phantom{\Bigg(}} $ & +0.36 \% \\ \hline
  $ \Ds{B_{S^P}} $ & $ \Ds{ m_t^2 \sqrt{\lambda} \frac{1}{x^2} (1 \!-\! y^2)} $ &
  $ \Ds{ m_t^2 \frac{1}{x^2} (1 \!-\! x^2) \phantom{\Bigg(}} $ &
  +0.37 \% \\ \hline
  $ \Ds{B_{I^P}} $ & $ \Ds{-\frac{1}{2} \sqrt{2} m_t^2 \sqrt{\lambda}
  \frac{1}{x}} $ &
  $ \Ds{-\frac{1}{2}  \sqrt{2}  m_t^2 \frac{1}{x} (1 \!-\! x^2)
  \phantom{\Bigg(}} $ &
  +0.29 \% \\ \hline
  $ \Ds{B_{A^P}} $ & $ \Ds{\frac{1}{2} \sqrt{2} m_t^2 \frac{1}{x} (1 \!-\!
  x^2 \!-\! y^2)} $ & $ \Ds{\frac{1}{2} \sqrt{2} m_t^2 \frac{1}{x} (1 \!-\! x^2)
  \phantom{\Bigg(}} $ & +0.24 \% \\ \hline
 \end{tabular}
 \end{center}
 \caption{Born term helicity structure functions $ B_i $ ($ i = U \!+\! L $,
  $ U^P \!+\! L^P $, $ U $, $ U^P $, $ L $, $ L^P $, $ F $, $ F^P $, $ S $,
  $ S^P $, $ I^P $, $ A^P $ for $ m_b \ne 0 $ and $ m_b=0 $. Fourth column
  gives the percentage increment when going from  $ m_b \ne 0 $ to
  $ m_b=0 $ including the phase space factor $ |\vec{q}| $. }
 \end{table}

 In the $ m_b = 0 $ case listed in column 3 of Table 1 one observes the simple
 patterns $ B_U = - B_{U^P} = - B_F = B_{F^P} $, $ B_L = B_{L^P} $ and
 $ B_{I^P}  = - B_{A^P} $.
 This pattern results from the fact that a massless $ b $-quark
 emerging from a $ (V-A) $ vertex is purely left-handed. Since from angular
 momentum conservation one has $ \lambda_t = \lambda_W - \lambda_b $ with
 $ \lambda_b=-1/2 $ one has the constraint $ \lambda_t - \lambda_W = 1/2 $. This
 implies that only the helicity configurations $ (\lambda_t = -1/2; \lambda_W =
 -1) $ and $ (\lambda_t = +1/2; \lambda_W = 0) $ are non-vanishing. A quick look
 at the relations (\ref{system3anfang}--\ref{HAP}) allows one to readily verify
 the $ m_b = 0 $ pattern in Table~1. For $ m_b \ne 0 $ there is a leakage into
 right-handed bottom mesons resulting in a breaking of the above pattern as can
 be observed in the $ m_b \ne 0 $ column of Table~1. As noted in the Introduction
 these simple patterns are also not valid at $ O(\alpha_s) $ even for massless
 bottom mesons because of the additional gluon emission including an anomalous
 spin-flip contribution \cite{spinflip}. When the relevant $ m_b=0 $ Born term
 helicity structure functions from Table 1 are substituted in (\ref{DiffRate})
 we reproduce the angular decay distribution as written down in \cite{c8}.
 
 For completeness we have also included the two Born term scalar helicity
 structure functions $ B_S $ and $ B_{S^P} $ in Table~1. They are obtained by use
 of the scalar projector $ \IP_S = q^{\mu} q^{\nu} / m_W^2 $. That they are
 identical to their longitudinal counterparts  $ B_L $ and $ B_{L^P} $ 
 even for $ m_b \neq 0 $ is a
 dynamical accident specific to the Born term level and does not hold true in
 general as e.g. evidenced by the $ O(\alpha_s) $ contributions to be discussed
 later on. These become equal to each other only in the limit $ m_t \rightarrow
 \infty $ as will be discussed in Sec.~7. The $ m_b \neq 0 $ Born term 
 equalities $ B_F = B_{U^P} $ and $ B_U = B_{F^P} $ can be seen to
 result from the fact that the double density matrix elements $ H_{++}^{--} $
 and $ H_{--}^{++} $ vanish at the Born term level due to angular momentum 
 conservation (see \ref{system3anfang}).   

 In Fig.3 we present a lego plot of the two-fold ($ m_b=0 $) Born
 term angular decay distribution in $ \cos \theta $ and $ \cos\theta_P $
 which results after taking the azimuthal average of Eq.(\ref{DiffRate}). 
 We have divided out the total Born term rate from the differential rate
 resulting in the hatted differential rate distribution as defined in 
 Eq.(\ref{hatDiffRate}). We have set $ P =1 $ in Fig.3. The lego plot shows that
 the  $ \cos \theta $  and $ \cos\theta_P $ variation of the two-fold 
 angular decay  distribution around its average value of 0.25 is quite strong.
 This will facilitate the experimental measurement of the structure
 functions $\Gamma_U$, $\Gamma_L$, $\Gamma_F$,  $\Gamma_{U^P}$,  $\Gamma_{L^P}$
 and $\Gamma_{F^P}$ .

 Finally, for the sake of definiteness we list the Born term rate in terms of the
 Born term function $ B_{U+L} $. One has

 \begin{equation}
  \Gamma_0 = \frac{G_F \, m_W^2 \, |\vec{q}\,|}
  {4 \sqrt{2} \, \pi \, m_t^2} \, |V_{tb}|^2 \, B_{U+L}. 
 \end{equation} 


\section{\bf One-loop contribution}

 The one-loop contributions to fermionic $(V-A)$ transitions have a long history.
 Since QED and QCD have the same structure at the one-loop level the history
 even dates back to QED times.

 Our reference will be the work of Gounaris and Paschalis \cite{gounaris}
 (see also \cite{schilcher}) who used a gluon mass regulator to regularize the
 gluon IR singularity. The one-loop amplitudes are defined by the covariant
 expansion ($ J^V_\mu = \bar{q}_b \gamma_\mu q_t , 
 J^A_\mu = \bar{q}_b \gamma_\mu \gamma_5 q_t $) 

 \alpheqn
 \begin{Eqnarray}  
 \label{formfactor} 
 \langle b(p_b) | J^V_{\mu} | t(p_t) \rangle & = &
  \bar{u}_b(p_b) \Big\{ \gamma_{\mu} F_1^V + p_{t, \mu} F_2^V +
  p_{b,\mu} F_3^V \Big\} u_t(p_t), \\ 
  \langle b(p_b) |J^A_{\mu} | t(p_t) \rangle & = &
  \bar{u}_b(p_b) \Big\{ \gamma_{\mu} F_1^A + p_{t, \mu} F_2^A +
  p_{b,\mu} F_3^A \Big\} \gamma_5 u_t(p_t).
 \end{Eqnarray}  
 \reseteqn

 \noindent In the Standard Model the appropriate current combination is given by
 $ J^V_\mu - J^A_\mu $. 

 We shall immediately take the limit $ m_b \rightarrow 0 $ of the one-loop
 expressions given in~\cite{gounaris} (see also Appendix C)\footnote{
 We have recalculated the one-loop results of Ref.~\cite{gounaris} and have
 found an acknowledged typo in the scalar form factors $ F_3(Q^2) $ and
 $ H_3(Q^2) $ of~\cite{gounaris}. The typo is corrected by replacing the factor
 $ (m_2 - m_1) / Q^2 $ in the last line of Eq.(A.8) of Ref.~\cite{gounaris} 
 by $ (m_2 - m_1) / (2 Q^2) $.}.
 Keeping only the finite terms and the relevant mass (M) ($ \ln y $ and
 $ \ln^2 y $) and infrared (IR) ($ \ln (\Lambda^2) $) singular logarithmic terms
 one obtains the rather simple result

 \alpheqn
 \begin{Eqnarray}
  F_1^V & = & \pp F_1^A = 1 \!-\! \frac{\alpha_s(q^2)}{4 \pi} C_F \Big(
  4 \!+\! \frac{1}{x^2} \ln(1 \!-\! x^2) \!+\!
  \ln \Big( \frac{y}{1 \!-\! x^2} \frac{\Lambda^4}{(1 \!-\! x^2)^2} \Big)
  \!+\! \\ & & \hspace{3.5cm} \!+\!
  2 \ln \Big( \frac{\Lambda^2}{y} \frac{1}{1 \!-\! x^2} \Big)
  \ln \Big( \frac{y}{1 \!-\! x^2} \Big) \!+\!  2 \Li (x^2) \Big), \nonumber \\
  F_2^V & = & - F_2^A = \frac{1}{m_t} \frac{\alpha_s(q^2)}{4 \pi} C_F \,
  \frac{2}{x^2} \Big(+ 1 + \frac{1 - x^2}{x^2} \ln (1 - x^2) \Big), \\
  F_3^V & = & - F_3^A = \frac{1}{m_t} \frac{\alpha_s(q^2)}{4 \pi} C_F \,
  \frac{2}{x^2} \Big(- 1 + \frac{2 \, x^2 \!-\! 1}{x^2} \ln (1 \!-\! x^2) \Big),
 \end{Eqnarray}
 \reseteqn

 \noindent where we have denoted the scaled gluon mass by
 $ \Lambda = m_g / m_t $. The dilog function $ \mbox{Li}_2(x) $ is defined by

 \begin{equation}
  \mbox{Li}_2(x) := - \int\limits_{0}^{x} \frac{\ln(1 - z)}{z} \, dz
 \end{equation} 

 Note that the one-loop contribution is purely real. This can be understood
 from an inspection of the one-loop Feynman diagram Fig.~1(b) which does not
 admit any nonvanishing physical two-particle cut. The fact that one has
 $ F^V_1 = F^A_1 $ and $ F^V_i = - F^A_i $ for $ i = 2,3 $ results from setting
 the $ b $-quark mass to zero. This can be seen by moving the chiral $ (1 - 
 \gamma_5) $ factor in the one-loop integrand numerator to the left. Because
 $ m_b $ is set to zero the Dirac numerator string will thus begin with
 $ \bar{u}_b (1 + \gamma_5) $ leading to the above pattern of relations between
 the loop amplitudes.
 We mention that the gluon mass regulator scheme can be converted to the 
 dimensional reduction scheme by the  replacement $ \log \Lambda^2 \rightarrow
 1 / \epsilon - \gamma_E + \log 4 \pi \mu^2/q^2 $ where $ 2 \epsilon
 = 4-N $, $ \gamma_E $ is the Euler-Mascharoni constant $ \gamma_E = 0.577
 \ldots, $ and $ \mu $ is the QCD scale parameter.


\section{\bf Tree graph contribution}

 The tree graph contribution results from the square of the real gluon emission
 graphs shown in Fig.1c and 1d. Omitting again the weak coupling factor
 $ V_{tb} \, g / \sqrt{2} $ for the time being the corresponding hadron tensor
 is given by

 \begin{Eqnarray} 
 \label{Hadrontensor}
  {\cal H}^{\mu \nu} & = & - 4 \pi \alpha_s \, C_F \,
  \frac{8}{(k \!\cdot\! p_t)(k \!\cdot\! p_b)} \Bigg\{ \! -
  \frac{k \!\cdot\! p_t}{k \!\cdot\! p_b} \Big[ (p_b \!\cdot\! p_b) \Big(
  k^{\mu} \, \bar{p}_t^{\nu} + k^{\nu} \, \bar{p}_t^{\mu} -
  k \!\cdot\! \bar{p}_t \, g^{\mu \nu} \Big) +
  \\ & & \hspace{-1.3cm} \rule{0mm}{6mm} + i \Big(
  \epsilon^{\alpha \beta \mu \nu} \, (p_b \!-\! k) \!\cdot\! \bar{p}_t -
  \epsilon^{\alpha \beta \gamma \nu} (p_b \!-\! k)^{\mu} \, \bar{p}_{t,\gamma} + 
  \epsilon^{\alpha \beta \gamma \mu} (p_b \!-\! k)^{\nu} \, \bar{p}_{t,\gamma} 
  \Big) k_{\alpha} \, p_{b,\beta} \Big] +
  \nonumber \\ & & \hspace{-1.3cm} \rule{0mm}{6mm} +
  \frac{k \!\cdot\! p_b}{k \!\cdot\! p_t} \Big[ (\bar{p}_t \!\cdot\! p_t)
  \Big( k^{\mu} \, p_b^{\nu} + k^{\nu} \, p_b^{\mu} -
  k \!\cdot\! p_b \, g^{\mu \nu} -
  i \, \epsilon^{\alpha \beta \mu \nu} k_{\alpha} \, p_{b,\beta} \Big) +
  \nonumber \\ & & \hspace{-1.3cm} \rule{0mm}{6mm} - 
  (\bar{p}_t \!\cdot\! k) \Big( (p_t \!-\! k)^{\mu} \, p_b^{\nu} +
  (p_t \!-\! k)^{\nu} \, p_b^{\mu} - (p_t \!-\! k) \!\cdot\! p_b \, g^{\mu \nu} -
  i \, \epsilon^{\alpha \beta \mu \nu}
  (p_t \!-\! k)_{\alpha} \, p_{b,\beta} \Big) \Big] +
  \nonumber \\ & & \hspace{-1.3cm} \rule{0mm}{6mm}  -
  (\bar{p}_t \!\cdot\! p_b) \Big( k^{\mu} \, p_b^{\nu} +
  k^{\nu} \, p_b^{\mu} - k \!\cdot\! p_b \, g^{\mu \nu} -
  i \, \epsilon^{\alpha \beta \mu \nu} k_{\alpha} \, p_{b,\beta} \Big) +
  (p_t \!\cdot\! p_b) \Big( k^{\mu} \, \bar{p}_t^{\nu} +
  k^{\nu} \, \bar{p}_t^{\mu} - k \!\cdot\! \bar{p}_t \, g^{\mu \nu} \Big) +
  \nonumber \\ & & \hspace{-1.3cm} \rule{0mm}{6mm} -
  (k \!\cdot\! p_b) \Big( p_t^{\mu} \, \bar{p}_t^{\nu} +
  p_t^{\nu} \, \bar{p}_t^{\mu} - p_t \!\cdot\! \bar{p}_t \, g^{\mu \nu} \Big) +
  (k \!\cdot\! p_t) \Big( (p_b \!+\! k)^{\mu} \, \bar{p}_t^{\nu} \!+\!
  (p_b \!+\! k)^{\nu} \, \bar{p}_t^{\mu} \!+\! (p_b \!+\! k) \!\cdot\! \bar{p}_t
  \, g^{\mu \nu} \Big) +
  \nonumber \\ & & \hspace{-1.3cm} \rule{0mm}{6mm} +
  (k \!\cdot\! \bar{p}_t)
  \Big( 2 p_b^{\mu} \, p_b^{\nu} - p_b \!\cdot\! p_b g^{\mu \nu} \Big) -
  i \Big(\epsilon^{\alpha \beta \mu \nu} \, (k \!\cdot\! \bar{p}_t) +
  \epsilon^{\alpha \beta \gamma \mu} \, k^{\nu} \bar{p}_{t,\gamma} -
  \epsilon^{\alpha \beta \gamma \nu} \, k^{\mu} \bar{p}_{t,\gamma} \Big)
  p_{b,\alpha} \, p_{t,\beta} +
  \nonumber \\ & & \hspace{-1.3cm} \rule{0mm}{6mm} +
  i \Big(\epsilon^{\alpha \beta \mu \nu} \, (p_t \!\cdot\! \bar{p}_t) +
  \epsilon^{\alpha \beta \gamma \mu} \, p^{\nu}_t \bar{p}_{t,\gamma} -
  \epsilon^{\alpha \beta \gamma \nu} \, p^{\mu}_t \bar{p}_{t,\gamma} \Big)
  k_{\alpha} \, p_{b,\beta} \Bigg\} + B^{\mu \nu} \cdot \Delta_{SGF}
  \nonumber
 \end{Eqnarray}
 \begin{equation} 
  \Delta_{SGF} := - 4 \pi \alpha_s \, C_F
  \Big( \frac{m_b^2}{(k \!\cdot\! p_b)^2} + \frac{m_t^2}{(k \!\cdot\! p_t)^2} -
  2 \frac{p_b \!\cdot\! p_t }{(k \!\cdot\! p_b)(k \!\cdot\! p_t)} \Big)
 \end{equation}

 \noindent where $ k $ is the 4-momentum of the emitted gluon. $ \Delta_{SGF} $
 is the IR-divergent {\it soft gluon function} and $ \bar{p}_t = p_t - m_t s_t $ as
 in Sec.4 \footnote{Contrary to the Born term case the polarization of the top
 quark cannot be accounted for by replacing {\it all} $ p_t $ momenta by their
 barred counterparts $ \bar{p}_t $.}.

 We have isolated the IR-singular part of the tree-graph contribution by
 splitting off a universal soft gluon factor which multiplies the lowest order
 Born term tensor $ B^{\mu \nu} $. This facilitates the treatment of the soft
 gluon  singularity to be regularized by a (small) gluon mass $ m_g $. Since the
 soft gluon factor is universal in that it multiplies the lowest order Born
 contribution, the requisite soft gluon integration has to be done only once and
 is identical for all eight structure functions. The result for the integrated
 soft gluon function is given in Sec.~8. Integrating only the soft gluon
 function $ \Delta_{SGF} $ and neglecting the finite part in
 Eq.(\ref{Hadrontensor}) amounts to what is called the soft gluon approximation.
 We emphasize that we always include the full tree-graph contribution
 (soft plus finite part) in our calculation. Also, we integrate over
 the full phase space of the gluon, and not only up to a given energy
 cut-off of the gluon.

 We have deliberately used a calligraphic notation for the tree graph hadron
 tensor ${\cal H}^{\mu \nu}$ in (\ref{Hadrontensor}) since ${\cal H}^{\mu \nu}$
 is {\it not} the hadron tensor $ H^{\mu \nu} $ defined in Sec.~2. In fact, the
 mass dimension of ${\cal H}^{\mu \nu}$ differs from that of $H^{\mu \nu}$. To
 relate the two hadron tensors one has to do the appropiate phase space
 integration on the tree graph hadron tensor.

 Next one makes use of the covariant projection operators and the covariant forms
 of the longitudinal and transverse polarization vectors defined in Sec.~2 to
 obtain the contributions to the three unpolarized and five polarized structure
 functions. Since we are aiming for a fully inclusive measurement regarding the
 $ X_b $ system the resulting expressions have to be integrated over the full
 two-dimensional phase space. As phase space variables we take the gluon energy
 $ k_0 $ and the W energy $ q_0 $ where the $ k_0 $ integration is done first.
 The phase space limits of the respective integrations are given by 

 \begin{equation}
 \label{phasenraumgrenzen1}
  k_{0,-} \le k_0 \le k_{0,+}
 \end{equation}
 and 
 \begin{equation} 
 \label{phasenraumgrenzen2}
 m_{W} \le q_0 \le \frac{m_t^2 + m_W^2 - (m_b + m_g)^2}{2 m_t},
 \end{equation}
 where
 \begin{equation}
 \label{phasenraumgrenzen3}
  k_{0,\pm} = \frac{(m_t - q_0)(M_{+}^2 - 2 q_0 m_t) \pm \sqrt{q_0^2 - m_W^2}
  \sqrt{(M_{-}^2 - 2 q_0 m_t)^2 - 4 m_g^2 m_b^2}}{2 (m_t^2 + m_W^2 - 2 q_0 m_t)}
 \end{equation}
 and
 \begin{equation}
 \label{phasenraumgrenzen4} 
 M_{\pm}^2 := m_t^2 + m_W^2 - m_b^2 \pm m_g^2.
 \end{equation}

 It is clear from Eq.(\ref{phasenraumgrenzen1}-\ref{phasenraumgrenzen4}) that
 the integration boundaries considerably simplify when the gluon mass is set to
 zero. In particular the second square root factor in the $ k_{0,\pm} $ boundary
 turns into a polynomial in $ q_0 $ which is an essential simplification for the
 second $ q_0 $ integration. This observation is at the core of our tree-level
 integration strategy exemplified by the partitioned form of
 Eq.~(\ref{Hadrontensor}). The soft gluon singularity has been isolated and
 brought into a simple form. The remaining part of the tree-graph contribution
 is IR finite and can be integrated without the gluon mass regulator.

 The integration over the gluon energy  $ k_0 $ ($ k_{0,-} \le k_0 \le k_{0,+} $)
 is simple and the results will not be presented here in explicit analytical
 form. Instead we present some representative results on the differential
 $ W $-boson energy distribution that result from the real gluon emission graphs
 Fig.~1c and 1d in graphical form in Figs.~4 and 5. 
 Fig.~4 shows the $W$-boson energy distribution for the total rate
 $ d \Gamma_{U+L} / dq_0 $. The energy distribution rises sharply from the
 lower energy limit, where the $ W $-boson is produced at rest, then increases
 rapidly over the intermediate range of $ W $-boson energies and finally
 rises sharply again towards the end of the spectrum, where the soft gluon
 singularity is located. In Fig.5 we show the same distribution for the partial
 rate into positive helicity W-bosons $ d \Gamma_{+} / dq_0 $ 
 $(\Gamma_+=\frac{1}{2}(\Gamma_U+\Gamma_F))$ for $ m_b=0 $ and
 for $ m_b \neq 0 $. As mentioned before there is no Born term  contribution to
 $d \Gamma_{+} / dq_0 $ for
 $ m_b = 0 $ and thus $d \Gamma_{+} / dq_0 $ possesses
 no IR singularity in this limit. The absence of the IR singularity in the 
 $ m_b = 0 $ case (dashed line) is quite apparent in Fig.~4. The distribution
 rises moderately fast from the lower end of the spectrum, then turns down
 over the intermediate range of energies and finally tends to zero at
 the end of the spectrum where the phase space closes. The $ m_b = 0 $ (dashed
 line) and  $ m_b \neq 0 $ (full line) distributions lie on top of each other
 for most of the lower part of the spectrum. Starting at around 4.8 GeV below
 the upper phase space boundary the two distributions begin to diverge from each
 other. Whereas the $ m_b = 0 $ curve turns down and goes to zero at the end of
 the spectrum the $ m_b \neq 0 $ curve starts to rise again and, in fact, tends
 to infinity at the end of the spectrum due to its IR singular behaviour.
 Note the huge differences in scale of the $ d \Gamma_{U+L} / dq_0 $ and the
 $ d \Gamma_{+} / dq_0 $ distributions which will be reflected in big
 differences in the total $ \alpha_s $-corrections for the two respective rates.

 The second integration over the energy of the $W$-boson is more difficult.
 Details can be found in Sec.~8 and in the Appendices. As it turns out the
 analytical $ m_b \neq 0 $ results are quite lengthy. We thus chose to present
 our $ m_b = 0 $ results first since they are sufficiently simple to be presented
 in compact form. They have been obtained by taking the $ m_b \rightarrow 0 $
 limit of our $ m_b \neq 0 $ results written down in Sec.~8. For practical
 purposes the $ m_b = 0 $ results are sufficiently accurate for top decays
 since $ m_b \neq 0 $ effects are generally quite small. This is particularly
 true if a running $b$ quark mass at the top mass scale is used. Quantitative
 results on the $ \alpha_s $ $ m_b \neq 0 $ corrections are given at the end of
 Sec.~8 as well as in~\cite{FGKM}.


\section{\bf  Complete \protect \boldmath $ O(\alpha_s) $ results
  for \protect \boldmath $ m_b = 0 $}

 We are now in the position to put together our $ m_b = 0 $ results. We add
 together the Born term results from Sec.~4, the one-loop results from Sec.~5 and
 the $ m_b \rightarrow 0 $ limit of the integrated tree graph results according
 to Sec.~8. The mass and infrared singular log terms cancel among the
 $ O(\alpha_s) $ one-loop and tree-graph contributions as they must according to
 the Lee-Nauenberg theorem and one remains with a finite result.
 We choose to present our results in terms of scaled rate functions defined by
 $ \hat{\Gamma}_{i} := \Gamma_{i} / \Gamma_{0} $ ($i$ = $ U+L $, $ U^P +L^P $,
 $ U $, $ L $, $ F $, $ S $ , $ U^P $, $ L^P $, $ F^P $, $ S^P $, $ I^P $,
 $ A^P $) with $ \Gamma_{0} = \Gamma_{U+L} (\mbox{Born}) $ given by
 ($x=m_W/m_t$)

 \begin{equation}
  \Gamma_{0} = \Gamma_{U+L}(Born) =
  \frac{G_F \, m_W^2 \, m_t}{8 \sqrt{2} \, \pi} 
  | V_{tb} |^2 \frac{(1 - x^2)^2 (1 + 2 x^2)}{x^2}.
 \end{equation}

 \noindent The angular decay distribution reads

 \begin{Eqnarray}
 \label{hatDiffRate}
  \frac{d \hat{\Gamma}}{d \cos \theta_p d \cos \theta d \phi} & = &
  \frac{1}{4 \, \pi} \Big\{ 
  \frac{3}{8} (\hat{\Gamma}_U + P \cos \theta_p \hat{\Gamma}_{U^P})
  (1 + \cos^2 \theta) + \hspace{5mm} \\ & & \hspace{-1.5cm} +
  \frac{3}{4} (\hat{\Gamma}_L + P \cos \theta_p \hat{\Gamma}_{L^P})
  \sin^2 \theta + \nonumber \\ & & \hspace{-1.5cm} + 
  \frac{3}{4} (\hat{\Gamma}_F + P \cos \theta_p \hat{\Gamma}_{F^P})
  \cos \theta + \nonumber \\ & & \hspace{-1.5cm} +
  \frac{3}{2 \sqrt{2}} \hat{\Gamma}_{I^P} P
  \sin \theta_p \sin 2 \theta \cos \phi + \nonumber \\ & & \hspace{-1.5cm} +
  \frac{3}{\sqrt{2}} \hat{\Gamma}_{A^P} P
  \sin \theta_p \sin \theta \cos \phi \Big\}, \nonumber
 \end{Eqnarray}

 \noindent where $ P $ is the degree of polarization of the top quark. As
 mentioned before one recovers the angular decay distribution written down in
 \cite{c8} when substituting the $ m_b=0 $ Born term expressions from Table~1 in 
 Eq.(\ref{hatDiffRate}). 

 The various reduced rates $\hat {\Gamma}_i$ are given by

 \begin{Eqnarray}
  \hat{\Gamma}_{U+L} & = & 1 \!+\!
  \frac{\alpha_s} {2 \pi} C_F \frac{x^2}{(1 \!-\! x^2)^2 (1 \!+\! 2 x^2)}
  \Bigg\{ \frac{(1 \!-\! x^2)(5 \!+\! 9 x^2 \!-\! 6 x^4)}{2 x^2} \!-\!
  \frac{2 (1 \!-\! x^2)^2 (1 \!+\! 2 x^2) \pi^2}{3 x^2}
  \nonumber \\ & & \hspace{-0.96cm} \,-\,
  \frac{(1 \!-\! x^2)^2 (5 \!+\! 4 x^2)}{x^2} \ln (1 \!-\! x^2) \!-\! 
  \frac{4 (1 \!-\! x^2)^2 (1 \!+\! 2 x^2)}{x^2} \ln(x) \ln(1 \!-\! x^2) \!-\!
  4 (1 \!+\! x^2) \!\times\! \hspace{5mm} \\ & & \hspace{-0.96cm} \,\times\, 
  (1 \!-\! 2 x^2) \ln(x) - \frac{4 (1 \!-\! x^2)^2 (1 \!+\! 2 x^2)}{x^2}
  \mbox{Li}_2(x^2) \Bigg\}, \nonumber \\[0.6cm]
  \hat{\Gamma}_{(U+L)^P} & = & \frac{1 \!-\! 2 x^2}{1 \!+\! 2 x^2} \!+\!
  \frac{\alpha_s} {2 \pi} C_F \frac{x^2}{(1 \!-\! x^2)^2 (1 \!+\! 2 x^2)}
  \Bigg\{ \!-\! \frac{(1 \!-\! x)^2 (15 \!+\! 2 x \!-\! 5 x^2 \!-\! 12 x^3 \!+\!
  2 x^4)}{2 x^2} + \nonumber \\ & & \hspace{-0.96cm} \,+\,
  \frac{(1 \!+\! 4 x^2) \pi^2}{3 x^2} \!-\! \frac{(1 \!-\! x^2)^2 
  (1 \!-\! 4 x^2)}{x^2} \ln(1 \!-\! x) \!-\! \frac{(1 \!-\! x^2)(3 \!-\! x^2)
  (1 \!+\! 4 x^2)}{x^2} \ln(1 \!+\! x) \!+\!
  \nonumber \\ & & \hspace{-0.96cm} \,-\,  
  \frac{4 (1 \!-\! x^2)^2 (1 \!-\! 2 x^2)}{x^2} \mbox{Li}_2(x) +
  \frac{4 (2 \!+\! 5 x^4 \!-\! 2 x^6)}{x^2} \mbox{Li}_2(-x) \Bigg\},
 \end{Eqnarray}

 \vspace{-1.0cm}

 \begin{Eqnarray}
  \hat{\Gamma}_U & = & \frac{2 x^2}{1 \!+\! 2 x^2} \!+\!
  \frac{\alpha_s} {2 \pi} C_F \frac{x^2}{(1 \!-\! x^2)^2 (1 \!+\! 2 x^2)}
  \Bigg\{ \!-\! (1 \!-\! x^2)(19 \!+\! x^2) + \frac{2 (5 \!+\! 5 x^2 \!-\!
  2 x^4) \pi^2}{3} \!+\! \nonumber \\ & - &
  2 \frac{(1 \!-\! x^2)^2(1 \!+\! 2 x^2)}{x^2} \ln (1 \!-\! x^2) -
  4 (5 \!+\! 7 x^2 \!-\! 2 x^4) \ln(x) - 2 (1 \!-\! x)^2 \times 
  \nonumber \\ & \times &
  \frac{(5 \!+\! 7 x^2 \!+\! 4 x^3)}{x} \ln(x) \ln(1 \!-\! x) +
  \frac{2 (1 \!+\! x)^2(5 \!+\! 7 x^2 \!-\! 4 x^3)}{x} \ln(x) \ln(1 \!+\! x) +
  \nonumber \\ & - &
  \frac{2(1 \!-\! x)^2 (5 \!+\! 4 x \!+\! 15 x^2 \!+\! 8 x^3)}{x} \mbox{Li}_2(x)
  \!+\! \frac{2(1 \!+\! x)^2 (5 \!-\! 4 x \!+\! 15 x^2 \!-\!8 x^3)}{x} 
  \mbox{Li}_2(-x) \Bigg\}, \\[0.6cm]
  \hat{\Gamma}_{L} & = & \frac{1}{1 \!+\! 2 x^2} +
  \frac{\alpha_s} {2 \pi} C_F \frac{x^2}{(1 \!-\! x^2)^2 (1 \!+\! 2 x^2)}
  \Bigg\{ \frac{ (1 \!-\! x^2 )(5 \!+\! 47 x^2 \!-\! 4 x^4)}{2 x^2} -
  \frac{2 \pi^2}{3} \times \nonumber \\ & \times &
  \frac{(1 \!+\! 5 x^2 \!+\! 2 x^4)}{x^2} \!-\! \frac{3 (1 \!-\! x^2)^2}{x^2}
  \ln (1 \!-\! x^2) + 16 (1 \!+\! 2 x^2) \ln(x) -
  2 (1 \!-\! x)^2 \!\times\! \nonumber \\ & \times &
  \frac{2 \!-\! x \!+\! 6 x^2 \!+\! x^3}{x^2} \ln (1 \!-\! x) \ln(x) - 
  \frac{2 (1 \!+\! x)^2 (2 \!+\! x \!+\! 6 x^2 \!-\! x^3)}{x^2} \ln(x)
  \ln (1 \!+\! x) + \nonumber \\ & - &
  \frac{2 (1 \!-\! x)^2 (4 \!+\! 3 x \!+\! 8 x^2 \!+\! x^3)}{x^2}
  \mbox{Li}_2(x) - \frac{2(1 \!+\! x)^2 (4 \!-\! 3 x \!+\! 8 x^2 \!-\! x^3)}
  {x^2} \mbox{Li}_2(-x) \Bigg\}, \\[0.6cm]
  \hat{\Gamma}_{F} & = & \frac{-2 x^2}{1 \!+\! 2 x^2} \!+\!
  \frac{\alpha_s} {2 \pi} C_F \frac{x^2}{(1 \!-\! x^2)^2 (1 \!+\! 2 x^2)}
  \Bigg\{ - 2 (1 \!-\! x)^2(3 \!-\! 4 x) \,+\, \frac{2 (2 \!+\! x^2) \pi^2}{3} +
  \nonumber \\ & + & 
  \frac{2 (1 \!-\! x^2)^2 (1 \!+\! 2 x^2)}{x^2} \ln(1 \!-\! x) \,+\,
  \frac{2 (1 \!-\! x^2) (1 \!-\! 9 x^2 \!+\! 2 x^4)}{x^2} \ln (1 \!+\! x) +
  \nonumber \\ & + & 
  8 (1 \!-\! x^2)^2 \, \mbox{Li}_2(x) \,+\, 8 (1 \!+\! 3 x^2 \!-\! x^4) \,
  \mbox{Li}_2(-x) \Bigg\}, \\[0.6cm]
  \hat{\Gamma}_{S} & = & \frac{1}{1 \!+\! 2 x^2} \!+\!
  \frac{\alpha_s} {2 \pi} C_F \frac{x^2}{(1 \!-\! x^2)^2 (1 \!+\! 2 x^2)}
  \Bigg\{ \frac{9 (1 \!-\! x^2)^2}{2 x^2} - \frac{2 (1 \!-\! x^2)^2 \pi^2}
  {3 x^2} + \nonumber \\ & + &
  \frac{(1 \!-\! x^2)^2 (2 \!-\! 5 x^2)}{x^4} \ln(1 \!-\! x^2) -
  4 (1 \!-\! x^2) \ln(x) - \frac{4 (1 \!-\! x^2)^2}{x^2} \ln(x)
  \ln(1 \!-\! x^2) + \nonumber \\ & - &
  \frac{4 (1 \!-\! x^2)^2}{x^2} \, \mbox{Li}_2(x^2) \Bigg\}, \\[0.6cm] 
  \hat{\Gamma}_{U^P} & = & \frac{- 2 x^2}{1 \!+\! 2 x^2} + 
  \frac{\alpha_s}{2 \pi} C_F \frac{x^2}{(1 \!-\! x^2)^2 (1 \!+\! 2 x^2)}
  \Bigg\{ \!-\! \frac{(1 \!-\! x)^2 (12 \!-\! 55 x \!+\! 6 x^2 \!-\! x^3)}{x} -
  \frac{10 \pi^2}{3} \times \nonumber \\ & \times &
  (2 \!+\! x^2) + \frac{2 (1 \!-\! x^2)^2 (1 \!+\! 2 x^2)}{x^2}
  \ln (1 \!-\! x) + \frac{2 (1 \!-\! x^2)(7 \!+\! 21 x^2 \!+\! 2 x^4)}{x^2}
  \ln (1 \!+\! x) \!+\! \nonumber \\ & + &
  8 (1 \!-\! x^2)^2 \mbox{Li}_2(x) - 8 ( 11 \!+\! 3 x^2 \!+\! x^4 ) 
  \mbox{Li}_2(-x) \Bigg\}, \\[0.6cm]
  \hat{\Gamma}_{L^P} & = & \frac{1}{1 \!+\! 2 x^2} +
  \frac{\alpha_s} {2 \pi} C_F \frac{x^2}{(1 \!-\! x^2)^2 (1 \!+\! 2 x^2)}
  \Bigg\{ -(15 \!-\! 22 x \!+\! 105 x^2 \!-\! 24 x^3 \!+\! 4 x^4) \times
  \nonumber \\ & \times &
  \frac{(1 \!-\! x)^2}{2 x^2} \!+\! \frac{(1 \!+\! 24 x^2 \!+\! 10 x^4) \pi^2}
  {3 x^2} - \frac{3 (1 \!-\! x^2)^2}{x^2} \ln(1 \!-\! x) -
  \frac{(1 \!-\! x^2)(17 \!+\! 53 x^2)}{x^2} \times \nonumber \\ & \times &
  \ln (1 \!+\! x) - \frac{4 (1 \!-\! x^2)^2}{x^2} \mbox{Li}_2(x) + 
  \frac{4 (2 \!+\! 22 x^2 \!+\! 11 x^4)}{x^2} \mbox{Li}_2(-x) \Bigg\}, \\[0.6cm]
  \hat{\Gamma}_{F^P} & = & \frac{2 x^2}{1 \!+\! 2 x^2} \!+\!
  \frac{\alpha_s}{2 \pi} C_F \frac{x^2}{(1 \!-\! x^2)^2 (1 \!+\! 2 x^2)}
  \Bigg\{ 2 (1 \!-\! x^2) (4 \!+\! x^2) \!-\!
  \frac{2 (1 \!+\! x^2 \!+\! 2 x^4) \pi^2}{3} + \nonumber \\ & - &
  \frac{2 (1 \!-\! x^2)^2 (1 \!+\! 2 x^2)}{x^2} \ln (1 \!-\! x^2) - 
  4 (2 \!-\! 5 x^2 \!-\! 2 x^4) \ln(x) - \ln (x) \ln (1 \!-\! x) 
  \times \nonumber \\ & \times &
  \frac{4 (1 \!-\! x)^2 (1 \!+\! 3 x \!+\! 2 x^2 \!+\! 2 x^3)}{x} \!+\!
  \frac{4 (1 \!+\! x)^2 (1 \!-\! 3 x \!+\! 2 x^2 \!-\! 2 x^3)}{x} \ln (x) 
  \ln (1 \!+\! x) + \nonumber \\ & - & 
  \frac{4 (1 \!-\! x )^2 (1 \!+\! 5 x \!+\! 6 x^2 \!+\! 4 x^3)}{x}
  \mbox{Li}_2(x) + \frac{4 (1 \!+\! x)^2 (1 \!-\! 5 x \!+\! 6 x^2 \!-\! 4 x^3)}
  {x} \mbox{Li}_2(-x) \Bigg\}, \hspace{0.2cm} \\[0.6cm]
  \hat{\Gamma}_{S^P} & = & \frac{1}{1 \!+\! 2 x^2} \!+\!
  \frac{\alpha_s}{2 \pi} C_F \frac{x^2}{(1 \!-\! x^2)^2 (1 \!+\! 2 x^2)}
  \Bigg\{ - \frac{(1 \!-\! x)^2 (11 \!-\! 6 x \!-\! 7 x^2)}{2 x^2} +
  \frac{(1 \!+\! 2 x^2) \pi^2}{3 x^2} + \nonumber \\ & + & 
  \frac{(1 \!-\! x^2)^2(2 \!-\! 5 x^2)}{x^4} \ln(1 \!-\! x) +
  \frac{(1 \!-\! x^2)(2 \!-\! 9 x^2 \!+\! x^4)}{x^4} \ln(1 \!+\! x) -
  \frac{4 (1 \!-\! x^2)^2}{x^2} \times \nonumber \\ & \times &
  \mbox{Li}_2 (x) + \frac{4 (2 \!+\! x^4)}{x^2} \mbox{Li}_2 (-x)
  \Bigg\}, \\[0.6cm]
  \hat{\Gamma}_{I^P} & = & \frac{-x}{\sqrt{2} (1 \!+\! 2 x^2)} \!+\!
  \frac{\alpha_s}{2 \pi} C_F \frac{x^2}{(1 \!-\! x^2)^2 (1 \!+\! 2 x^2)}
  \Bigg\{ \frac{(1 \!-\! x)^2 (12 \!-\! 7 x \!+\! 12 x^2)}{\sqrt{2} x} -
  \frac{\pi^2}{6 \sqrt{2}} \times \nonumber \\ & \times &
  \frac{(5 \!+\! 19 x^2 \!+\! 2 x^4)}{x} + \frac{(1 \!-\! x^2)^2
  (1 \!+\! 5 x^2)}{2 \sqrt{2} x^3} \ln(1 \!-\! x) + \frac{(1 \!-\! x^2)
  (1 \!+\! 30 x^2 \!+\! 21 x^4)}{2 \sqrt{2} x^3}
  \times \nonumber \\ & \times &
  \ln(1 \!+\! x) + \frac{2 \sqrt{2} (1 \!-\! x^2)^2}{x} \mbox{Li}_2 (x) -
  \frac{\sqrt{2} (7 \!+\! 15 x^2 \!+\! 4 x^4)}{x} \mbox{Li}_2 (-x)
  \Bigg\}, \\[0.6cm]
  \hat{\Gamma}_{A^P} & = & \frac{x}{\sqrt{2} (1 \!+\! 2 x^2)} \!+\!
  \frac{\alpha_s}{2 \pi} C_F \frac{x^2}{(1 \!-\! x^2)^2 (1 \!+\! 2 x^2)}
  \Bigg\{ \frac{(1 \!-\! x^2)(1 \!+\! 2 x^2)}{\sqrt{2} x} -
  \frac{\pi^2}{6 \sqrt{2}} \times \nonumber \\ & \times &
  \frac{(3 \!-\! 5 x^2 \!+\! 6 x^4)}{x} -
  \frac{(1 \!-\! x^2)^2 (1 \!+\! 5 x^2)}{2 \sqrt{2} x^3} \ln(1 \!-\! x^2) -
  \frac{x (5 \!-\! 11 x^2)}{\sqrt{2}} \ln(x) + \nonumber \\ & - &
  \frac{(1 \!-\! x)^2(3 \!+\! 7 x \!+\! 6 x^2)}{\sqrt{2} x} \ln(x)
  \ln(1 \!-\! x) - \frac{(1 \!+\! x)^2 (3 \!-\! 7 x \!+\! 6 x^2)}{\sqrt{2} x}
  \ln(x) \ln(1 \!+\! x) \!+\! \nonumber \\ & - & 
  \frac{(1 \!-\! x)^2 (7 \!+\! 15 x \!+\! 10 x^2)}{\sqrt{2} x} \mbox{Li}_2 (x) -
  \frac{(1 \!+\! x)^2 (7 \!-\! 15 x \!+\! 10 x^2)}{\sqrt{2} x} \mbox{Li}_2 (-x)
  \Bigg\}.
\end{Eqnarray}

 As mentioned in the Introduction the results for the total rate $ (U+L) $
 agree with the analytical results given in~\cite{c15,c16,c17,c18} and
 in~\cite{fischer99}.
 The six (mass zero) diagonal structure functions $ U,L,F $ and $ U^P,L^P,F^P $
 have already been listed in~\cite{fischer99}. They had been checked against the
 corresponding numerical results given in~\cite{c19,c20,c21}. The results on
 the non-diagonal structure functions $ A^P $ and $ I^P $ are new.
 As concerns the unpolarized transverse structure functions explicit expressions
 for the two linear combinations $ T_{+} = \frac{1}{2} (U + F) $ and
 $ T_{-} = \frac{1}{2} (U - F) $ relevant for the interpretation
 of the CDF measurement \cite{cdf} have been given in \cite{FGKM}. 

 We have also included $ O(\alpha_s) $ results on the unpolarized and
 polarized scalar structure functions $ \hat{\Gamma}_{S} $ and 
 $ \hat{\Gamma}_{S^P} $. They determine the $ m_b = 0 $ unpolarized
 and polarized decay of the top quark into a charged Higgs 
 ($ t \rightarrow b + H^+ $) as it occurs e.g. in the
 Two-Higgs-Doublet-Model ($ 2HDM $). This can be seen as follows. The
 scalar projection of the Standard Model (SM) left-chiral current
 structure $ \gamma^\mu P_L $ determines the coupling of the SM
 Goldstone boson, i.e. $ \slashq P_L \rightarrow (m_t P_R - m_b P_L) $.
 This would be the coupling structure of the charged Higgs in the $ 2HDM $
 when the ratio of vacuum expectation values is taken to be one. It is 
 then evident that, for $ m_b=0 $, the scalar structure functions
 $ \hat{\Gamma}_{S} $ and $ \hat{\Gamma}_{S^P} $ describe the decay
 $ t \rightarrow b + H^+$ in the $ 2HDM $ irrespective of the value of
 the ratio of vacuum expectation values.
 The unpolarized scalar structure function $ \hat{\Gamma}_{S} $ has been
 checked against the result of~\cite{czarnecki}. The result on the
 polarized scalar structure function $ \hat{\Gamma}_{S^P} $ is new.  

 Before turning to the numerical evaluation of the various contributions we
 would like to discuss the large $ m_t $ limit of the various helicity structure
 functions. As expected from the statements of the Goldstone boson equivalence
 theorem the longitudinal and scalar contributions $ L, L^P, S $ and $ S^P $
 dominate in this limit. In fact, setting $ x = 0 $ one finds

 \begin{Eqnarray} 
  \hat{\Gamma}_L & = & \hat{\Gamma}_S = 1 +
  \frac{\alpha_s}{2 \pi} \, C_F \,
  \Big( \frac{5}{2} - \frac{2}{3} \pi^2 \Big), \\
  \hat{\Gamma}_{L^P} & = & \hat{\Gamma}_{S^P} = 1 +
  \frac{\alpha_s}{2 \pi} \, C_F \, 
  \Big( - \frac{15}{2} + \frac{1}{3} \pi^2 \Big).
 \end{Eqnarray}

 That $ \hat{\Gamma}_L = \hat{\Gamma}_S $ and $ \hat{\Gamma}_{L^P} =
 \hat{\Gamma}_{S^P} $ for $ m_t \rightarrow \infty $ can be understood from the
 fact that the longitudinal and scalar polarisation vectors $ \epsilon^{\mu}(0) $
 and $ \epsilon^{\mu}(S) $ become equal to each other in this limit since the
 longitudinal polarisation vector then simplifies to  $ \epsilon^{\mu}(0) =
 q^\mu / m_W + O (m_W/q_0) $. The same observation is also at the heart of the
 proof of the Goldstone boson equivalence theorem. As concerns the tree graph
 contribution the statement that $ \epsilon^{\mu}(0) = q^\mu / m_W + O(m_W/q_0) $
 is certainly not true for all of three-body phase space as e.g. close to the
 phase space point where the $ W $-boson is at rest. The contribution from this
 phase space region to the three-body rate, however, becomes negligibly small
 when $ m_t \rightarrow \infty $.  

 We now turn to our numerical results. As numerical input values we take $
 m_t = 175 \mbox{ GeV} $  and $ m_W = 80.419 \mbox{ GeV} $. The strong coupling
 constant is evolved from $ \alpha_s(M_Z) = 0.1175 $ to $ \alpha_s(m_t) =
 0.1070 $ using two-loop running.
 The results are presented such that the reduced Born term rates are factored
 out from the reduced rates. This way of presenting the results allows one to
 quickly assess the size of the radiative corrections. One has

 \alpheqn
 \begin{Eqnarray}
  \hat{\Gamma}_{U+L^{\phantom{P}}}
                     & = & \pp 1 - 0.0854, \\ \rule{0mm}{5mm}
  \hat{\Gamma}_{U^{\phantom{P}}}
                     & = & \pp 0.297 \, (1 - 0.0624), \\ \rule{0mm}{5mm}
  \hat{\Gamma}_{L^{\phantom{P}}}
                     & = & \pp 0.703 \, (1 - 0.0951), \\ \rule{0mm}{5mm}
  \hat{\Gamma}_{F^{\phantom{P}}}
                     & = &   - 0.297 \, (1 - 0.0687), \\ \rule{0mm}{5mm}
  \hat{\Gamma}_{(U+L)^P} & = & \pp    0.406 \, (1 - 0.1162), \\ \rule{0mm}{5mm}
  \hat{\Gamma}_{U^P} & = &   - 0.297 \, (1 - 0.0689), \\ \rule{0mm}{5mm}
  \hat{\Gamma}_{L^P} & = & \pp 0.703 \, (1 - 0.0962), \\ \rule{0mm}{5mm}
  \hat{\Gamma}_{F^P} & = & \pp 0.297 \, (1 - 0.0639), \\ \rule{0mm}{5mm}
  \hat{\Gamma}_{I^P} & = &   - 0.228 \, (1 - 0.0810), \\ \rule{0mm}{5mm}
  \hat{\Gamma}_{A^P} & = & \pp 0.228 \, (1 - 0.0820), \\ \rule{0mm}{5mm}
  \hat{\Gamma}_{S^{\phantom{P}}}
                     & = & \pp 0.703 \, (1 - 0.0895), \\ \rule{0mm}{5mm}
  \hat{\Gamma}_{S^P} & = & \pp 0.703 \, (1 - 0.0922).
 \end{Eqnarray}
 \reseteqn  

 The radiative corrections to the unpolarized and polarized rate functions are
 sizeable. They range from $ -6.2 \, \%  $ for $ \hat{\Gamma}_U $ to
 $ -11.6 \, \% $ for $ \hat{\Gamma}_{(U+L)^P} $ compared to the rate
 correction of $ -8.5 \, \% $. The radiative corrections to the longitudinal and
 scalar contributions are largest. The radiative corrections all tend to go in
 the same direction. This is an indication that the bulk of the radiative
 corrections come from phase space regions close to the IR/M singular region
 where the radiative corrections are universal. When normalizing the rate
 functions to the total rate, as is appropriate for the definition of
 polarization observables, the size of the radiative corrections to the
 polarization observables is much reduced. For example, the $ O(\alpha_s) $
 radiative corrections decrease the ratio $ \Gamma_L/\Gamma_{U+L} $ by 
 $ 1.1 \, \% $ and increase the ratio  $ \Gamma_U/\Gamma_{U+L} $ and
 the magnitude of the ratio $ \Gamma_F /\Gamma_{U+L} $ by $ 2.5 \, \% $ and
 $ 1.8 \, \% $, resp., relative to their Born term ratios. The relative ratio
 $ \Gamma_U/\Gamma_L $ is increased by $ 3.6 \, \% $. The values of the
 radiative corrections to the polarization observables are, however, large
 enough that they must be included in a meaningful comparison of future high
 precision data with the theoretical predictions of the Standard Model.

 The combination $ (\hat{\Gamma}_U + \hat{\Gamma}_F)/2 $ determines the decay
 of an unpolarized top quark into a right-handed $ W $-boson. This combination
 vanishes at the Born term
 level for $ m_b=0 $ as Eqs.~(39) and (41) show. Adding up the corresponding
 numerical values of the $ O(\alpha_s) $ contributions in Eq.~(51) one finds that
 the right-handed $ W $-boson occurs only with $ 0.094 \, \% $ probability.
 The $ m_b \ne 0 $ effect in the Born term alone 
 already amounts to $ 0.036 \, \% $ (see Table~1).

 Altogether the $ O(\alpha_s) $ and the Born term  $ m_b \neq 0 $ corrections to
 the transverse-plus rate occur only at the sub-percent level. It is safe to
 say that, if top quark decays reveal a violation of the Standard Model (SM)
 $(V-A)$ current structure that exceeds the 1\% level, the violations must
 have a non-SM origin.
 In this context it is interesting to note that a possible $ (V+A) $ admixture
 to the SM $ t \rightarrow b $ current is already severely bounded indirectly to
 below 5\% by existing data on $ b \rightarrow s + \gamma $ decays \cite{fy94,
 cm94, gp00}.
   
 The rate combination $ (\hat{\Gamma}_U + \hat{\Gamma}_F)/2 $ is in fact not the
 only combination that vanishes at the Born term level for $ m_b=0 $.
 Considering the fact that one must have $ \lambda_W - \lambda_t = -1/2 $ at the
 Born term level the only surviving Born term level rate expressions are
 $ \hat{\Gamma}_{--}^{--}$, $ \hat{\Gamma}_{\mbox{oo}}^{++} $ and 
 $ \hat{\Gamma}_{-\mbox{o}}^{-+}$ as alluded to before in Sec.~4. The notation
 employed for the reduced rates follows the notation used in
 Eq.~(\ref{system3anfang}). The remaining rate expressions vanish at the Born
 term level but become populated at $ O(\alpha_s) $. They are

 \begin{Eqnarray}
 \label{small-number}
 \hat{\Gamma}_{++}^{++} & = &
 \frac{1}{4} (\hat{\Gamma}_{U} + \hat{\Gamma}_{F} +
 \hat{\Gamma}_{U^P} + \hat{\Gamma}_{F^P}) = 0.000 \, 833, \nonumber \\
 \hat{\Gamma}_{\mbox{oo}}^{--} & = &
 \frac{1}{2}(\hat{\Gamma}_{L} - \hat{\Gamma}_{L^P}) = 0.000 \, 389, \nonumber \\
 \hat{\Gamma}_{+\mbox{o}}^{+-} & = &
 (\hat{\Gamma}_{I^P} + \hat{\Gamma}_{A^P}) =- 0.000 \, 236, \\
 \hat{\Gamma}_{++}^{--} & = &
 \frac{1}{4} (\hat{\Gamma}_{U} + \hat{\Gamma}_{F} -
 \hat{\Gamma}_{U^P} - \hat{\Gamma}_{F^P}) = 0.000 \, 093, \nonumber \\
 \hat{\Gamma}_{--}^{++} & = &
 \frac{1}{4} (\hat{\Gamma}_{U} - \hat{\Gamma}_{F} +
 \hat{\Gamma}_{U^P} - \hat{\Gamma}_{F^P}) = 0.000 \, 120 . \nonumber 
 \end{Eqnarray}
 
 \noindent As remarked on before the latter two reduced rates
 $ \hat{\Gamma}_{++}^{--} $ and $ \hat{\Gamma}_{--}^{++} $ vanish at the Born
 term level even for $ m_b \neq 0 $ since the net helicity of these transitions
 $ |\lambda_W - \lambda_t| = 3/2 $ exceeds that of the $ b $ quark
 $ |\lambda_b| = 1/2 $.
 
 The four reduced rates $ \hat{\Gamma}_{++}^{++} $,
 $ \hat{\Gamma}_{\mbox{oo}}^{--} $, $ \hat{\Gamma}_{++}^{--} $ and
 $ \hat{\Gamma}_{--}^{++} $ are positive definite quantities since
 they result from squares of helicity amplitudes. Contrary to these
 $ \hat{\Gamma}_{+\mbox{o}}^{+-} $ is an interference contribution
 and thus can be negative as it in fact is. In Eq.~(\ref{small-number})
 we have also included the numerical values for the above five structure
 function combinations  resulting from the (tree graph) $ \alpha_s $
 corrections. They are all very small at the sub per mille level.

 In Sec.4 (Fig.3) we have shown a lego plot of the Born term two-fold angular
 decay distribution in $ \cos \theta $ and $ \cos \theta_P $. In order 
 to be able to exhibit the size of the $ \alpha_s $ corrections we show in Fig.6
 a contour plot of the same two-fold angular decay distribution with
 and without radiative corrections, again setting $ P = 1 $.
 The radiative corrections are not very large in the upper two quadrants and
 become largest in the lower left quadrant of the contour plot when both
 $ \cos \theta $ and $ \cos \theta_P $ tend to one.

 Instead of analyzing the three-fold or two-fold angular decay
 distributions one can also consider single angle decay distributions.
 They are obtained by integrating over the two respective complementary 
 decay angles. For the $ \cos \theta $ distribution one obtains

 \begin{equation}
  \frac{d \widehat{\Gamma}}{d \cos \theta} =
  \frac{3}{8} (\widehat{\Gamma}_U +2 \widehat{\Gamma}_L)
  (1 + \alpha_{\theta} \cos \theta + \beta_{\theta} \cos^2 \theta) ,
 \end{equation}

 \noindent where

 \begin{Eqnarray}
  \alpha_{\theta} & = & 2 \, \frac{\widehat{\Gamma}_F}
  {\widehat{\Gamma}_U + 2 \widehat{\Gamma}_L} \qquad
  \Bigg( = - \frac{2 x^2}{1 + x^2} = -0.349 \Bigg), \\
  \beta_{\theta} & = & \phantom{2 \,} 
  \frac{\widehat{\Gamma}_U - 2 \widehat{\Gamma}_L}
  {\widehat{\Gamma}_U + 2 \widehat{\Gamma}_L} \qquad
  \Bigg( = - \frac{1 - x^2}{1 + x^2} = -0.651 \Bigg).
 \end{Eqnarray}

 \noindent We have added the analytical and numerical Born term results for the
 asymmetry parameters in brackets using $ x^2=0.211 $. The $ O(\alpha_s) $
 values for the asymmetry parameters are $ \alpha_\theta = -0.357 $ and
 $ \beta_\theta = -0.641 $, i.e. the $ \alpha_s $ corrections raise the
 magnitude of $ \alpha_\theta $ by $ 2.3 \,\% $ and lower the magnitude
 of $ \beta_\theta $ by $ 1.5 \,\% $. In Fig.7 we show the $ \cos \theta $
 distribution both for the Born term case and the radiatively corrected case.
 There is a pronounced forward-backward asymmetry. In the forward direction
 the differential Born term rate drops to zero. As discussed before the
 $ O(\alpha_s)$ rate does not vanish in the forward direction due to real
 gluon emission. However, the radiative corrections are so small that the
 nonvanishing of the  $ O(\alpha_s)$ rate in the forward direction cannot
 be discerned at the scale of the plot. In absolute terms the radiative
 corrections are largest for $ \cos \theta \approx 0 $ because of the large
 size of the radiative corrections to the longitudinal rate
 $ \hat{\Gamma}_{L} $. Note that $ \alpha_\theta $ is not the conventional
 forward-backward asymmetry parameter which is defined by

 \begin{equation}
  \alpha_{FB} = \frac
  {d \Gamma (0 \ge \theta \ge \frac{\pi}{2}) -
   d \Gamma (\frac{\pi}{2} \ge \theta \ge \pi)}
  {d \Gamma (0 \ge \theta \ge \frac{\pi}{2}) +
   d \Gamma (\frac{\pi}{2} \ge \theta \ge \pi)} =
  \frac{3}{4} \frac{\widehat{\Gamma}_F}{\widehat{\Gamma}_{U+L}} \quad
  \Bigg( = - \frac{3}{2} \frac{x^2}{1 + 2 x^2} = - 0.223 \Bigg).
 \end{equation} 
 
 \noindent The $ \alpha_s $ corrections raise $ \alpha_{FB} $ by $ 1.7\,\% $
 in magnitude.
 
 For the $ \cos \theta_P $ distribution one obtains

 \begin{equation}
  \frac{d \widehat{\Gamma}}{d \cos \theta_p} =
  \frac{1}{2} (\widehat{\Gamma}_{U+L})
  (1 + P \alpha_{\theta_p} \cos \theta_p),
  \end{equation} 
 
 \noindent where

 \begin{equation}
  \alpha_{\theta_p} = \frac{\widehat{\Gamma}_{(U+L)^P}}
  {\widehat{\Gamma}_{U+L}} \qquad
  \Bigg(= \frac{1 - 2 x^2}{1 + 2 x^2} = 0.406 \Bigg).
 \end{equation}

 \noindent The $ \alpha_s $-corrections lower
 $ \alpha_{\theta_P} $ by $ 3.4\,\% $.

 Finally, the $ \phi $ distribution reads

 \begin{equation}
  \frac{d \widehat{\Gamma}}{d \phi} =
  \frac{1}{2 \pi} (1 +P \gamma_{\phi} \cos \phi),
 \end{equation}

 \noindent where

 \begin{equation}
  \gamma_{\phi} = \frac{3 \pi^2}{8 \, \sqrt{2}}
  \frac{\widehat{\Gamma}_{A^P}}{\widehat{\Gamma}_{U+L}} \qquad
  \Bigg( = \frac{3 \pi^2}{16} \frac{x}{1 +2 x^2} = 0.597 \Bigg).
 \end{equation}

 \noindent The $ \cos \phi $ dependent contribution from $ \hat{\Gamma}_{I_P} $
 has dropped out because of having integrated over the full range of 
 $ \cos \theta $. If desired the contribution of $ \hat{\Gamma}_{I_P} $
 to the $ \phi $ distribution can be retained if one integrates only over half
 the range of $ \cos\theta $. The $ \alpha_s $-corrections raise $ \gamma_\phi $
 by the small amount of $ 0.32\,\% $. In Fig.8 we show the $ \phi $ distribution
 both for the Born term case and the radiatively corrected case setting $ P=1 $.  

 \section{\bf  Complete \protect \boldmath $ O(\alpha_s) $ results for
 \protect \boldmath $ m_b \neq 0 $}

 Differing from the presentation of our $ m_b =0 $ results in Sec.~7 we shall
 present our $ m_b \ne 0 $ results in a form where each of the separate
 contributions to the rate remains identified. In particular we do not explicitly
 cancel the IR terms coming from the one-loop and tree graph contributions.
 We thus write

 \begin{Eqnarray} 
 \label{complete}
  \Gamma_{i}^{QCD} & = & \Gamma_i(\mbox{\scriptsize Born}) \!+\!
  \frac{m_t^3 |V_{tb}|^2 G_F}{8 \, \sqrt{2} \, \pi}
  \Bigg\{ \sum\limits_{\tau} \kappa_{i, \, \tau} \, F_{\tau} \Bigg\} +
  \frac{4}{\sqrt{\lambda}} \Gamma_i(\mbox{\scriptsize Born}) S(\Lambda) -
  \frac{\alpha_s}{4 \, \pi}
  \frac{m_t^3 |V_{tb}|^2 \, G_F \, x^2}{8 \sqrt{2} \, \pi}
  \hspace{1cm} \\ & & \hspace{-1.5cm} \times \Bigg\{
  \sum\limits_{n=-1,0} \rho_{(n), \, i} \,\, {\cal R}_{(n)} +
  \sum\limits_{m,n} \rho_{(m,n), \, i} \,\, {\cal R}_{(m,n)} +
  \sum\limits_{n=0,1} \sigma_{(n), \, i} \,\, {\cal S}_{(n)} +
  \sum\limits_{m,n} \sigma_{(m,n), \, i} \,\, {\cal S}_{(m,n)}
  \Bigg\} \nonumber
 \end{Eqnarray}
 
 The first term in Eq.~(\ref{complete}) represents the Born term contribution
 which is given by  

 \begin{equation}
 \label{complete1}
  \Gamma_{i}(\mbox{Born}) =
  \frac{G_F \, m_W^2 \, |V_{tb}|^2}
  {8 \, \sqrt{2} \, \pi \, m_t}
  \, \sqrt{\lambda} \, B_i
 \end{equation}

 \noindent where the $ B_i $ are the Born term rates listed in Table 1.
 The Born term contribution $ \Gamma_{i}(\mbox{Born}) $ also appears as
 a factor in the third term where it multiplies the soft gluon factor
 $S(\Lambda)$. The index $i$ runs over the various structure function
 labels $ i = U+L $, $ U^P+L^P $, $ U $, $ U^P $, $ L $, $ L^P $, $ F $,
 $ F^P $, $ S $, $ S^P $, $ I^P $ and $ A^P $.

 The second term in Eq.~(\ref{complete}) represents the one-loop contribution
 which is obtained by folding the one-loop amplitude in Appendix C with the
 Born term amplitude and then doing the appropriate projection onto the various
 structure functions. The appropriate coefficient functions $ \kappa_{i,\tau} $
 are listed in Table 2. The coefficient functions $ \kappa_{i,\tau} $
 multiply the $ \alpha_s $ one-loop amplitudes $ F_{\tau} = F_1^V, F_2^V, F_3^V,
 F_1^A, F_2^A, F_3^A $ which are listed in Appendix C. We label the one-loop
 amplitudes consecutively by the index $ \tau = 1,\ldots ,6 $. Note that
 Table~2 contains only the vector current coefficient functions
 $ \kappa_{i,\tau} \, (\tau=1,2,3) $. The axial vector coefficient functions
 labelled by $ \tau = 4,5,6 $ can be easily obtained from the vector current
 coefficient functions by the substitution
 
 \begin{equation}
 \label{complete2}
  \kappa_{F_1^A} =   \kappa_{F_1^V} \Big|_{y \rightarrow -y}, \quad
  \kappa_{F_2^A} = - \kappa_{F_2^V} \Big|_{y \rightarrow -y}, \quad
  \kappa_{F_3^A} = - \kappa_{F_3^V} \Big|_{y \rightarrow -y} .
 \end{equation}

 \begin{table}
 \renewcommand{\arraystretch}{1.2}
 \begin{center} 
  \begin{tabular}{|l|c|c|c|}\hline
   i & $ \kappa_{F_1^V},i $ & $ \kappa_{F_2^V},i $ &
   $ \kappa_{F_3^V},i $ \\ \hline \hline
   $ U \!+\! L $ &
   $ \Ss{\sqrt{\lambda} \, ((1 - y)^2 - x^2) ((1 + y)^2 + 2 x^2)} $ &
   $ \Ss{\frac{1}{2} \, m_t \, \sqrt{\lambda^3} \, (1 + y)} $ &
   $ \Ss{\frac{1}{2} \, m_t \, \sqrt{\lambda^3} \, (1 + y)} $ \\ \hline
   $ U^P \!+\! L^P $ &
   $ \Ss{\lambda \, (1 - 2 x^2 - y^2)} $ &
   $ \Ss{\frac{1}{2} \, m_t \, \lambda \, ((1 + y)^2 - x^2) (1 - y)} $ &
   $ \Ss{\frac{1}{2} \, m_t \, \lambda \, ((1 + y)^2 - x^2) (1 - y)} $ \\ \hline
   $ U $ &
   $ \Ss{2 \, \sqrt{\lambda} \, ((1 - y)^2 - x^2) \, x^2} $ &
   $ 0 $ & $ 0 $ \\ \hline
   $ U^P $ &
   $ - 2 \, x^2 \, \lambda $ &
   $ 0 $ & $ 0 $ \\ \hline
   $ L $ &
   $ \Ss{\sqrt{\lambda} \, ((1 - y)^2 - x^2) (1 + y)^2} $ &
   $ \Ss{\frac{1}{2} \, m_t \, \sqrt{\lambda^3} \, (1 + y)} $ &
   $ \Ss{\frac{1}{2} \, m_t \, \sqrt{\lambda^3} \, (1 + y)} $ \\ \hline
   $ L^P $ &
   $ \Ss{\lambda \, (1 - y^2)} $ &
   $ \Ss{\frac{1}{2} \, m_t \, \lambda \, ((1 + y)^2 - x^2)^2 (1 - y)} $ &
   $ \Ss{\frac{1}{2} \, m_t \, \lambda \, ((1 + y)^2 - x^2)^2 (1 - y)} $
   \\ \hline
   $ F $ &
   $ \Ss{- 2 \, \lambda \, x^2} $ &
   $ 0 $ & $ 0 $ \\ \hline
   $ F^P $ &
   $ \Ss{2 \, \sqrt{\lambda} \, ((1 - y)^2 - x^2) \, x^2} $ &
   $ 0 $ & $ 0 $ \\ \hline
   \renewcommand{\arraystretch}{1.2}
   $ S $ &
   $ \Ss{\sqrt{\lambda} \, ((1 + y)^2 - x^2) (1 - y)^2} $ &
   $ \begin{array}{c}
     \Ss{\frac{1}{2} \, m_t \, \sqrt{\lambda} \, ((1 + y)^2 - x^2) \times} \\
     \Ss{(1 + x^2 - y^2) (1 - y)}
     \end{array} $ &
   $ \begin{array}{c}
     \Ss{\frac{1}{2} \, m_t \, \sqrt{\lambda} \, ((1 + y)^2 - x^2) \times} \\
     \Ss{(1 - x^2 - y^2) (1 - y)}
     \end{array} $ \\ \hline
   \renewcommand{\arraystretch}{1.5}
   $ S^P $ &
   $ \Ss{\lambda \, (1 - y^2)} $ &
   $ \Ss{\frac{1}{2} \, m_t \, \lambda \, (1 + x^2 - y^2) (1 + y)} $ &
   $ \Ss{\frac{1}{2} \, m_t \, \lambda \, (1 - x^2 - y^2) (1 + y)} $ \\ \hline
   $ I^P $ &
   $ \Ss{- \frac{1}{\sqrt{2}} \, \lambda \, x} $ &
   $ \Ss{- \frac{1}{4 \, \sqrt{2}} \, m_t \, \lambda \, 
     ((1 + y)^2 - x^2) \, x} $ &
   $ \Ss{- \frac{1}{4 \, \sqrt{2}} \, m_t \, \lambda \, 
     ((1 + y)^2 - x^2) \, x} \rule[-2.5mm]{0mm}{3mm} $ \\ \hline
   $ A^P $ &
   $ \Ss{\frac{1}{\sqrt{2}} \, \sqrt{\lambda} \,
     ((1 - y)^2 - x^2) (1 + y) \, x} \rule[-2.5mm]{0mm}{3mm} $ &
   $ \Ss{\frac{1}{4 \, \sqrt{2}} \, m_t \, \sqrt{\lambda^3} \, x} $ &
   $ \Ss{\frac{1}{4 \,  \sqrt{2}} \, m_t \, \sqrt{\lambda^3} \, x} $ \\ \hline
  \end{tabular}
 \end{center}
 \renewcommand{\arraystretch}{1}
 \caption{Coefficient functions $ \kappa_{i, \, \tau} $ that determine the
  contributions of the $ \alpha_s $ vector current one-loop amplitudes to
  the different rates $ \Gamma_i $ ($x=m_W/m_t$, $y=m_b/m_t$,
  $\lambda=1+x^4+y^4 - 2 x^2 y^2 - 2 x^2 - 2 y^2)$.}
 \end{table}

 The third term in Eq.~(\ref{complete}) contains the result of integrating the
 soft gluon function $ \Delta_{SGF} $ in Eq.~(\ref{Hadrontensor}). The result
 depends on the (small) IR regularisation parameter $ \Lambda = m_g/m_t $ as
 indicated in the argument of the soft gluon factor $ S(\Lambda) $. The
 universal soft gluon factor $ S(\Lambda) $ is obtained by explicit integration
 and reads  

 \begin{Eqnarray} 
 \label{complete3}
  S(\Lambda) & = & - \frac{\alpha_s}{4 \, \pi} \, C_F \,
  \Bigg\{ (1 \!-\! x^2 \!+\! y^2) \Bigg[
  2 \, \Li(1 - w_1 w_{\mu}) + \Li(1 - w_1^2) -
  \Li \Bigg( \! 1 - \frac{w_1}{w_{\mu}} \Bigg) +
  \nonumber \\ & + &
  \frac{1}{4} \ln^2 (w_1 w_{\mu}) + \ln (w_1 w_{\mu}) \, 
  \Big\{ \ln \Big( \frac{\lambda \, w_1}{x \, y \, \Lambda} \Big) +
  \frac{1}{2} \Big\} \Bigg] + 2 \, \sqrt{\lambda} \, \Big\{
  \ln \Big( \frac{\lambda}{x \, y \, \Lambda} \Big) - 2 \Big\} + \nonumber \\ & + & 
  \ln \Bigg( \frac{w_1}{w_{\mu}} \Bigg) -
  2 \, y^2 \ln (w_1) \Bigg\}, 
 \end{Eqnarray}

 \noindent where as in \cite{c14} we have used the abbreviations
 
 \begin{equation} 
  \label{w1wmu}
  w_1 = \frac{x}{y} \cdot \frac{1 - x^2 + y^2 - \sqrt{\lambda}}
  {1 + x^2 - y^2 + \sqrt{\lambda}}, \qquad
  w_{\mu} = \frac{x}{y} \cdot \frac{1 - x^2 + y^2 - \sqrt{\lambda}}
  {1 + x^2 - y^2 - \sqrt{\lambda}}.
 \end{equation}

 In the limit $ y \rightarrow 0 $ one has

 \begin{Eqnarray} 
  S(\Lambda) & = & - \frac{\alpha_s}{4 \, \pi} \, C_F \,
  \Bigg\{ (1 - x^2) \Bigg[\frac{\pi^2}{3} - 4 + \ln^2 y - 2 \ln \Lambda +
  (1 + 2 \ln \Lambda) \ln \Big( \frac{1 - x^2}{y} \Big) + \nonumber \\ & - &
  2 \ln \Big( \frac{x}{1 - x^2} \Big) +
  \ln(1 - x^2) \ln \Big( \frac{x^2}{1 - x^2} \Big) +
  \Li (x^2) \Bigg] + \ln x^2 \Bigg\}. 
 \end{Eqnarray}

 \noindent In agreement with the Lee-Nauenberg theorem the logarithmic
 dependence on the IR regularisation parameter $ \Lambda $ can be seen to cancel
 between the loop and the soft gluon contributions for each of the ten structure
 functions.

 The fourth term in Eq.~(\ref{complete}) finally contains the result of
 integrating the finite piece in the tree graph contribution
 Eq.~(\ref{Hadrontensor}), again after having done the appropriate projections.
 The result is given in terms of a set of standard integrals
 $ {\cal R}_{(n)} $, $ {\cal R}_{(m,n)} $, $ {\cal S}_{(n)} $ and
 $ {\cal S}_{(m,n)} $ which are listed in Appendix A. Appendix B gives the
 values of the coefficient functions $ \rho_{(n), \, i} $,
 $ \rho_{(m,n), \, i} $, $ \sigma_{(n), \, i} $ and $ \sigma_{(m,n), \, i} $
 that multiply the standard set of integrals in the various helicity structure
 functions. In Table 3 we have listed the range of values of the parameters
 $ m $ and $ n $ that characterize the different types of tree graph integrals.

 At this point it is perhaps appropiate to offer an excuse to the potential user of
 our $ m_b \ne 0 $ results that our results are presented in a multiply nested
 form to be collected from Eqs.(\ref{complete}, \ref{complete1}, \ref{complete2},
 \ref{complete3}, \ref{w1wmu}), Table~2 and Appendices A, B and C. Contrary to
 the $ m_b=0 $ results where a closed form representation was possible a
 presentation of unnested closed form expressions  for $ m_b \ne 0 $ would
 require an extraordinary amount of space because of the presence of many
 different log and dilog functions and products thereof. Codes of the relevant
 expressions can be obtained from the authors on request.
 
  When we have evaluated Eq.~(\ref{complete}) numerically the IR factors
  proportional
 to $\ln \Lambda $ in the one-loop and tree graph contributions were set
 to zero by hand. The numerical evaluation of the remaining part is 
 quite stable numerically. In particular the limit $ m_b \rightarrow 0 $
 is numerically quite smooth. This is demonstrated in Fig. 9 where we plot
 the bottom mass dependence of the total rate. Note that the $ O(\alpha_s) $
 rate shows less dependence on the bottom mass than the Born term rate.

 \begin{table}[htbp]
 \begin{center}
  \renewcommand{\arraystretch}{1.2}
  \begin{tabular}{|l|c|c|c|c|} \hline
   i & $ \rho^{1}_{(n)},i $ & $ \rho^{0}_{(m,n)},i $ &
   $ \sigma^{1}_{(n)},i $ & $ \sigma^{0}_{(m,n)},i $
   \rule[-3mm]{0mm}{5mm} \\ \hline \hline
   $ U \!+\! L $ & $ - $ & $ (-2,-1)...(0,-1) $ & $ - $ &
   $ (0,0),(1,0) $ \\ \hline
   $ U^P \!+\! L^P $ & $ -1,0 $ & $ (-2,0)...(1,0) $ &
   $ 0,1 $ & $ (0,0),(0,1)...(2,1) $ \\ \hline
   $ U $ & $ - $ & $ (-2,1)...(2,1) $ & $ - $ &
   $ (0,2)...(3,2) $ \\ \hline
   $ U^P $ & $ -1,0 $ & $ (-2,2)...(3,2) $ &
   $ 0,1 $ & $ (0,0),(0,3)...(4,3) $ \\ \hline
   $ L $ & $ - $ & $ (-2,1)...(2,1) $ & $ - $ &
   $ (0,2)...(3,2) $ \\ \hline
   $ L^P $ & $ -1,0 $ & $ (-2,2)...(3,2) $ &
   $ 0,1 $ & $ (0,0),(0,3)...(4,3) $ \\ \hline
   $ F $ & $ -1,0 $ & $ (-2,0)...(1,0) $ &
   $ 0,1 $ & $ (0,0),(0,1)...(2,1) $ \\ \hline
   $ F^P $ & $ - $ & $ (-2,1)...(2,1) $ & $ - $ &
   $ (0,2)...(3,2) $ \\ \hline   
   $ S $ & $ - $ & $ (-2,-1)...(0,-1) $ & $ - $ &
   $ (0,0),(1,0) $ \\ \hline
   $ S^P $ & $ -1,0 $ & $ (-2,0)...(1,0) $ &
   $ 0,1 $ & $ (0,0),(0,1)...(2,1) $ \\ \hline
   $ I^P $ & $ -1,0 $ & $ (-2,2)...(2,2) $ &
   $ 0,1 $ & $ (0,0),(0,3)...(3,3) $ \\ \hline
   $ A^P $ & $ - $ & $ (-2,1)...(1,1) $ & $ - $ &
   $ (0,2)...(2,2) $ \\ \hline
  \end{tabular}
  \renewcommand{\arraystretch}{1}
 \end{center}
 \caption{Range of values of powers $ m,n $ in the different
  basic tree graph integrals}
 \end{table}

 The quality of the $ m_b = 0 $ approximation has been discussed before at the
 Born level. For example, at the Born term level the total rate is decreased by
 $ 0.27 \, \% $ when going from $ m_b = 0 $ to $ m_b = 4.8 \mbox{ GeV} $. Using
 the $ O(\alpha_s) $ $ m_b \ne 0 $ results from this section one finds that the
 $ m_b \ne 0 $ corrections to the total $ O(\alpha_s) $ rate reduce the rate by 
 $ 0.16 \, \% $ compared to the Born term reduction of $ 0.27 \, \% $,
 i.e. the $ m_b \ne 0 $ corrections to the $ \alpha_s $-contribution
 alone tend to counteract the $ m_b \ne 0 $ effect in the Born
 term in the total rate (see also Fig. 9). The $ m_b \ne 0 $ corrections from the
 $ \alpha_s $-contributions alone are surprisingly large considering the fact that the factor
 multiplying the $ \alpha_s $-corrections $ C_F \, \alpha_s/(2 \pi) = 0.023 $
 is a rather small number. This can be understood in part by noting that the
 $ \alpha_s $-contributions contain terms proportional to $ (m_b^2/m_W^2)
 \ln(m_b^2/m_t^2) = -0.026 $ which is not a very small number. A further
 discussion of  $ m_b \ne 0 $ effects for the $ \alpha_s $-contributions can be
 found in \cite{FGKM}. Noteworthy is a large 20\% correction to the
 $O(\alpha_s)$ transverse-plus rate $ \hat{\Gamma}_+ $ due to $m_b$ effects
 \cite{FGKM}. That the bottom quark mass effect is so large in $ \hat{\Gamma}_+ $
 can be appreciated in part by looking at the differential distribution in Fig.~5. 
 We emphasize again that the mass effect may have been overestimated
 due to using a fixed pole mass, rather than a running mass which is smaller at
 the top mass scale.
    

\section{\bf Summary and conclusion }

 We have obtained analytical expressions for the $ O(\alpha_s) $ radiative
 corrections to the three unpolarized and five polarized structure functions that
 govern the decay of a polarized top quark. Although bottom quark mass effects
 are quite small in top quark decays we have retained the full bottom mass
 dependence in our calculation. In the limit $ m_b \rightarrow 0 $ the analytical
 results considerably simplify leading to compact expressions for the eight
 structure functions which are listed in the main text. The full mass dependence
 of our analytical results is written down in Sec.~8 and in the Appendices A and
 B. These finite mass results will prove useful for the theoretical description
 of $ b \rightarrow c $ bottom meson and bottom baryon decays (see e.g. \cite{fischer2000}).

 For top quark decays the radiative corrections to the structure functions
 range from $ -6.2 \, \% $ to $ -11.6 \, \% $ where the radiative corrections to
 the unpolarized longitudinal structure functions $ \hat{\Gamma}_L $ and
 the polarized structure function $ \hat{\Gamma}_{(U+L)^P} $ are largest. These corrections are to be
 compared with the correction to the total rate which is $ -8.5 \, \% $.
 The radiative corrections to the structure functions all go in the same
 directions indicating that the bulk of the radiative corrections derive from
 contributions close to the IR/M region of phase space where the radiative
 corrections are universal. Nevertheless the span of values of the radiative
 corrections exceeds $ 5 \, \% $ and must be taken into account in a future
 comparison with precision experiments. The radiative corrections to rate
 combinations that vanish at the Born term level have been found to be rather
 small. In particular, the $\alpha_s$-correction to the normalized rate of an
 unpolarized top into positive helicity $W$-bosons amounts to only 0.1\%.
 As discussed in Sec.~7, the minuteness of the $\alpha_s$-contribution to
 positive helicity $W$-bosons is of relevance when discussing a possible
 $(V+A)$-admixture to the Standard Model current.

 We have also determined the $ O(\alpha_s) $ corrections to unpolarized and
 polarized $ q_1 \rightarrow q_2 $ scalar current transitions. For $ t
 \rightarrow b $ transitions these scalar current transitions are relevant
 for top quark decays into a bottom quark and a charged Higgs as they occur in
 the two-Higgs doublet model. For $ b \rightarrow c $ transitions these
 transition matrix elements are needed e.g. for the description of the
 semi-inclusive decays of the $ B $-mesons and the $ \Lambda_b $ into
 spin-zero $ D_s $ mesons~\cite{fischer2000,aleksan}.

 In this paper we have only studied the first order QCD corrections to the
 structure functions in polarized top decays. For the total rate one obtains a
 correction of $ -8.5 \, \% $. Second order QCD corrections to the rate are
 expected to amount to $ -2.6 \, \% $ \cite{CHSS} while electroweak corrections
 are known to increase the rate by $ +1.7 \, \% $ \cite{c14,EMMS}. For a high
 precision comparison of theory and experiment of the structure functions it
 would therefore be desirable to calculate the two-loop $ O(\alpha_s^2) $ and the
 electroweak one-loop corrections to the eight structure functions. While the
 two-loop QCD corrections to the structure functions are very difficult and are
 therefore not likely to be done in the next few years the calculation of the
 one-loop electroweak corrections to the eight structure functions is presently
 under way \cite{dgkm}. Finite width corrections will also have to be accounted
 for. They lower the total width by 1.56\% \cite{dgkm,jk93} and affect the
 different partial helicity rates by differing amounts \cite{dgkm}. 

 We would like to conclude this paper with a speculative note concerning a
 possible top quark mass measurement from an angular decay analysis using the
 fact that the structure functions are top mass dependent.  
 This suggestion is much in the spirit of the suggestion of Grunberg {\it et al.}
 who advocated a similar measurement of heavy quark masses in the context of
 $ e^{+} e^{-} $-annihilations~\cite{grunberg}. Assume that the percentage
 measurement errors on a $ L/(U+L) $ and $ U/(U+L) $ measurement are $ \delta_L $
 and $ \delta_U $, respectively. The percentage error on the mass measurement
 will be denoted by $ \delta $, i.e. we write $ m_t = \bar{m}_t(1 \pm \delta) $
 where $ \bar{m}_t $ is some given central value of the top mass. From the
 dependence of the respective Born term ratios on the mass ratio $ x = m_W/m_t $
 (assuming that the $ W $-mass is fixed) one finds that the percentage error on
 the top mass measurement is given by $ \delta = \delta_L (1+2 x_0^2)/(4x_0^2) $
 and by $ \delta = \delta_U (1+2 x_0^2)/2 $, resp., where we write
 $ x^2 = x_0^2 ( 1 \mp 2 \delta) $ with $ x_0 = m_W / \bar{m}_t $.
 If we take $ m_t = 175 \mbox{ GeV} $ as central value ($ x_0^2 = 0.211 $) this
 would imply that an 1\% error on the angular structure function measurement
 would  allow one to determine the top quark mass with $ 1.7 \, \% $ and
 $ 0.7 \, \% $ accuracy, depending on whether the angular measurement was done on
 the longitudinal $ (L) $ or on the unpolarized-transverse $ (U) $ (or for that
 matter $ (F) $) mode. Since the radiative corrections change the ratios
 $ \Gamma_L/\Gamma_{U+L} $ and $ \Gamma_U/\Gamma_{U+L} $ by $ 1.1 \, \% $ and
 $ 2.4 \, \% $, respectively, it is clear that one has to use the full
 $ O(\alpha_s) $ results for the angular structure functions if such
 experimental accuracies can be reached. This is illustrated in Figs.~10 and 11
 where we plot the top mass dependence of $ \Gamma_L/\Gamma_{U+L} $ and
 $ \Gamma_U/\Gamma_{U+L} $ for the Born term case and the $ O(\alpha_s) $ case
 for $ m_b = 0 $.
 Note that the  $ O(\alpha_s) $ curves are horizontally displaced from the Born
 term curves by approximately 3 and 3.4 GeV, resp., meaning that one would make
 the correponding mistakes in the top mass determination from a measurement of
 the angular structure functions if the Born curves were used instead of the
 radiatively corrected ones. The present TEVATRON RUN I uncertainties on the
 top mass are around $ 4 \, \% $ which is anticipated to be  improved to
 $ 1.7 \, \% $ during the initial stages of TEVATRON RUN II. It remains to be
 seen whether a mass determination based on angular measurements as proposed here
 can compete with the conventional method using invariant mass reconstruction.

 \clearpage


 {\bf Acknowledgements:}
 M.~Fischer and M.C.~Mauser were partly supported by the DFG (Germany) through
 the Graduiertenkolleg ``Teilchenphysik bei hohen und mittleren Energien'' and
 its successor ``Eichtheorien'' at the University of Mainz. M.C.~Mauser was also
 supported by the BMBF (Germany) under contract 05HT9UMB/4. S.~Groote and
 J.G.~K\"orner acknowledge partial support by the BMBF (Germany) under contract
 06MZ865 and S.G. acknowledges support by the DFG. We would like to thank
 B.~Lampe for initial participation in the project, H.~Spiesberger for 
 illuminating discussions and A.~Arbuzov for checking some of the formula. 


\begin{appendix}

\section{Integrals}
 
 In this Appendix we catalogue the basic set of tree graph integrals that are
 needed in our $ m_b \neq 0 $ calculation and give their analytical results.


\subsection{Basic integrals}

 In the first step of the tree graph integration one integrates over the gluon
 energy $ k_0 $. After having done the integration on the gluon energy it proves
 to be convenient to perform a shift in the $ W $-energy $ q_0 $ integration
 variable by introducing the variable $ z = 1 + x^2 - 2 q_0 / m_t $. One then
 encounters the following set of integrals

 \alpheqn
 \begin{Eqnarray}
  {\cal R}_{(m,n)} & := & 
  \int\limits_{y^2 + \epsilon^{\prime}_{2}}^{(1 - x)^2 - \epsilon^{\prime}_{1}}
  \hspace{-5mm} \frac{z^m \, dz}{\sqrt{\lambda^n(1,x^2,z)}}, \quad
  {\cal R}_{(n)} := 
  \int\limits_{y^2 + \epsilon^{\prime}_{2}}^{(1 - x)^2 - \epsilon^{\prime}_{1}}
  \hspace{-5mm} \frac{dz}{(z - y^2) \, \sqrt{\lambda^n(1,x^2,z)}}, \\
  {\cal S}_{(m,n)} & := & 
  \int\limits_{y^2 + \epsilon^{\prime}_{2}}^{(1 - x)^2 - \epsilon^{\prime}_{1}}
  \hspace{-5mm} \frac{z^m \, dz}{\sqrt{\lambda^n(1,x^2,z)}} \ln \Big( \frac
  {1-x^2+z+\sqrt{\lambda(1,x^2,z)}}{1-x^2+z-\sqrt{\lambda(1,x^2,z)}} \Big), \\
  {\cal S}_{(n) \phantom{m,}} & := & 
  \int\limits_{y^2 + \epsilon^{\prime}_{2}}^{(1 - x)^2 - \epsilon^{\prime}_{1}}
  \hspace{-5mm} \frac{dz}{(z - y^2) \, \sqrt{\lambda^n(1,x^2,z)}} \ln \Big(\frac
  {1-x^2+z+\sqrt{\lambda(1,x^2,z)}}{1-x^2+z-\sqrt{\lambda(1,x^2,z)}} \Big).
 \end{Eqnarray}
 \reseteqn

 \noindent where $\lambda(1,x^2,z)= 1 + x^4 + z^2 - 2 x^2 z - 2 x^2 - 2 z $.
 The required range of values of the parameters $ m $ and $ n $ are
 listed in Table~3. The cut-off parameters $ \epsilon^{\prime}_{1} $ and
 $ \epsilon^{\prime}_{2} $ are needed to account for the spurious singularities
 which are artificially introduced by partial fractioning the integrands. The
 spurious singularities cancel as they must when all contributions to a
 particular helicity structure function are summed.
  
 In order to get rid of the square roots the final substitution $ z =: 1 \!+\!
 x^2 \!-\! x (r + 1)/r $ is introduced. The variable $ r $ has to be
 integrated in the interval $ [1 \!+\! \epsilon_1, \eta \!-\! \epsilon_2] $,
 where

 \begin{equation}
  \eta = (1 + x^2 - y^2 + \sqrt{\lambda})/2x
  \qquad \mbox{and} \qquad  \lambda = \lambda(1,x^2,y^2)
 \end{equation}

 \noindent as before. The spurious cut-off parameters $ \epsilon_{1} $
 and $ \epsilon_{2} $ replace the above cut-off parameters
 $ \epsilon^{\prime}_{1} $ and $ \epsilon^{\prime}_{2} $ and cancel in all
 final expressions.
 
 In order to keep our results at a manageable length we introduce the following
 set of auxiliary functions

 \alpheqn
 \begin{Eqnarray} 
  {\cal L}_1 & := & \ln \Big( \frac{\eta-x}{\eta \, (1-\eta \, x)} \Big), \qquad
  {\cal L}_2 := \ln \Big( \frac{\eta \, (\eta-x)}{1-\eta\,x} \Big),
  \hspace{6.7cm} \\
  {\cal L}_3 & := & \ln \Big( \frac{(1-x)^2-y^2}{x} 
  \frac{(1-x)^2}{\epsilon_1^2 \, y^2} \Big), \qquad
  {\cal L}_4 := \ln \Big( \frac{(1+x)^2-y^2}{x} \frac{(1-x)^2}{4y^2} \Big), \\
  {\cal L}_5 & := & \ln \Big( \frac{1-x}{y} \Big), \qquad
  {\cal L}_6 := \ln \Big( \frac{\eta\,(1-x)}{\eta-x} \Big), \\[5mm]
  {\cal N}_0 & := & \Li (\eta\,x) + 
  \Li \Big( \frac{x}{\eta} \Big) - 2 \, \Li (x), \qquad
  {\cal N}_1 := \Li (\eta\,x) - \Li \Big( \frac{x}{\eta} \Big) \\
  {\cal N}_2 & := &  - \ln (\eta) \ln (1+x) +
  \ln \Big( \frac{\eta-x}{(\eta-1)(1+x)} \Big)
  \ln \Big( \frac{\eta-x}{\eta\,(1-\eta\,x)} \Big) + \\ & - &
  \Li \Big( \frac{1}{\eta} \Big) +
  \Li \Big( \frac{(\eta^2-1)\,x}{\eta-x} \Big) +
  \Li \Big( \frac{1-\eta\,x}{\eta-x} \Big), \nonumber \\
  {\cal N}_3 & := & - \ln (\eta) \ln (1-x) - 
  \ln \Big( \frac{(\eta+1)(1-x)}{\eta-x} \Big)
  \ln \Big( \frac{\eta-x}{\eta\,(1-\eta\,x)} \Big) + \\ & - &
  \Li \Big( \!\!- \frac{1}{\eta} \Big) +
  \Li \Big( \frac{(\eta^2-1)\,x}{\eta-x} \Big) +
  \Li \Big( \!\!- \frac{1-\eta\,x}{\eta-x} \Big), \nonumber
 \end{Eqnarray}

 and

 \begin{equation}
  \beta_{+}(n) := (x-1)^n + (x+1)^n, \qquad
  \beta_{-}(n) := (x-1)^n - (x+1)^n, 
 \end{equation}
 \begin{equation}
  \beta(n) := \frac{(x-1)^n}{\eta-1} - \frac{(1+x)^n}{\eta+1}.
 \end{equation}
 \reseteqn
 
 In the following we list our analytical results for the various types of
 integrals that are needed in our calculation.


 \subsection{Integrals of type R \protect\boldmath $ \!_{\Ss (m,n)} $}

 \alpheqn
 \begin{Eqnarray} 
 {\cal R}_{(-2,-1)} & = & \frac{\lambda^{1/2} }{y^2} +
 \frac{{\cal L}_2-{\cal L}_1}{2} - \frac{1+x^2}{1-x^2} \frac{{\cal L}_2+
 {\cal L}_1}{2}, \hspace{7.7cm} \\ 
 {\cal R}_{(-1,-1)} & = & - \lambda^{1/2} -
 (1+x^2) \frac{{\cal L}_2-{\cal L}_1}{2} + (1-x^2) \frac{{\cal L}_2+
 {\cal L}_1}{2},\\
 {\cal R}_{(0,-1) \pp} & = & \frac{1}{2} (1+x^2-y^2) \,
 \lambda^{1/2} - x^2 ({\cal L}_2-{\cal L}_1), \\
 {\cal R}_{(1,-1) \pp} & = & - \frac{1}{3} \lambda^{3/2} +
 (1+x^2) \left( \frac{1}{2} (1+x^2-y^2) \lambda^{1/2} - 
 x^2 ({\cal L}_2-{\cal L}_1) \right), \\[3mm]
 {\cal R}_{(-2,0) \pp} & = & \frac{1}{y^2} - \frac{1}{(1-x)^2}, \quad
 {\cal R}_{(-1,0)}^{0} =2 {\cal L}_5, \\
 {\cal R}_{(0,0) \pp \pp} & = & (1-x)^2 - y^2, \quad\
 {\cal R}_{(1,0)} = \frac{(1-x)^4}{2} - \frac{y^4}{2}, \quad
 {\cal R}_{(2,0)} = \frac{(1-x)^6}{3} - \frac{y^6}{3}, \\[3mm]
 {\cal R}_{(-2,1) \pp} & = & \frac{1}{(1-x^2)^2} \Bigg(
 \frac{\lambda^{1/2}}{y^2} + \frac{1+x^2}{1-x^2}
 \frac{{\cal L}_2+{\cal L}_1}{2}\Bigg), \quad
 {\cal R}_{(-1,1)} = \frac{1}{1-x^2} \frac{{\cal L}_2+{\cal L}_1}{2}, \\
 {\cal R}_{(0,1) \pp \pp} & = & \frac{{\cal L}_2-{\cal L}_1}{2}, \quad
 {\cal R}_{(1,1)} = - \lambda^{1/2} + (1+x^2)
 \frac{{\cal L}_2-{\cal L}_1}{2}, \\
 {\cal R}_{(2,1) \pp \pp} & = & - \frac{1}{2} y^2 \lambda^{1/2} -
 \frac{3}{2} (1+x^2) \, \lambda^{1/2} + (1+4x^2+x^4)
 \frac{{\cal L}_2-{\cal L}_1}{2}, \\
 {\cal R}_{(3,1) \pp \pp} & = & - \frac{1}{3} \lambda^{3/2} +
 \frac{3}{2} (1+x^2)(1+x^2-y^2) \lambda^{1/2} - (3+x^2)(1+3x^2)
 \lambda^{1/2} + \\ & + & (1+x^2)(1+8x^2+x^4) \frac{{\cal L}_2-{\cal L}_1}{2},
 \nonumber \\[3mm]
 {\cal R}_{(-2, 2) \pp} & = &
 \frac{1}{4x} \frac{{\cal L}_3}{(1-x)^4} - \frac{1}{4x} \frac{{\cal L}_4}
 {(1+x)^4}+ \frac{1}{(1-x^2)^2}
 \left( \frac{1}{y^2} - \frac{1}{(1-x)^2} \right), \\  
 {\cal R}_{\Ss{(-1, 2) \pp}} & = &
 \frac{1}{4x} \frac{{\cal L}_3}{(1-x)^2} - \frac{1}{4x}
 \frac{{\cal L}_4}{(1+x)^2}, \quad
 {\cal R}_{(0, 2)} = \frac{1}{4x} ({\cal L}_3-{\cal L}_4), \\
 {\cal R}_{(1, 2) \pp \pp} & = &
 \frac{(1-x)^2}{4x} {\cal L}_3 - \frac{(1+x)^2}{4x} {\cal L}_4 - 
 \frac{1}{2x} \beta_{-}(2) \, {\cal L}_5, \\
 {\cal R}_{(2, 2) \pp \pp} & = &
 \frac{(1-x)^4}{4x} {\cal L}_3 - \frac{(1+x)^4}{4x} {\cal L}_4 -
 \frac{1}{2x} \beta_{-}(4) \, {\cal L}_5 + \left( (1-x)^2 - y^2 \right), \\
 {\cal R}_{(3, 2) \pp \pp} & = &
 \frac{(1-x)^6}{4x} {\cal L}_3 - \frac{(1+x)^6}{4x} {\cal L}_4 -
 \frac{1}{2x} \beta_{-}(6) \, {\cal L}_5 \!+\!
 3 \left( (1\!-\!x)^2 \!-\! y^2 \right) (1 \!+\! x^2) - \frac{1}{2} \lambda,\\
 {\cal R}_{(4, 2) \pp \pp} & = &
 \frac{(1-x)^8}{4x} {\cal L}_3 - \frac{(1+x)^8}{4x} {\cal L}_4 -
 \frac{1}{2x} \beta_{-}(8) \, {\cal L}_5 + \left( (1-x^2) - y^2 \right) 
 \times \\ & \times & \left( \frac{1}{3} (1+x+x^2-y^2)^2 +
 (6+17x^2+6x^4) \right)- 2(1+x^2) \lambda. \nonumber
 \end{Eqnarray}
 \reseteqn


 \subsection{Integrals of type R \protect\boldmath $ \!_{\Ss (n)} $}

 \alpheqn
 \begin{Eqnarray} 
  {\cal R}_{(-1)} & = & -
  \lambda^{1/2} - (1+x^2-y^2) \frac{{\cal L}_2-{\cal L}_1}{2} + \lambda^{1/2}
  \ln \Bigg( \frac{\lambda^{1/2}}{x} \frac{\eta}{\epsilon_2} \Bigg), 
  \hspace{5.3cm} \\ 
  {\cal R}_{(0) \pp} & = &
  \frac{1}{2} \ln \Bigg( \frac{(1-x)^2-y^2}{(1+x)^2-y^2} \Bigg) +
  \ln \Bigg( \frac{\eta}{\epsilon_2} \Bigg),
 \end{Eqnarray}
 \reseteqn


 \subsection{Integrals of type S \protect\boldmath $ \!_{\Ss (m,n)} $}

 \alpheqn
 \begin{Eqnarray} 
 {\cal S}_{(0,0)} & = & \lambda^{1/2} - x^2 ({\cal L}_2-{\cal L}_1) - y^2
 {\cal L}_1, \hspace{9.4cm} \\ 
 {\cal S}_{(1,0)} & = & \frac{1}{4} (1+5x^2+y^2) \lambda^{1/2} -
 (2+x^2) x^2 \frac{{\cal L}_2-{\cal L}_1}{2} - y^4 \frac{{\cal L}_1}{2}, \\[3mm]
 {\cal S}_{(0,1)} & = & {\cal N}_0, \quad
 {\cal S}_{(1,1)} = (1+x^2) {\cal N}_0 - \lambda^{1/2} {\cal L}_1 +
 2 (1-x^2) {\cal L}_5 - \left( (1-x)^2-y^2 \right), \\ 
 {\cal S}_{(2,1)} & = & (1+4x^2+x^4) {\cal N}_0 -
 \frac{1}{2} (3+3x^2+y^2) \lambda^{1/2} {\cal L}_1 + 
 3(1-x^4) {\cal L}_5 + \\ & - & \frac{1}{4} \left( (1-x)^2 - y^2 \right)
 \left( (1-x)^2+4+8x^2+y^2 \right), \nonumber \\[3mm]
 {\cal S}_{(0,2)} & = & - \frac{1}{2x} ({\cal N}_2-{\cal N}_3), \\
 {\cal S}_{(1,2)} & = & - \frac{(1+x)^2}{2x} {\cal N}_2 +
 \frac{(1-x)^2}{2x} {\cal N}_3 + {\cal N}_1, \\
 {\cal S}_{(2,2)} & = & - \frac{(1+x)^4}{2x} {\cal N}_2 +
 \frac{(1-x)^4}{2x} {\cal N}_3 + 2(1+x^2) {\cal N}_1 +
 \lambda^{1/2} - x^2 ({\cal L}_2-{\cal L}_1) - y^2 {\cal L}_1, \\
 {\cal S}_{(3,2)} & = & - \frac{(1+x)^6}{2x} {\cal N}_2 +
 \frac{(1-x)^6}{2x} {\cal N}_3 + (3+x^2)(1+3x^2) {\cal N}_1 + \\ & + &
 \frac{1}{4} (9+13x^2+y^2) \lambda^{1/2} - (6+5x^2)x^2
 \frac{{\cal L}_2-{\cal L}_1}{2} - y^2 \left( 4(1+x^2)+y^2 \right)
 \frac{{\cal L}_1}{2}, \nonumber \\[3mm]
 {\cal S}_{(0,3)} & = & \frac{1}{4x} \Big\{
 \frac{2}{1-x} + \frac{1}{1-x} {\cal L}_3 - \frac{1}{1+x} {\cal L}_4 -
 \frac{\beta(0)}{x} {\cal L}_1 + \frac{\beta_{+}(0)}{x} {\cal L}_6 \Big\}, \\
 {\cal S}_{(1,3)} & = & \frac{1}{4x} \Big\{
 2(1-x) + (1-x) {\cal L}_3 - (1+x) {\cal L}_4 - \frac{\beta(2)}{x} {\cal L}_1 +
 \frac{\beta_{+}(2)}{x} {\cal L}_6 \Big\}, \\
 {\cal S}_{(2,3)} & = & \frac{1}{4x} \Big\{
 2(1-x)^3 + (1-x)^3 {\cal L}_3 - (1+x)^3 {\cal L}_4 - \frac{\beta(4)}{x}
 {\cal L}_1 + \frac{\beta_{+}(4)}{x} {\cal L}_6 + 4x {\cal N}_0 \Big\}, \\
 {\cal S}_{(3,3)} & = & \frac{1}{4x} \Big\{
 2(1-x)^5 + (1-x)^5 {\cal L}_3 - (1+x)^5 {\cal L}_4 - \frac{\beta(6)}{x}
 {\cal L}_1 + \frac{\beta_{+}(6)}{x} {\cal L}_6 + \\ & + &
 12 x (1 + x^2) {\cal N}_0 + 8 x (1 - x^2) {\cal L}_5 -
 4x \lambda^{1/2} {\cal L}_1 -  4x \left( (1-x)^2-y^2) \right) \Big\},
 \nonumber \\
 {\cal S}_{(4,3)} & = & \frac{1}{4x} \Big\{
 2(1-x)^7 + (1-x)^7 {\cal L}_3 - (1+x)^7 {\cal L}_4 -
 \frac{\beta(8)}{x} {\cal L}_1 + \frac{\beta_{+}(8)}{x} {\cal L}_6 + \\ & + &
 24 x (1 + 3 x^2 + x^4) {\cal N}_0 +
 28 x (1 - x^4) {\cal L}_5 - 2x(7+7x^2+y^2) \lambda^{1/2} {\cal L}_1 +
 \nonumber \\ & - &  2x \left( (1-x)^2-y^2) \right) (7+9x^2) + x \lambda \Big\}
 \rule{0mm}{6mm} \nonumber 
 \end{Eqnarray}
 \reseteqn


 \subsection{Integrals of type S \protect\boldmath $ \!_{\Ss (n)} $}

 \alpheqn
 \begin{Eqnarray}
  {\cal S}_{(0)} & = & - \frac{{\cal L}_1^2}{2} +
  {\cal L}_1 \ln \Big( \frac{\lambda^{1/2}}{x} \frac{\eta}{\epsilon_2} \Big) +
  ({\cal L}_2-{\cal L}_1) \ln y + \\ & + &
  \Li (\eta \, x) - \Li \Big( \frac{x}{\eta} \Big) - 
  2 \Li \Big( \frac{ (\eta^2-1) x}{\eta-x} \Big), 
  \hspace{7.8cm} \nonumber \\
  {\cal S}_{(1)} & = & \frac{1}{\lambda^{1/2}} \Big\{ \!\!-
  \frac{{\cal L}_1^2}{2} + {\cal L}_1 \ln \Big( \frac{\lambda^{1/2}}{x} 
  \frac{1}{\epsilon_2} \Big) + 2 \Li \Bigg( \!\!- \frac{1}{\eta} \Bigg) -
  2 \Li \Big( \!\!- \frac{1-\eta \, x}{\eta-x} \Big) \Big\},
 \end{Eqnarray}
 \reseteqn

 Of all the many integrals listed in A.2--A.5 the total rate calculation done
 before in~\cite{c14,c15,c16,c17,c18,ghinculov} requires only the five basic
 integrals $ {\cal R}_{(-2,-1)} $, $ {\cal R}_{(-1,-1)} $,
 $ {\cal R}_{(0,-1) \pp} $, $ {\cal S}_{(0,0)} $ and $ {\cal S}_{(1,0)} $
 compared to the 33 basic integrals that are needed for the full calculation.
 This may serve as a measure of the additional labour that is incurred when
 one calculates the complete set of structure functions as done in this paper.
 

 \section{Coefficient functions \protect \boldmath $ \rho_{(n)} $,
 $ \rho_{(m,n)} $, $ \sigma_{(n)} $ and $ \sigma_{(m,n)} $}

 In this Appendix we list the values of the various coefficient functions
 $ \rho_{(n),i} $, $ \rho_{(m,n),i} $, $ \sigma_{(n),i} $ and
 $ \sigma_{(m,n),i} $ ($ i = U \!+\! L $, $ U^P \!+\! L^P $, $ U $, $ U^P $,
 $ L $, $ L^P $, $ F $, $ F^P $, $ S $, $ S^P $, $ I^P $ and $ A^P $) that
 multiply the basic set of integrals listed in Appendix A as spelled out
 in the rate expression Eq.~(\ref{complete}). The coefficient
 functions involve polynomials in $ x^2 $ and $ y^2 $ which we sort by 
 increasing powers of $ y^2 $. For reasons of conciseness we drop
 the suffix $ i $ denoting the particular type of structure function in the
 following listing. The contributions are collected in terms of powers of
 $ y^2 $.
 

 \subsection{Total rate \protect \boldmath $ i= U \!+\! L $}

 \alpheqn
 \begin{Eqnarray}
  \rho_{(-2,-1)} & = & -
  \frac{y^2 (1 - x^2)((1 + 2 x^2) + y^2)}{x^2}, \\
  \rho_{(-1,-1)} & = & \pp \frac{(1 - x^2)(1 + 2 x^2) +
  (4 - 3 x^2) y^2 + 3 y^4}{x^2}, \hspace{6.05cm} \\
  \rho_{(0,-1) \pp} & = & - \frac{(3 - 2 x^2) + 3 y^2}{x^2}, \\
  \sigma_{(0,0) \phantom{++}} & = & - 2
  \frac{y^2 ((1 + 2 x^2) + y^2)}{x^2}, \\
  \sigma_{(1,0) \phantom{++}} & = & \pp 2
  \frac{(1 + 2 x^2) + y^2}{x^2},
 \end{Eqnarray}
 \reseteqn

 \subsection{Polarized total rate \protect \boldmath $ i= U^P \!+\! L^P $}

 \alpheqn
 \begin{Eqnarray}
  \rho_{(-2,0)} & = & -
  \frac{y^2 (1 - x^2)^2 ((1 - 2 x^2) - y^2)}{x^2}, \\
  \rho_{(-1,0)} & = & \pp \frac{(1 - 4 x^2 + 5 x^4 - 2 x^6) +
  (5 - 4 x^2 - 5 x^4) y^2 - 2 (3 - x^2) y^4}{x^2}, \\
  \rho_{(0,0) \pp} & = & \pp \frac{2 (1 - x^2 - 6 x^4) -
  3 y^2 + y^4}{x^2}, \\
  \rho_{(1,0) \phantom{+}} & = & -
  \frac{(7 - 6 x^2) - 7 y^2}{x^2}, \\
  \rho_{(-1) \phantom{,0}} & = & \pp 8
  \frac{\sqrt{\lambda} ((1 - 2 x^2) - y^2)}{x^2}, \\
  \rho_{( 0) \phantom{+,0}} & = & - 8
  \frac{\lambda ((1 - 2 x^2) - y^2)}{x^2}, \\
  \sigma_{(0,0) \pp} & = & - 4
  \frac{\sqrt{\lambda} ((1 - 2 x^2) - y^2)}{x^2}, \\
  \sigma_{(0,1) \pp} & = & - 2 \frac{4 x^2 (1 - 2 x^2)(1 - x^2) +
  (7 - 5 x^2 - 6 x^4) y^2 - (9 + x^2) y^4 + 2 y^6}{x^2}, \hspace{2.3cm} \\
  \sigma_{(1,1) \pp} & = & \pp 2 \frac{(3 - x^2 + 6 x^4) -
  (2 - x^2) y^2 - y^4}{x^2}, \\
  \sigma_{(2,1) \pp} & = & \pp 2
  \frac{(1 - 2 x^2) - y^2}{x^2}, \\
  \sigma_{(0) \phantom{-,1}} & = & - 4 \frac{\sqrt{\lambda}
  (1 - x^2 + y^2)((1 - 2 x^2) - y^2)}{x^2}, \\
  \sigma_{(1) \phantom{-,1}} & = & \pp 4 \frac{\lambda
  (1 - x^2 + y^2)((1 - 2 x^2) - y^2)}{x^2},
 \end{Eqnarray}
 \reseteqn


 \subsection{Longitudinal rate \protect \boldmath $ i = L $}

 \alpheqn
 \begin{Eqnarray}
  \rho_{(-2,1)} & = & -
  \frac{y^2 (1 + y^2) (1 - x^2)^3}{x^2}, \hspace{9.55cm} \\
  \rho_{(-1,1)} & = & \pp \frac{(1 - x^2)((1 - x^2)^2 +
  (6 + x^2 - 3 x^4) y^2 + (5 + 3 x^2) y^4)}{x^2}, \\
  \rho_{(0,1) \pp} & = & - \frac{(5 - 2 x^2 - 7 x^4 + 4 x^6) +
  (12 - 33 x^2 + x^4) y^2 + (7 + x^2) y^4}{x^2}, \\
  \rho_{(1,1) \pp} & = & \pp \frac{(7 - 31 x^2 + 4 x^4) +
  (10 + x^2) y^2 + 3 y^4}{x^2}, \\
  \rho_{(2,1) \pp} & = & - 3 \frac{1 + y^2}{x^2}, \\
  \sigma_{(0,2) \pp} & = & - 2 \frac{y^2 ((1 + 10 x^2 - 11 x^4) +
  (1 + x^2)^2 y^2)}{x^2}, \\
  \sigma_{(1,2) \pp} & = & \pp 2 \frac{(1 + 10 x^2 - 11 x^4) +
  (3 - 4 x^2 + x^4) y^2 + 2 (1 + x^2) y^4}{x^2}, \\
  \sigma_{(2,2) \pp} & = & - 2 \frac{2 (1 - 3 x^2) +
  (3 + 2 x^2) y^2 + y^4}{x^2}, \\
  \sigma_{(3,2) \pp} & = & \pp 2 \frac{1 + y^2}{x^2},   
 \end{Eqnarray}
 \reseteqn

 \subsection{Polarized longitudinal rate \protect \boldmath $ i = L^P $}

 \alpheqn
 \begin{Eqnarray}
  \rho_{(-2,2)} & = & -
  \frac{y^2 (1 - y^2) (1 - x^2)^4}{x^2}, \hspace{9.5cm} \\
  \rho_{(-1,2)} & = & \pp \frac{(1 - x^2)^2 ((1 - x^2)^2 +
  (7 - 2 x^2 + 3 x^4) y^2 - 4 (2 + x^2) y^4)}{x^2}, \\
  \rho_{(0,2) \pp} & = & \pp 2 \frac{2 x^2 (3 - x^2)(1 - x^2)^2 -
  (7 - 5 x^4 + 6 x^6) y^2 + (7 - 18 x^2 + 3 x^4) y^4}{x^2}, \\
  \rho_{(1,2) \pp} & = & - 2 \frac{(5 + 10 x^2 + 13 x^4 - 4 x^6) -
  (9 + 48 x^2 + 11 x^4) y^2 + 2 (2 + x^2) y^4}{x^2}, \\
  \rho_{(2,2) \pp} & = & \pp \frac{(16 - 20 x^2 - 4 x^4) -
  (17 + 20 x^2) y^2 + y^4}{x^2}, \\
  \rho_{(3,2) \pp} & = & - 7 \frac{1 - y^2}{x^2}, \\
  \rho_{(-1) \phantom{,2}} & = & \pp 8
  \frac{\sqrt{\lambda} (1 - y^2)}{x^2}, \\
  \rho_{(0) \phantom{+,2}} & = & - 8
  \frac{ \lambda (1 - y^2)}{x^2}, \\
  \sigma_{(0,0) \pp} & = & - 4
  \frac{\sqrt{\lambda} (1 - y^2)}{x^2}, \\
  \sigma_{(0,3) \pp} & = & - 2 \frac{(1 - x^2)
  (4 x^2 (1 - x^2)^2 + (7 - 10 x^2 + 7 x^4 - 4 x^6) y^2)}{x^2} + \\ & + & \pp
  2 \frac{(1 - x^2)((9 + 8 x^2 - 5 x^4) y^4 -
  2 (1 - x^2) y^6)}{x^2}, \nonumber \\
  \sigma_{(1,3) \pp} & = & \pp 2 \frac{(3 - 5 x^2 + 17 x^4 - 15 x^6) +
  (12 - 9 x^2 + 18 x^4 + 11 x^6) y^2}{x^2} + \\ & - & \pp
  2 \frac{(19 + 14 x^2 + 11 x^4) y^4 - 4 (1 + x^2) y^6}{x^2}, \nonumber \\
  \sigma_{(2,3) \pp} & = & - 2 \frac{(5 - 14 x^2 - 7 x^4) +
  (4 + 21 x^2 + 11 x^4) y^2 - (11 + 7 x^2) y^4 + 2 y^6}{x^2}, \\
  \sigma_{(3,3) \pp} & = & \pp 2
  \frac{(1 + 3 x^2) + 5 x^2 y^2 - y^4}{x^2}, \\
  \sigma_{(4,3) \pp} & = & \pp 2 \frac{1 - y^2}{x^2}, \\
  \sigma_{(0) \phantom{,-1}} & = & - 4
  \frac{\sqrt{\lambda} (1 - y^2) (1 - x^2 + y^2)}{x^2}, \\
  \sigma_{(1) \phantom{,-1}} & = & \pp 4
  \frac{\lambda (1 - y^2) (1 - x^2 + y^2)}{x^2},
 \end{Eqnarray}
 \reseteqn


 \subsection{Unpolarized-transverse rate \protect \boldmath $ i = U $}

 \alpheqn
 \begin{Eqnarray}
  \rho_{(-2,1)} & = & - 2 y^2 (1 - x^2)^3, \hspace{11cm} \\
  \rho_{(-1,1)} & = & \pp 2 (1 - x^2) ((1 - x^2)^2 -
  (1 - 5 x^2) y^2 - 2 y^4), \rule{0mm}{5mm} \\
  \rho_{(0,1) \pp} & = & \pp 2 ((1 - 6 x^2 + 5 x^4) -
  3 (5 - x^3) y^2 - 2 y^4), \rule{0mm}{5mm} \\
  \rho_{(1,1) \pp} & = & \pp 2 ((17 - 5 x^2) + y^2), \rule{0mm}{5mm} \\
  \rho_{(2,1) \pp} & = & \pp 2, \rule{0mm}{5mm} \\
  \sigma_{(0,2) \pp} & = & \pp 4 y^2
  ((5 - 4 x^2 - x^4) + 2 y^2), \rule{0mm}{5mm} \\
  \sigma_{(1,2) \pp} & = & - 4
  ((5 - 4 x^2 - x^4) - 2 (2 + x^2) y^2), \rule{0mm}{5mm} \\
  \sigma_{(2,2) \pp} & = & - 4 ((6 + 2 x^2) + y^2), \rule{0mm}{5mm} \\
  \sigma_{(3,2) \pp} & = & \pp 4, \rule{0mm}{5mm}
 \end{Eqnarray}
 \reseteqn

 \subsection{Polarized unpolarized-transverse rate
  \protect \boldmath $ i = U^P $}

 \alpheqn
 \begin{Eqnarray}
  \rho_{(-2,2)} & = & \pp 2 y^2 (1 - x^2)^4, \hspace{11cm} \\
  \rho_{(-1,2)} & = & - 2 (1 - x^2)^2 ((1 - x^2)^2 +
  2 (1 + 3 x^2) y^2 - 2 y^4), \rule{0mm}{5mm} \\
  \rho_{(0,2) \pp} & = & - 4 ((1-x^2)^2 (3+x^2) -
  2 (1 + 3 x^4) y^2 - 2 (5 - x^2) y^4), \rule{0mm}{5mm} \\
  \rho_{(1,2) \pp} & = & \pp 4 ((9 + 10 x^2 + 5 x^4) -
  (27 + 5 x^2) y^2 + y^4), \rule{0mm}{5mm} \\
  \rho_{(2,2) \pp} & = & \pp 2 (10 (1 - x^2) + 3 y^2), \rule{0mm}{5mm} \\
  \rho_{(3,2) \pp} & = & \pp 6, \rule{0mm}{5mm} \\
  \rho_{(-1) \phantom{,1}} & = & - 16 \sqrt{\lambda}, \rule{0mm}{5mm} \\
  \rho_{(0) \phantom{+,1}} & = & \pp 16 \lambda, \rule{0mm}{5mm} \\
  \sigma_{(0,0) \pp} & = & \pp 8 \sqrt{\lambda}, \rule{0mm}{5mm} \\
  \sigma_{(0,3) \pp} & = & \pp 4 (1 - x^2) (4 x^2 (1 - x^2)^2 +
  (1 + 4 x^2 - 5 x^4) y^2 - 2 (4 - x^2) y^4), \rule{0mm}{5mm} \\
  \sigma_{(1,3) \pp} & = & \pp 4 ((1 - x^4)(3 - 11x^2) +
  (9 - 22 x^2 - 11 x^4 ) y^2 - 2 (1 - 2 x^2) y^4), \rule{0mm}{5mm} \\
  \sigma_{(2,3) \pp} & = & - 4 ((13 + 11 x^4) -
  (15 + 7 x^2) y^2 + 2 y^4), \rule{0mm}{5mm} \\
  \sigma_{(3,3) \pp} & = & - 4 ((1 - 5 x^2) + y^2), \rule{0mm}{5mm} \\
  \sigma_{(4,3) \pp} & = & - 4, \rule{0mm}{5mm} \\
  \sigma_{(0) \phantom{+,1}} & = & \pp 8 \sqrt{\lambda}
  (1 - x^2 + y^2), \rule{0mm}{5mm} \\
  \sigma_{(1) \phantom{+,1}} & = & - 8 \lambda
  (1 - x^2 + y^2) \rule{0mm}{5mm} ,
 \end{Eqnarray}
 \reseteqn


 \subsection{Scalar rate \protect \boldmath $ i = S $}

 \alpheqn
 \begin{Eqnarray}
  \rho_{(-2,-1)} & = & -
  \frac{y^2 (1 + y^2) (1 - x^2)}{x^2}, \hspace{9.5cm} \\
  \rho_{(-1,-1)} & = & \pp \frac{(1 + y^2)((1 - x^2) + 3 y^2)}{x^2}, \\
  \rho_{(0,-1) \pp} & = & - 3 \frac{(1 + y^2)}{x^2}, \\ 
  \sigma_{(0,0) \pp \pp} & = & - 2 \frac{y^2 (1 + y^2)}{x^2}, \\
  \sigma_{(1,0) \pp \pp} & = & \pp 2 \frac{(1 + y^2)}{x^2},
 \end{Eqnarray}
 \reseteqn

 \subsection{Polarized scalar rate \protect \boldmath $ i = S^P $}

 \alpheqn
 \begin{Eqnarray}
  \rho_{(-2,0)} & = & -
  \frac{y^2 (1 - y^2)(1 - x^2)^2}{x^2}, \hspace{9.55cm} \\
  \rho_{(-1,0)} & = & \pp \frac{(1 - y^2)
  ((1 - x^2)^2 + 2 (3 - x^2) y^2)}{x^2}, \\
  \rho_{(0,0) \pp} & = & \pp \frac{(1 - y^2)
  (2 (1 + 5 x^2) - y^2)}{x^2}, \\
  \rho_{(1,0) \pp} & = & - 7 \frac{(1 - y^2)}{x^2}, \\
  \rho_{(-1) \phantom{,1}} & = & \pp 8
  \frac{\sqrt{\lambda} (1 - y^2)}{x^2}, \\
  \rho_{(0) \phantom{+,1}} & = & - 8
  \frac{\lambda (1 - y^2)}{x^2}, \\
  \sigma_{(0,0) \pp} & = & - 4 \frac{\sqrt{\lambda} (1 - y^2)}{x^2}, \\
  \sigma_{(0,1) \pp} & = & - 2 \frac{(1 - y^2)
  (4 x^2 (1 - x^2) + (7 + 5 x^2) y^2 - 2 y^4)}{x^2}, \\
  \sigma_{(1,1) \pp} & = & \pp 2
  \frac{(1 - y^2)(3 (1 - x^2) + y^2)}{x^2}, \\
  \sigma_{(2,1) \pp} & = & \pp 2 \frac{(1 - y^2)}{x^2}, \\
  \sigma_{(0) \phantom{+,0}} & = & - 4
  \frac{\sqrt{\lambda} (1 - y^2) ((1 - x^2) + y^2)}{x^2}, \\
  \sigma_{(1) \phantom{+,0}} & = & \pp 4
  \frac{\lambda(1 - y^2)((1 - x^2) + y^2)}{x^2},
 \end{Eqnarray}
 \reseteqn


 \subsection{Forward-backward-asymmetric rate \protect \boldmath $ i = F $}

 \alpheqn
 \begin{Eqnarray}
  \rho_{(-2,0)} & = & - 2 y^2 (1 - x^2)^2, \hspace{11.1cm} \\
  \rho_{(-1,0)} & = & \pp 2 ((1 - x^2)^2 + 4 x^2 y^2), \rule{0mm}{5mm} \\
  \rho_{(0,0) \pp} & = & \pp 2 (4 (2 + x^2) - 7 y^2), \rule{0mm}{5mm} \\
  \rho_{(1,0) \pp} & = & - 2, \rule{0mm}{5mm} \\
  \rho_{(-1) \phantom{,0}} & = & \pp 16 \sqrt{\lambda}, \rule{0mm}{5mm} \\
  \rho_{(0) \phantom{+,0}} & = & - 16 \lambda, \rule{0mm}{5mm} \\
  \sigma_{(0,0) \pp} & = & - 8 \sqrt{\lambda}, \rule{0mm}{5mm} \\
  \sigma_{(0,1) \pp} & = & - 4 (4 x^2 (1 - x^2) +
  (1 + 5 x^2) y^2 - 2 y^4), \rule{0mm}{5mm} \\
  \sigma_{(1,1) \pp} & = & - 4 (3 (1 + x^2) - y^2), \rule{0mm}{5mm} \\
  \sigma_{(2,1) \pp} & = & \pp 4, \rule{0mm}{5mm} \\
  \sigma_{(0) \phantom{+,0}} & = & - 8 \sqrt{\lambda}
  ((1 - x^2) + y^2), \rule{0mm}{5mm} \\
  \sigma_{(1) \phantom{+,0}} & = & \pp 8 \lambda
  ((1 - x^2) + y^2), \rule{0mm}{5mm}
 \end{Eqnarray}
 \reseteqn

 \subsection{Polarized forward-backward-asymmetric rate
 \protect \boldmath $ i=F^P $}

 \alpheqn
 \begin{Eqnarray}
  \rho_{(-2,1)} & = & \pp 2 y^2 (1 - x^2)^3, \hspace{11.1cm} \\
  \rho_{(-1,1)} & = & - 2 (1 - x^2) ((1 - x^2)^2 -
  (1 - 5 x^2) y^2), \rule{0mm}{5mm} \\
  \rho_{(0,1) \pp} & = & - 2 ((1 - 6 x^2 + 5 x^4) -
  (11 + x^2) y^2), \rule{0mm}{5mm} \\
  \rho_{(1,1) \pp} & = & - 2 ((11 + x^2) + 5 y^2), \rule{0mm}{5mm} \\
  \rho_{(2,1) \pp} & = & \pp 10, \rule{0mm}{5mm} \\
  \sigma_{(0,2) \pp} & = & - 4 y^2 (1 - x^2) (5 + x^2), \rule{0mm}{5mm} \\
  \sigma_{(1,2) \pp} & = & \pp 4 ((1 - x^2) (5 + x^2) -
  2 x^2 y^2), \rule{0mm}{5mm} \\
  \sigma_{(2,2) \pp} & = & \pp 4 (2 x^2 + y^2), \rule{0mm}{5mm} \\
  \sigma_{(3,2) \pp} & = & - 4, \rule{0mm}{5mm}
 \end{Eqnarray}
 \reseteqn


 \subsection{Polarized longitudinal-transverse-interference rate \protect \boldmath $ i=I^P $}

 \alpheqn
 \begin{Eqnarray}
  \rho_{(-2,2)} & = & \pp \frac{\sqrt{2}}{2} \frac{y^2 (1 - x^2)^4}{x}, \\
  \rho_{(-1,2)} & = & - \frac{\sqrt{2}}{2} \frac{(1 - x^2)^2 ((1 - x^2)^2 +
  (3 + 5 x^2) y^2 + 2 y^4)}{x}, \hspace{5.2cm} \\
  \rho_{(0,2) \pp} & = & - \frac{\sqrt{2}}{2} \frac{(1 - x^2)^2 (5 + 3 x^2) -
  (25 - 38 x^2 + 29 x^4) y^2 + 8 (1 + x^2) y^4}{x}, \\
  \rho_{(1,2) \pp} & = & \pp \frac{\sqrt{2}}{2} \frac{(1 + 50 x^2 - 3 x^4) -
  (21 + 23 x^2) y^2 + 10 y^4}{x}, \\
  \rho_{(2,2) \pp} & = & \pp \frac{\sqrt{2}}{2} \frac{(5 + 7 x^2) - 2 y^2}{x}, \\
  \rho_{(-1) \phantom{,0}} & = & - 4 \sqrt{2} \frac{\sqrt{\lambda}}{x}, \\
  \rho_{(0) \phantom{+,0}} & = & \pp 4 \sqrt{2} \frac{\lambda}{x}, \\
  \sigma_{(0,0) \pp} & = & \pp 2 \sqrt{2} \frac{\sqrt{\lambda}}{x}, \\
  \sigma_{(0,3) \pp} & = & \pp \sqrt{2} \frac{(1 - x^2)
  (4 x^2 (1 - x^2)^2 - (1 - 9 x^2 + 8 x^4) y^2 + 2 (1 + 2 x^2) y^4}{x}, \\
  \sigma_{(1,3) \pp} & = & \pp \sqrt{2} \frac{(5 - 18 x^2 +5 x^4 + 8 x^6)
  - 2 (5 + x^2 + 8 x^4) y^2 + 2 (1 + 4 x^2) y^4}{x}, \\
  \sigma_{(2,3) \pp} & = & - \sqrt{2} \frac{4 (1 + 3 x^2 + x^4) -
  (11 + 8 x^2) y^2 + 4 y^4}{x}, \\
  \sigma_{(3,3) \pp} & = & - \sqrt{2} \frac{1}{x}, \\
  \sigma_{(0) \phantom{+,0}} & = & \pp 2 \sqrt{2} 
  \frac{\sqrt{\lambda} ((1 - x^2) + y^2)}{x}, \\
  \sigma_{(1) \phantom{+,0}} & = & - 2 \sqrt{2} 
  \frac{\lambda ((1 - x^2) + y^2)}{x},
 \end{Eqnarray}
 \reseteqn


 \subsection{Polarized parity-asymmetric rate \protect \boldmath $ i = A^P $}

 \alpheqn
 \begin{Eqnarray}
  \rho_{(-2,1)} & = & \pp \frac{\sqrt{2}}{2} \frac{y^2 (1 - x^2)^3}{x}, \\
  \rho_{(-1,1)} & = & - \frac{\sqrt{2}}{2}
  \frac{(1 - x^2) ((1 - x^2)^2 + 4 y^2)}{x}, \hspace{8.15cm} \\
  \rho_{(0,1) \pp} & = & \pp \frac{\sqrt{2}}{2}
  \frac{4 (1 - x^2) + (3 - 7 x^2) y^2}{x}, \\
  \rho_{(1,1) \pp} & = & - \frac{\sqrt{2}}{2} \frac{3 - 7 x^2}{x}, \\
  \sigma_{(0,2) \pp} & = & \pp \sqrt{2}
  \frac{y^2 (1 - x^2) (1 + 2 x^2)}{x}, \\
  \sigma_{(1,2) \pp} & = & - \sqrt{2}
  \frac{(1 - x^2)(1 + 2 x^2) + (1 - 2 x^2) y^2}{x}, \\
  \sigma_{(2,2) \pp} & = & \pp \sqrt{2} \frac{1 - 2 x^2}{x},
 \end{Eqnarray}
 \reseteqn


 \section{Loop integrals}

 In this Appendix we list the $m_b \neq 0$ one-loop amplitude corrections to the
 process $ t \rightarrow b + W^{+} $. They are determined from the vertex
 correction Fig.~1b and the appropriate wave function renormalization constants
 $ Z_2 $. We present our results in terms of the three vector current amplitudes
 $ F_{i}^{V} $ $ (i=1,2,3) $ and the three axial vector current amplitudes
 $ F_{i}^{A} $ $ (i=1,2,3) $ defined in Eq.~(\ref{formfactor}) in Sec.~5.
 Using the abbreviations in Eq.~(\ref{w1wmu}) with $ q^2 = m_W^2 $, one has

 \begin{Eqnarray}
  F_1^V & = & 1 + \frac{\alpha_s}{4 \pi} C_F \Bigg\{
  - \frac{m_t^2 + m_b^2 - q^2}{m_t^2 \sqrt{\lambda}} \Bigg[
  2 \Li (1 - w_1^2) - 2 \Li \Big( 1 - \frac{w_1}{w_{\mu}} \Big) + \\ & + &
  \frac{1}{2} \ln \Big( \frac{\Lambda^4}{m_b^2 m_t^2} \Big)
  \ln (w_1 w_{\mu}) + \ln \Big( \frac{w_1^3}{w_{\mu}} \Big)
  \ln \Big( \frac{w_{\mu} (1 - w_1^2)}{w_{\mu} - w_1} \Big) \Bigg] -
  \ln \Big( \frac{\Lambda^4}{m_b^2 m_t^2} \Big) + \nonumber \\ & - &
  \frac{m_t^2 - m_b^2}{2 q^2} \ln \Big( \frac{m_b^2}{m_t^2} \Big) -
  4 + \ln (w_1 w_{\mu}) \Big( \frac{m_t^2 \sqrt{\lambda}}{2 q^2} -
  \frac{(m_t + m_b)^2 - q^2}{m_t^2 \sqrt{\lambda}} \Big) \Bigg\} \nonumber \\
  F_2^V & = & \frac{\alpha_s}{4 \pi} C_F \frac{m_t - m_b}{q^2} \Bigg\{
  2 - \Big( \frac{m_t + 2 m_b}{m_t - m_b} -
  \frac{m_t^2 - m_b^2}{q^2} \Big) \ln \Big( \frac{m_b^2}{m_t^2} \Big) + \\ & - &
  \Big( \frac{m_t^2 \sqrt{\lambda}}{q^2} - \frac{m_b}{m_t - m_b}
  \frac{q^2 + (m_t - m_b)(3 m_t + m_b)}{m_t^2 \sqrt{\lambda}} \Big)
  \ln (w_1 w_{\mu}) \Bigg\} \nonumber \\
  F_3^V & = & F_3^V (m_t,m_b) = F_2^V (m_b,m_t).
 \end{Eqnarray}

 As before the IR singularity is regularized by a small gluon mass $ m_g $.
 The axial vector amplitudes $ F_i^A $ can be obtained from the vector 
 amplitudes by the replacement $ m_t \rightarrow - m_t $,\ i.e. one has
 $ F_i^A(m_t) = F_i^V(-m_t) $ $ (i=1,2,3) $. Our one-loop amplitudes are
 linearly related to the one-loop amplitudes given in \cite{gounaris}.
 The two sets of one-loop amplitudes agree with each other after correcting
 for a typo in \cite{gounaris} mentioned in Sec.~5.

\end{appendix}

\newpage \thispagestyle{empty} \strut\vspace{3truecm}

\begin{figure}[h]
  \centering \leavevmode
  \put(0,28){a)}
  \psfig{file=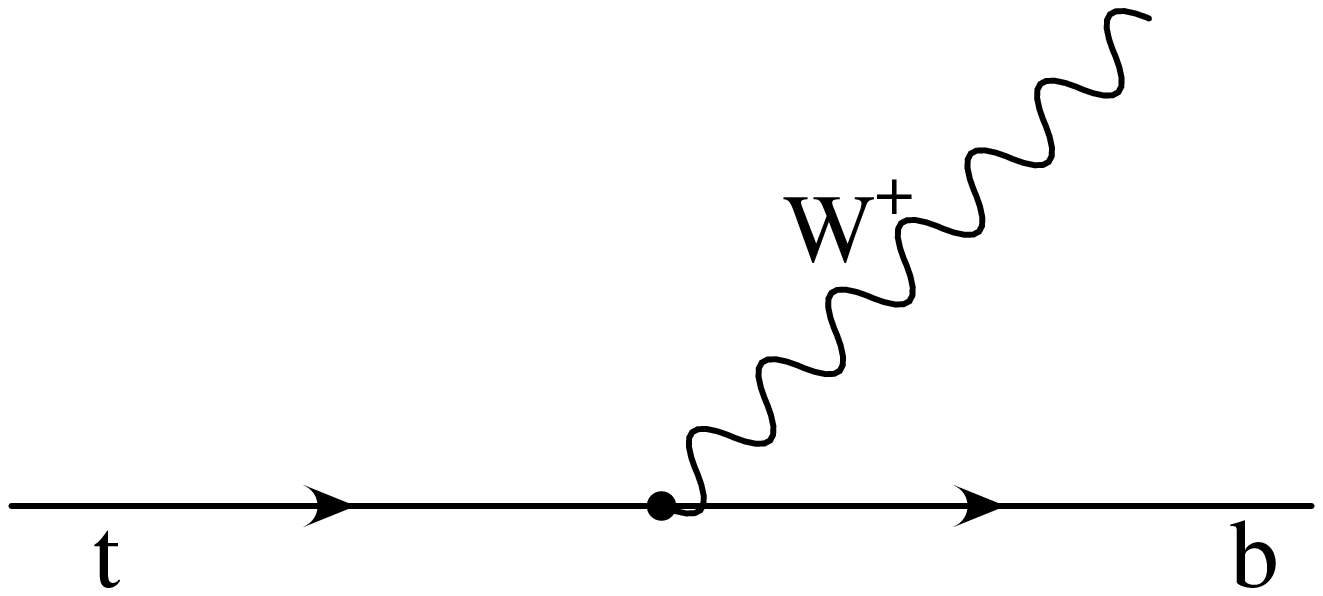,width=7.4cm,clip=}
  \put(0,28){b)}
  \psfig{file=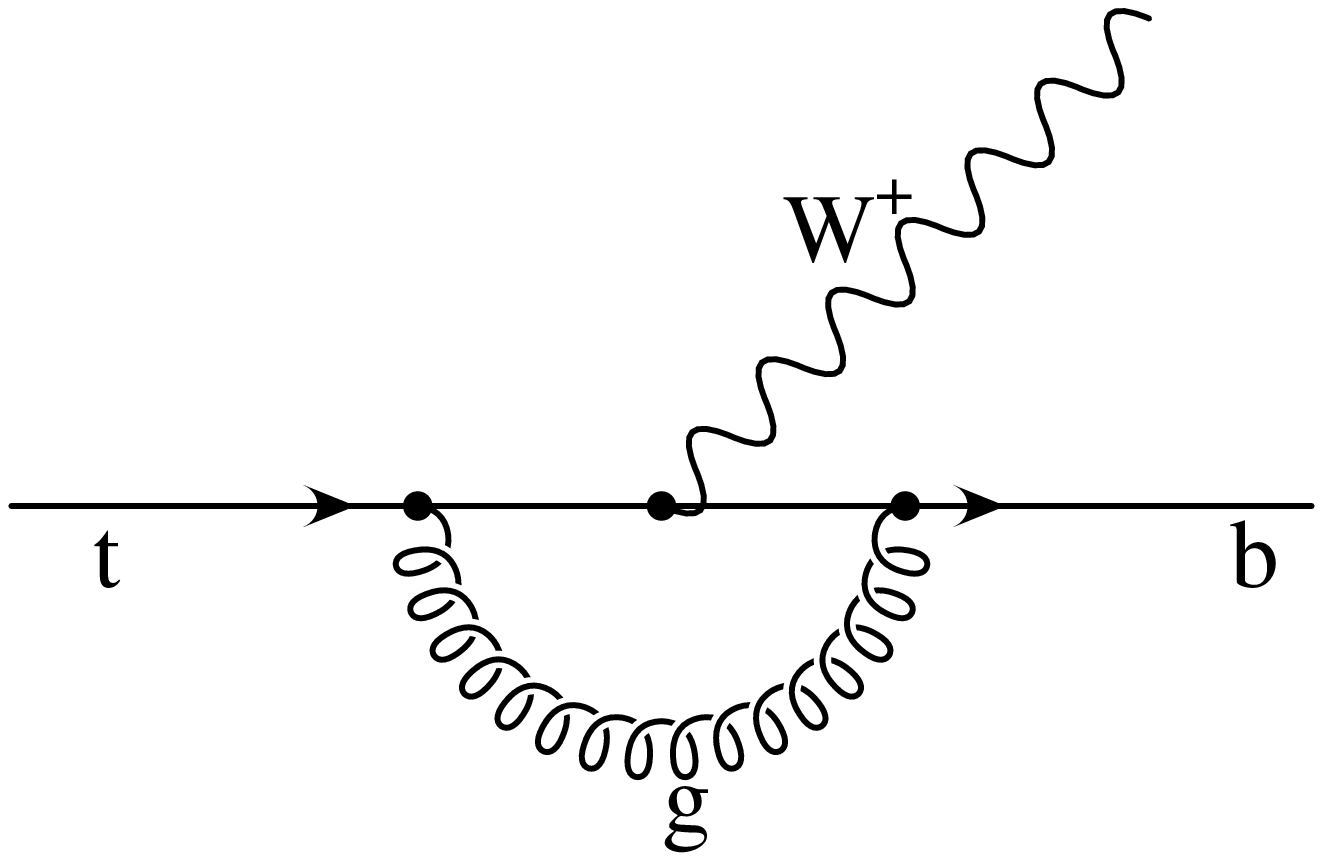,width=7.4cm,clip=} \newline
  \put(0,28){c)}
  \psfig{file=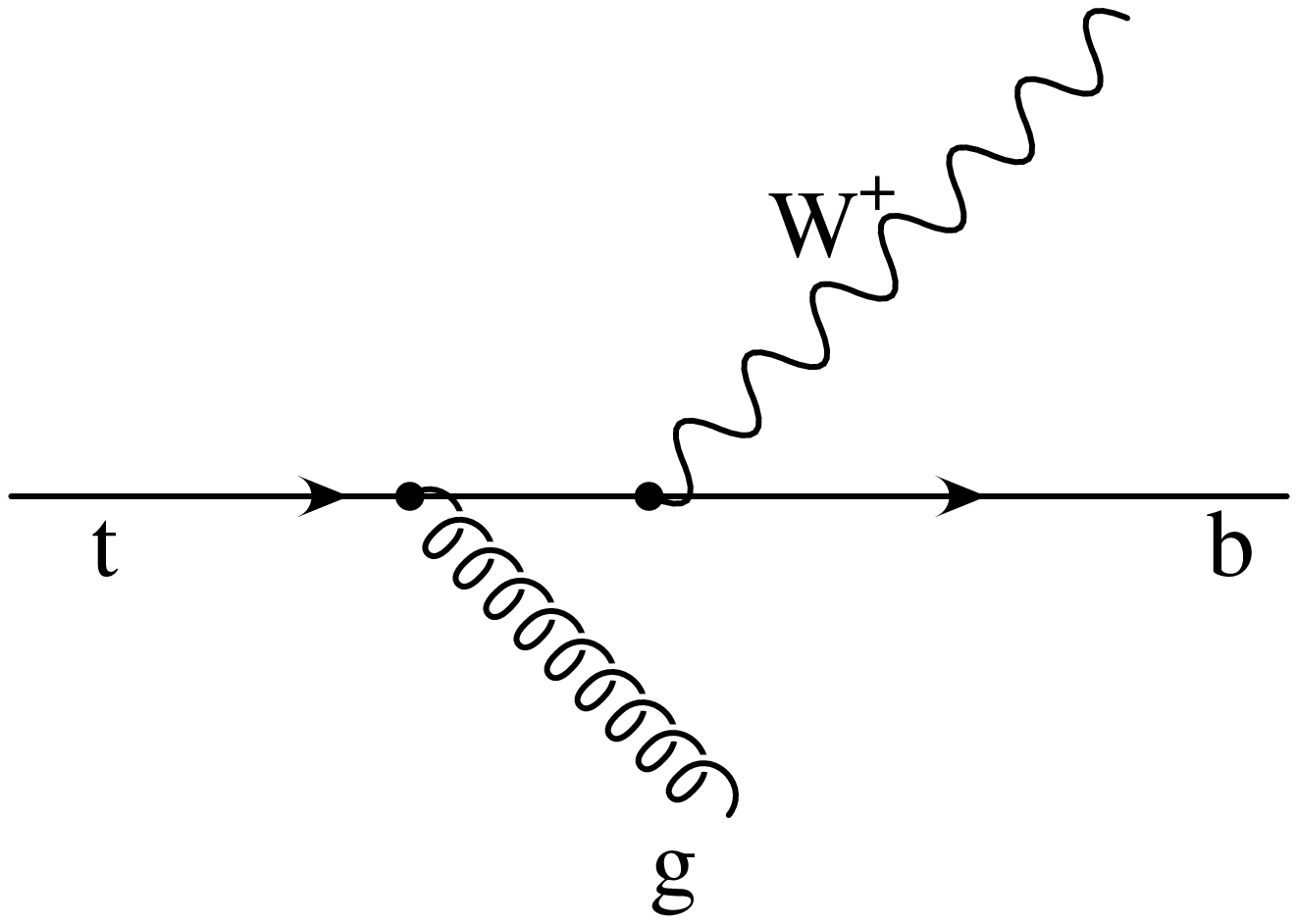,width=7.4cm,clip=}
  \put(0,28){d)}
  \psfig{file=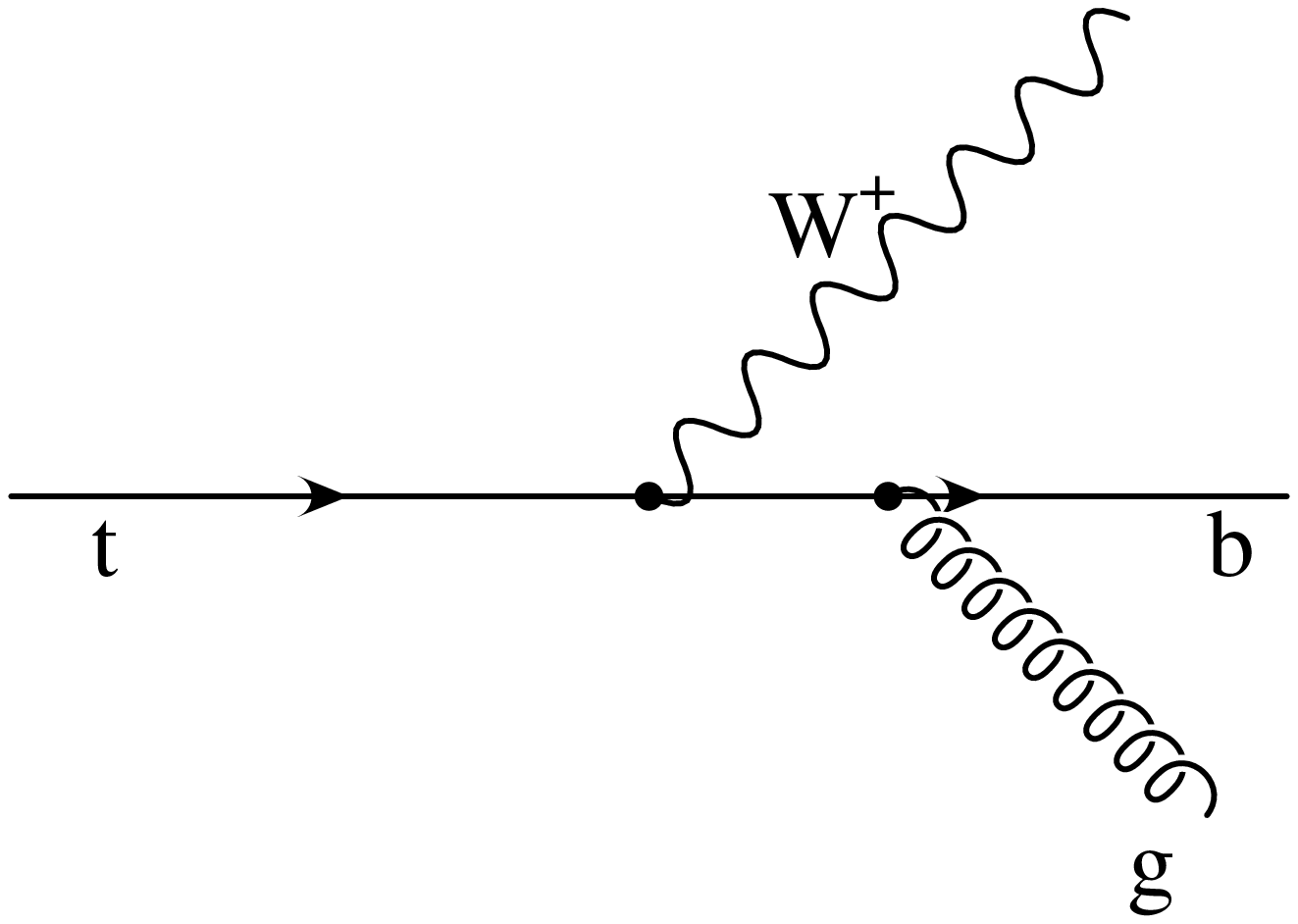,width=7.4cm,clip=}
\end{figure}

\vspace{1truecm} \centerline{\Large\bf Figure 1}

\begin{quote}
 Leading order Born term contribution (a) and $ O( \alpha_s ) $
 contributions (b,c,d) to $ t \!\rightarrow\! b \!+\! W^{+} $.
\end{quote}


\newpage \thispagestyle{empty} \strut\vspace{6truecm}

\begin{figure}[h]
  \centering \leavevmode 
  \psfig{file=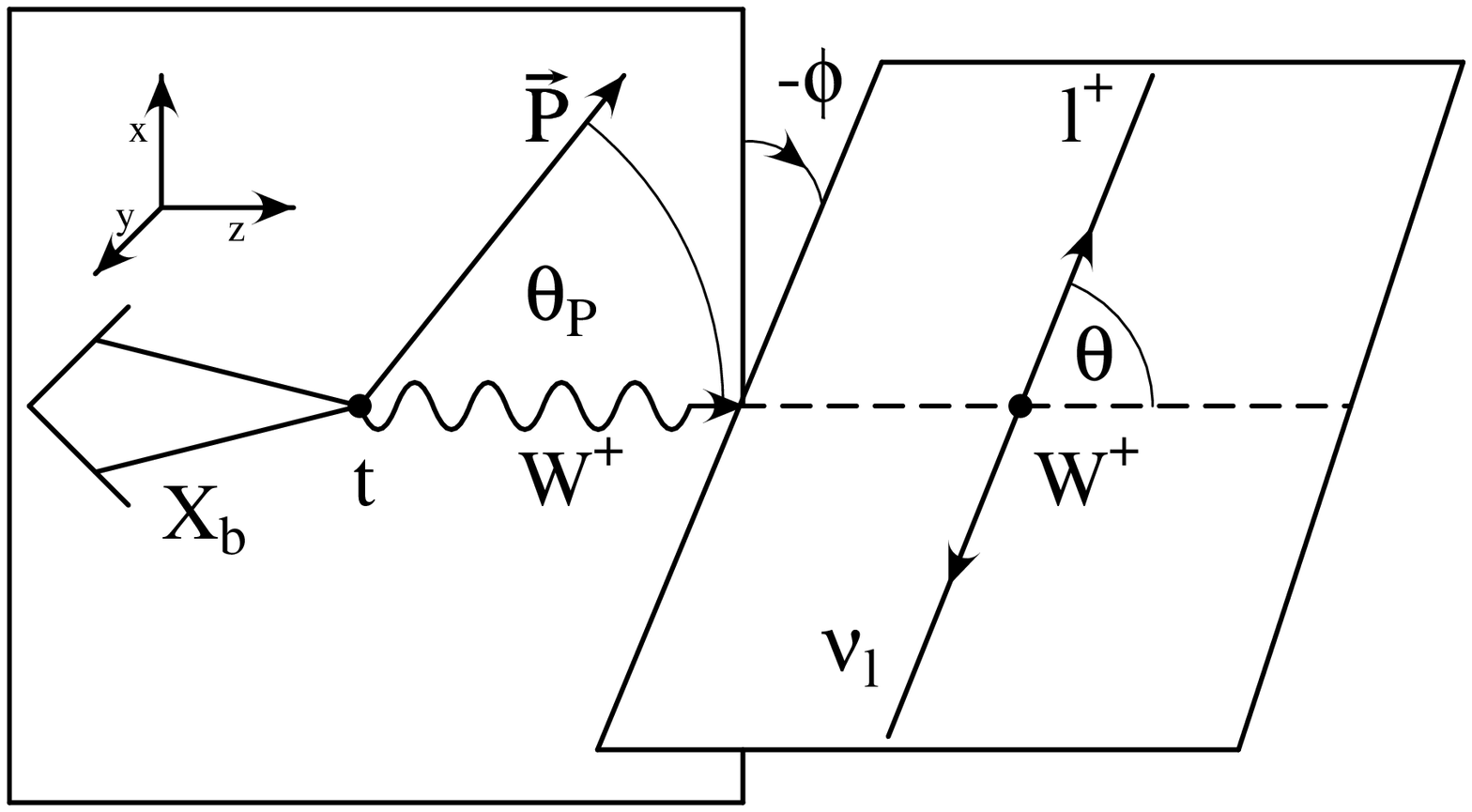,width=15cm,clip=}
\end{figure}

\vspace{1truecm} \centerline{\Large\bf Figure 2}

\begin{quote}
 Definition of the polar angles $ \theta $ and $ \theta_P $, and the azimuthal
 angle $ \phi $. $\vec{P}$ is the polarization vector of the top quark. 
\end{quote}


\newpage \thispagestyle{empty}

\begin{center}
\begin{picture}(120,120)
 \put(112,030.7){$ \mathbf{\theta} $}
 \put(047,014.0){$ \mathbf{\theta} $}
 \includegraphics[width=12cm,clip=]{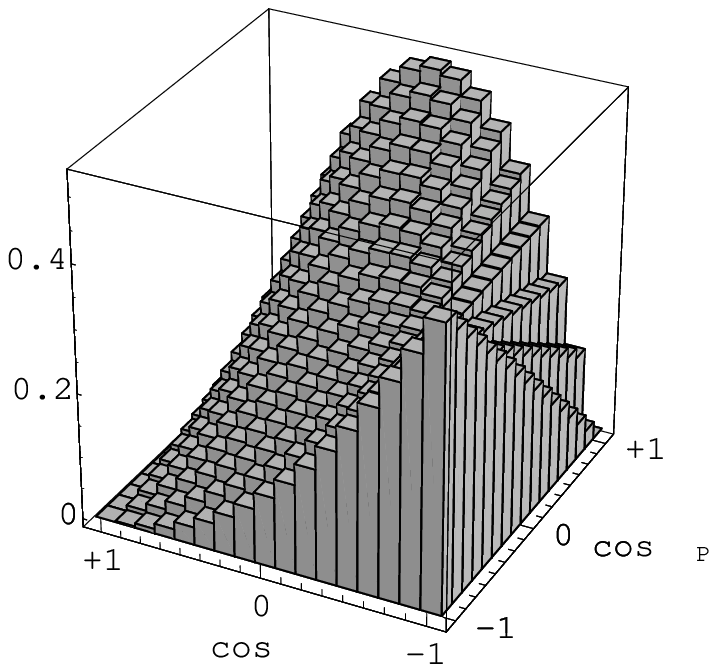}
\end{picture}
\end{center}

\vspace{1truecm} \centerline{\Large\bf Figure 3}

\begin{center}
 Born term Lego plot of the two-fold angular decay distribution \\
 $ d \widehat{\Gamma} / d \! \cos \theta d \! \cos \theta_p $ with $ P = 1 $.
\end{center}

\clearpage


\newpage \thispagestyle{empty} 
 
\centerline{\includegraphics[width=11.8cm,clip=]{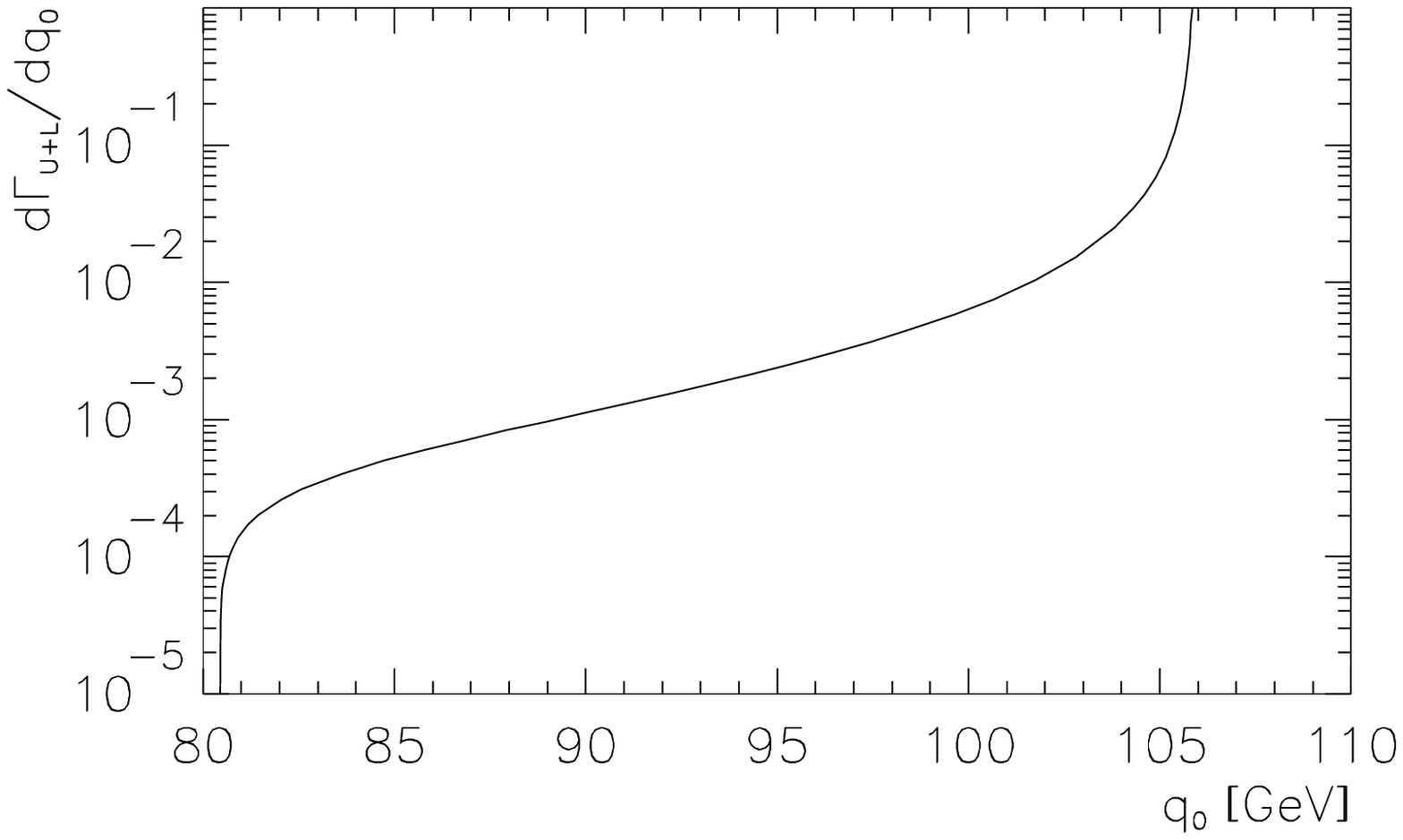}}

\vspace{0.4truecm} \centerline{\Large\bf Figure 4}

\begin{quote}
 Differential $ W $-boson energy distribution $ d\Gamma_{U+L}/dq_0 $
 for the total rate resulting from $ O(\alpha_s) $ gluon emission
 ($ m_b = 4.8 \mbox{ GeV} $).
\end{quote}

\centerline{\includegraphics[width=11.8cm,clip=]{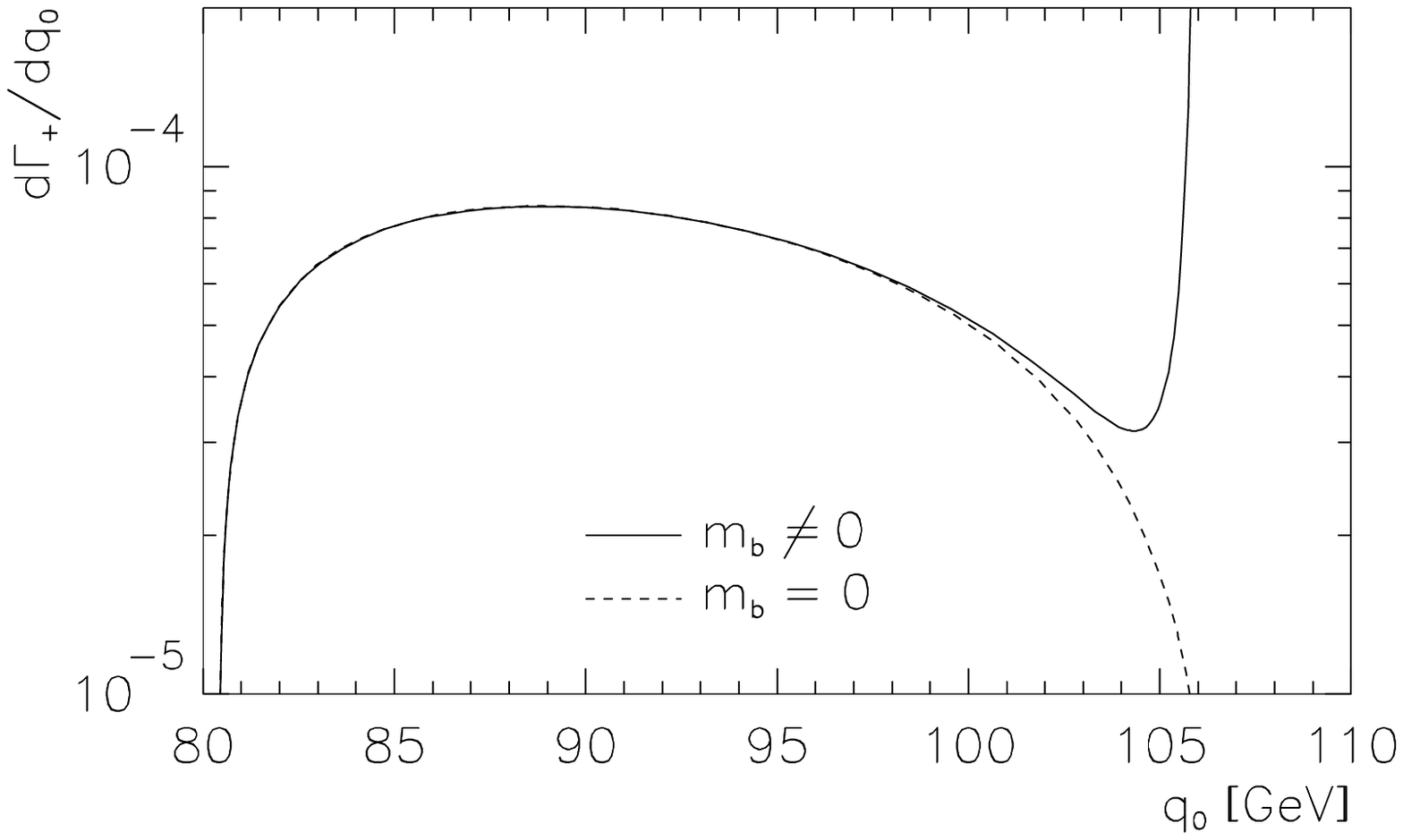}}

\vspace{0.4truecm} \centerline{\Large\bf Figure 5}

\begin{quote}
 Differential $ W $-boson energy distribution $d \Gamma_+/dq_0$ for the partial
 rate into positive helicity $ W $-bosons resulting from $ O(\alpha_s) $ gluon
 emission for $ m_b = 4.8 \mbox{ GeV} $ (solid line) and for $ m_b = 0 $ (dashed
 line).
\end{quote}

\clearpage


\newpage \thispagestyle{empty}

\centerline{\includegraphics[width=12cm,clip=]{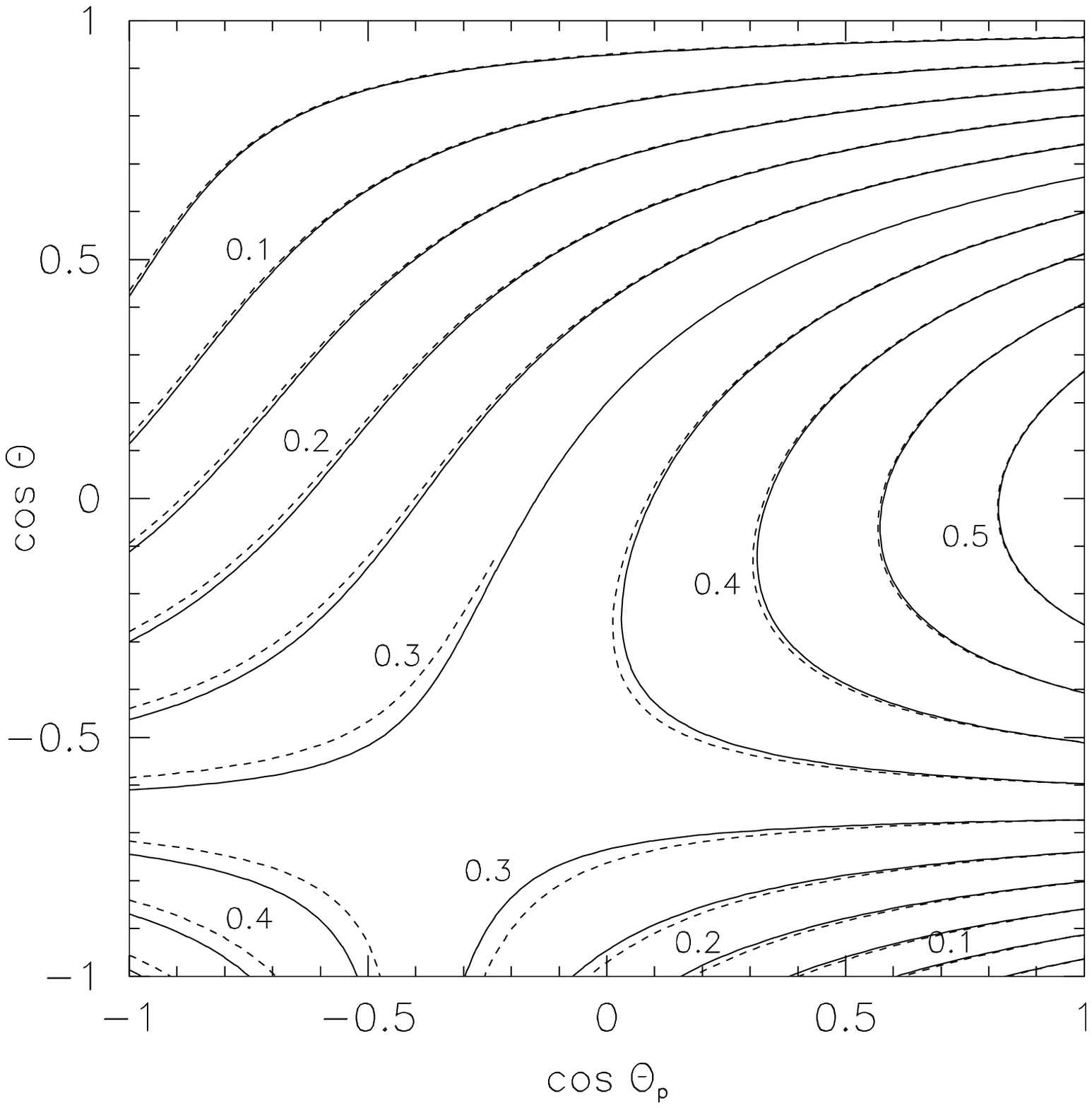}}

\vspace{1truecm} \centerline{\Large\bf Figure 6}

\begin{quote}
 Contours of the decay distribution of a fully polarized ($ P = 1 $)
 top quark in the $ \cos \theta_p $ - $ \cos \theta $ plane for $ m_b = 0 $.
 The full lines are the distribution including the $ O(\alpha_s) $ corrections. 
\end{quote}

\clearpage


\newpage \thispagestyle{empty}

\begin{center}
\begin{picture}(120,120)
 \put(000,054.5){\rotatebox{90}{$ \widehat{} $}}
 \includegraphics[width=12cm,clip=]{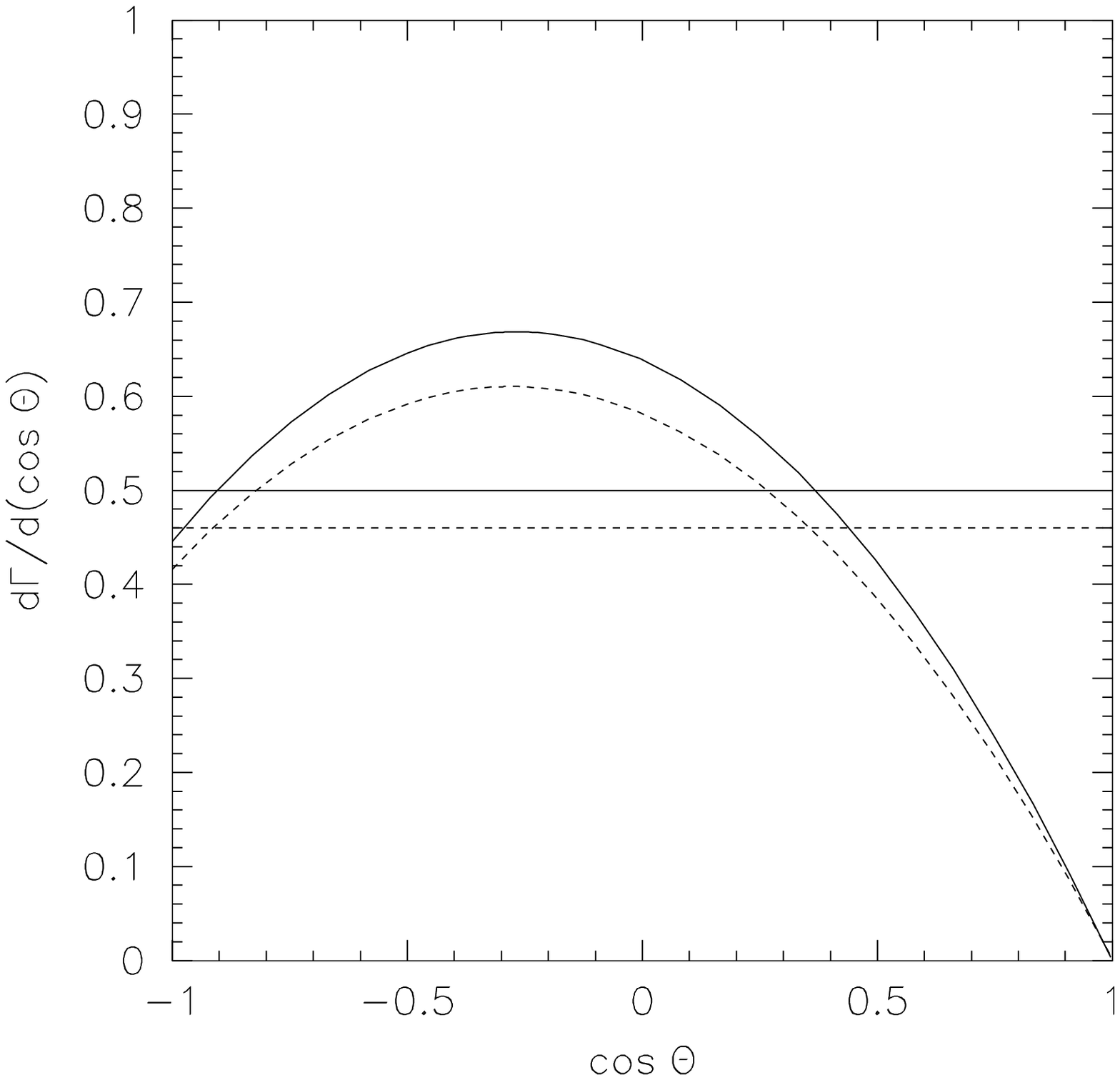}
\end{picture}
\end{center}

\vspace{1truecm} \centerline{\Large\bf Figure 7}

\begin{quote}
 Charged lepton polar angular distribution in the $ W $ rest frame for
 $ m_b = 0 $ (Born term: full line; $ O(\alpha_s) $: dashed line). Also
 shown are average values of the decay distribution.
\end{quote}

\clearpage


\newpage \thispagestyle{empty}

\begin{center}
\begin{picture}(120,120)
 \put(000,059){\rotatebox{90}{$ \widehat{} $}}
 \includegraphics[width=12cm,clip=]{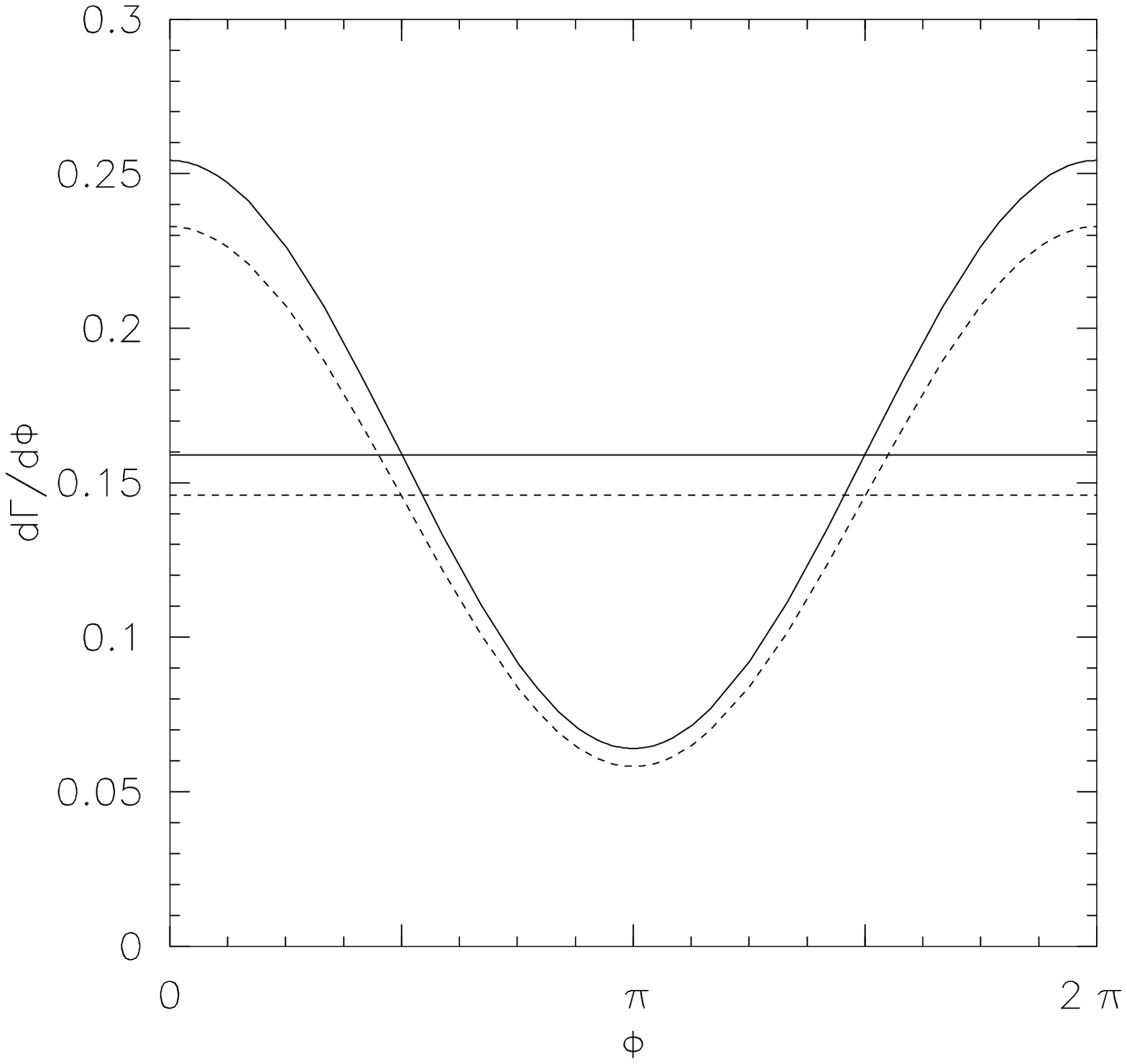}
\end{picture}
\end{center}

\vspace{1truecm} \centerline{\Large\bf Figure 8}

\begin{quote}
 Azimuthal distribution of normalized rate for $ m_b = 0 $
 (Born term: full line; $ O(\alpha_s) $: dashed line). Also shown
 are average values of the decay distribution.
\end{quote}

\clearpage


\newpage \thispagestyle{empty}

\centerline{\includegraphics[width=9.0cm,clip=]{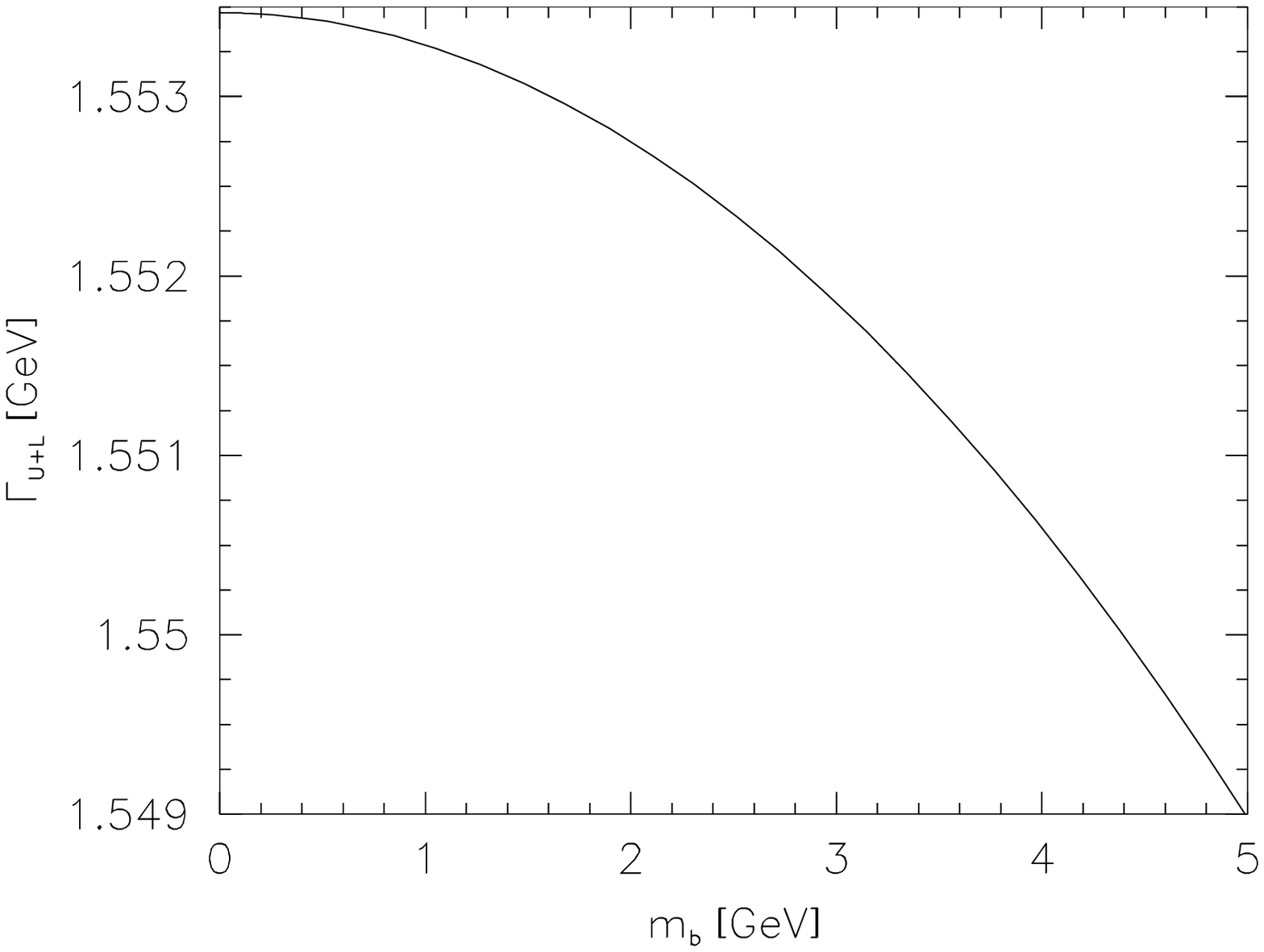}}
\centerline{\includegraphics[width=9.0cm,clip=]{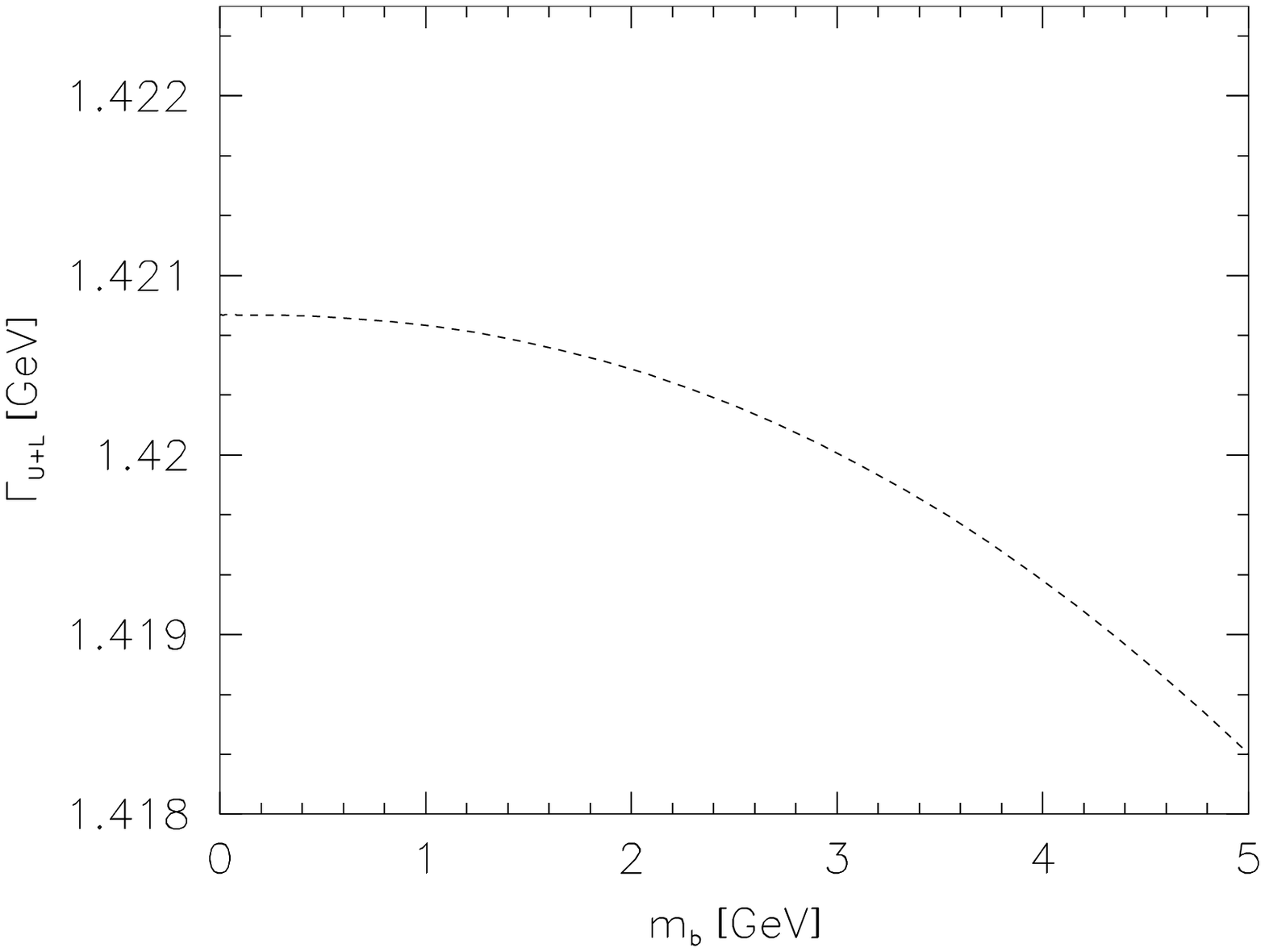}}

\vspace{0.5truecm} \centerline{\Large\bf Figure 9}

\begin{center}
 Bottom mass dependence of the total rate $ \Gamma_{U + L} $ \\
 (Born term: full line; $ O(\alpha_s) $: dashed line).
\end{center}

\clearpage


\newpage \thispagestyle{empty}

\centerline{\includegraphics[width=9.0cm,clip=]{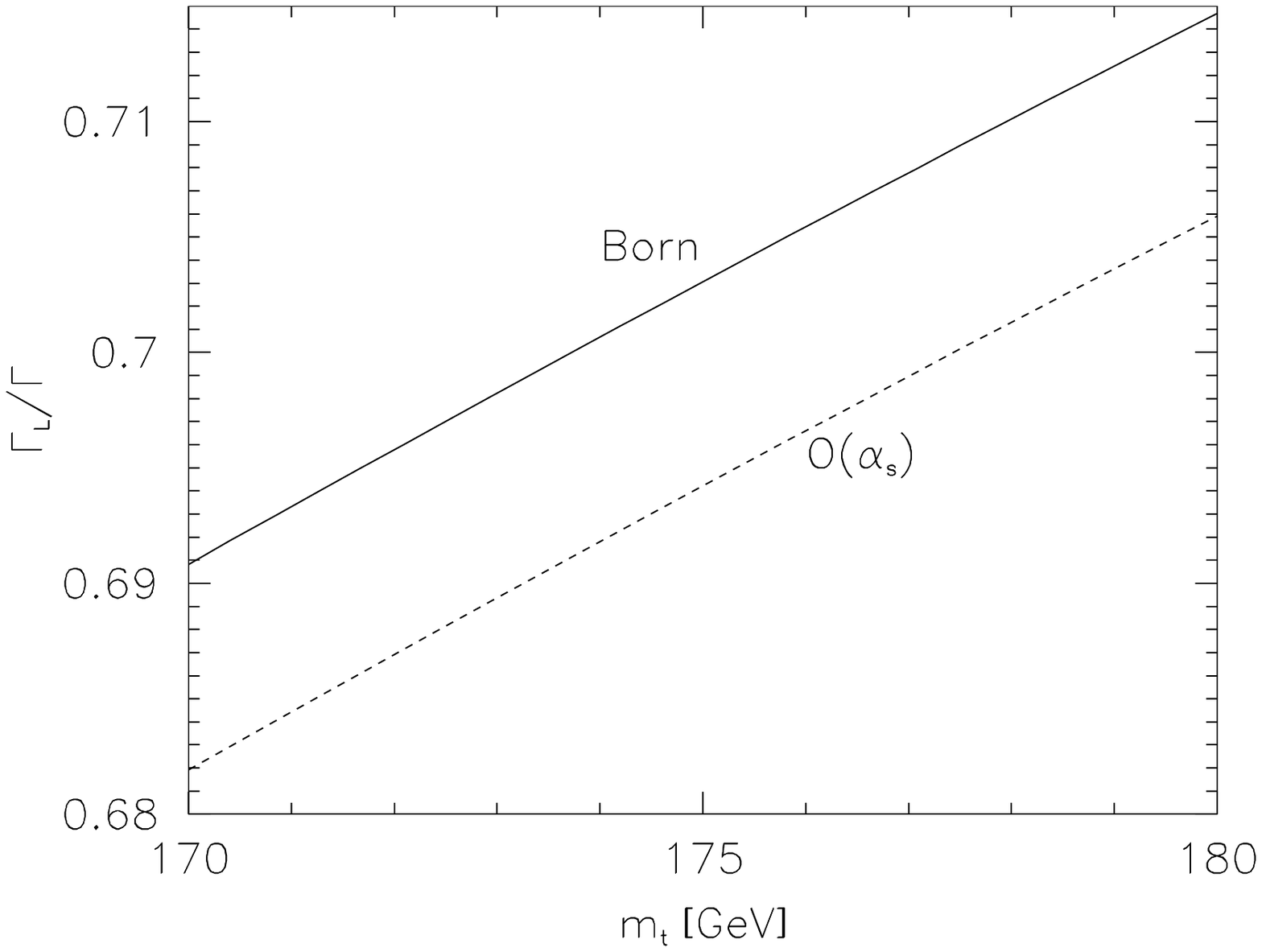}}

\vspace{0.5truecm} \centerline{\Large\bf Figure 10}

\begin{center}
 Top mass dependence of the rate ratio $ \Gamma_L / \Gamma_{U + L} $
 for $ m_b = 0 $ \\
 (Born term: full line; $ O(\alpha_s) $: dashed line).
\end{center}

\centerline{\includegraphics[width=9.0cm,clip=]{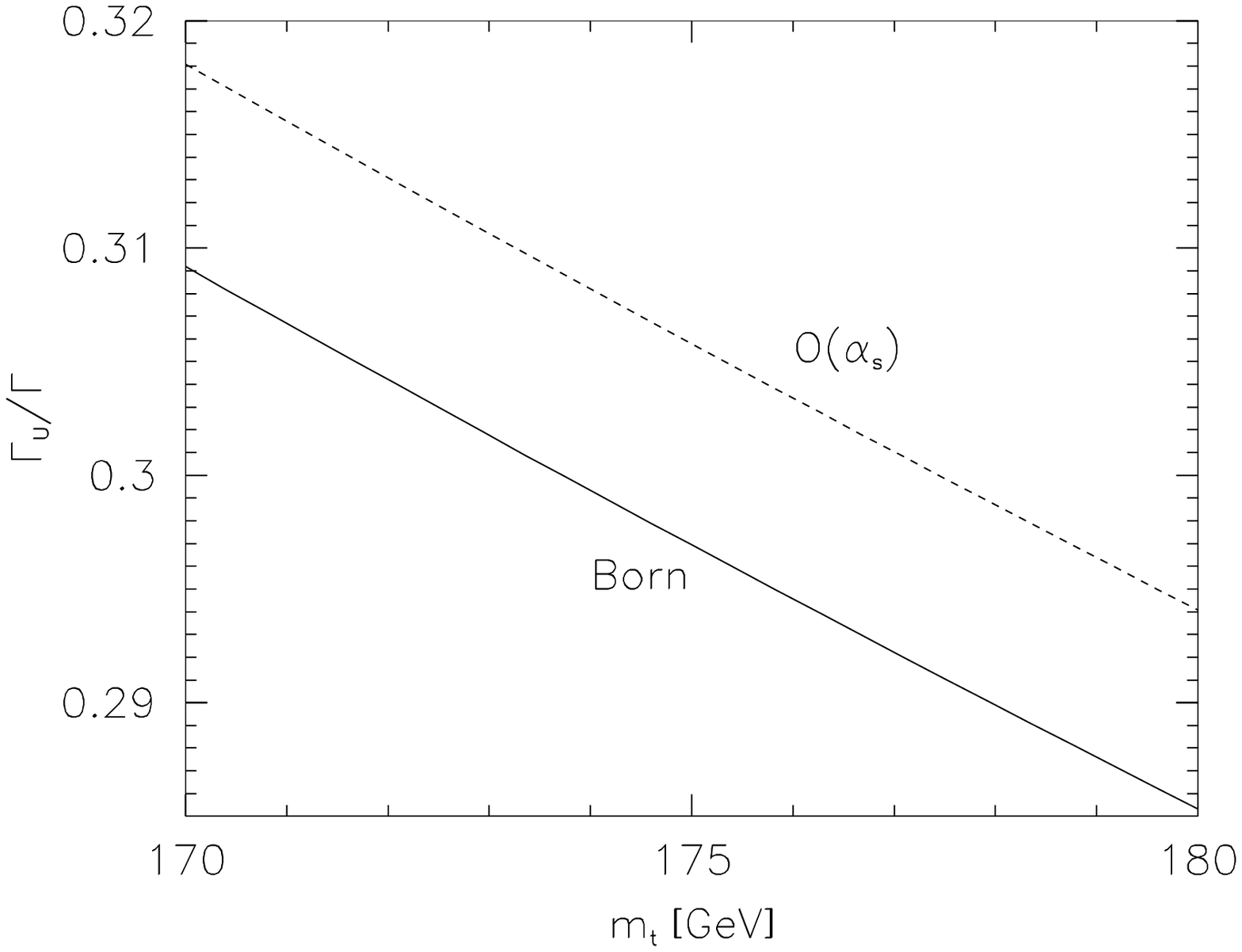}}

\vspace{0.5truecm} \centerline{\Large\bf Figure 11}

\begin{center}
 Top mass dependence of the rate ratio $ \Gamma_U / \Gamma_{U + L} $
 for $ m_b = 0 $ \\
 (Born term: full line; $ O(\alpha_s) $: dashed line).
\end{center}

\clearpage


\end{document}